\RequirePackage{lineno} 
\documentclass[twocolumn,epjc3]{svjour3}[12/10/2013]

\usepackage{atlasphysics}
\usepackage{amsmath}
\usepackage{amssymb}
\usepackage{graphicx}
\usepackage{placeins}
\usepackage{xspace}
\usepackage[hyperindex,breaklinks]{hyperref}
\hypersetup{colorlinks=true,linkcolor={blue},urlcolor={blue},citecolor={blue}}

\usepackage{preprintcover}  
\PreprintCoverPaperTitle{Muon reconstruction efficiency and momentum resolution of the ATLAS experiment in proton--proton collisions at $\sqrt{s}=7$~TeV in 2010}  
\PreprintIdNumber{CERN-PH-EP-2013-154}  
\PreprintCoverAbstract{This paper presents a study of the performance of the muon reconstruction in the analysis of proton--proton collisions at $\sqrt{s}=7$~TeV at the LHC, recorded by the ATLAS detector in 2010.
This performance is described in terms of reconstruction and isolation efficiencies and momentum resolutions for different classes of reconstructed muons.
The results are obtained from an analysis of $J/\psi$ meson and $Z$ boson decays to dimuons, reconstructed from a data sample corresponding to an integrated luminosity of 40~\ipb.
The measured performance is compared to Monte Carlo predictions and deviations from the predicted performance are discussed.
}  
\PreprintJournalName{EPJC}  

\include{macros}

\newcommand{\Muid}{\rm{chain~2}\xspace}
\newcommand{\Staco}{\rm{chain~1}\xspace}

\newcommand{\TisolRfour}{T_{\mathrm{isol}}^{\Delta R<0.4}}
\newcommand{\TisolRthree}{T_{\mathrm{isol}}^{\Delta R<0.3}}
\newcommand{\mySect}{Sect.}   
\newcommand{\mySects}{Sects.} 
\newcommand{\myFig}{Fig.}     
\newcommand{\myFigs}{Figs.}   

\newcommand{\myEqs}{Eqs.}

\begin{document}
\title{Muon reconstruction efficiency and momentum resolution of the ATLAS experiment in proton--proton collisions at $\sqrt{s}=7$~TeV in 2010}
\author{The ATLAS Collaboration}
\maketitle
\begin{abstract} 
This paper presents a study of the performance of the muon reconstruction in the analysis of proton--proton collisions at $\sqrt{s}=7$~TeV at the LHC, recorded by the ATLAS detector in 2010. 
This performance is described in terms of reconstruction and isolation efficiencies and momentum resolutions for different classes of reconstructed muons.
The results are obtained from an analysis of $J/\psi$ meson and $Z$ boson decays to dimuons, reconstructed from a data sample corresponding to an integrated luminosity of 40~\ipb. 
The measured performance is compared to Monte Carlo predictions and deviations from the predicted performance are discussed.
\end{abstract}

\section{Introduction}

One of the main components of the ATLAS detector is its huge Muon Spectrometer (MS).
It is based on the use of three very large air core toroidal magnets, each containing eight superconducting coils, and three measuring planes of high-precision chambers. This system is designed for efficient muon detection even in the presence of very high particle backgrounds and for excellent muon momentum resolution up to very high momenta of $\sim{}1$\,TeV. This unprecedented stand-alone performance of the ATLAS muon spectrometer is due to the large field integral (ranging between 2 and 6\,Tm for most of the detector), the very low multiple scattering  in the material of the air core toroids (1.3 units of radiation length over a large fraction of the acceptance in the barrel toroid), the very high precision measurements along the muon trajectory (chamber resolution $35\,\mu$m) and the extreme alignment precision of the measuring planes (30 $\mu$m).

The other very important component of the muon identification and measurement in ATLAS is the inner detector (ID) that complements the performance of the MS at momenta below $\sim\!100\,$GeV. In ATLAS the very efficient muon detection and high momentum resolution, with nominal relative momentum resolutions of $<3.5\%$ up to transverse momenta $\pt\sim200$~GeV and $<10\%$ up to $\pt\sim1$~TeV, are obtained by a combination of measurements from the ID and the MS~{\cite[p.162]{CSCBook}}. The complementarity of these measurements can be exploited to provide measurements of the muon reconstruction efficiencies in both tracking systems. In this paper, the muon reconstruction efficiencies are measured using dimuon decays of $J/\psi$ mesons to access the region $\pt<10$~GeV and dimuon decays of $Z$ bosons to access the region $20$~GeV$<\pt<100$~GeV. The efficiency determination in the region $10$~GeV$<\pt<20$~GeV is not possible due to the limited sample of muons with $\pt$ higher than 10 GeV in the $J/\psi$ decays and difficulties in controlling the backgrounds in the sample of $Z$ decays that lead to muons with $\pt$ smaller than 20 GeV.
For these analyses, one of the decay muons is reconstructed in both detector systems and the other is reconstructed by just one of the systems in order to probe the efficiency of the other. This method (known as {\it tag-and-probe}, and described in more detail in \mySect~\ref{sec::tag_and_probe_method}) is applied to the ATLAS proton--proton ($pp$) collision data recorded at the Large Hadron Collider (LHC) in 2010 at a centre-of-mass energy of 7~TeV. 

Muon isolation criteria are used to select muons in many physics analyses, and measurements of the isolation efficiency performed using $Z\to\mu^+\mu^-$ decays are described in \mySect~\ref{sec::iso_eff}. 
The invariant mass distributions from these data are also used to extract the muon momentum resolutions.
The analysed data sample corresponds to the full 2010 $pp$ dataset with an integrated luminosity of 40~\ipb
\cite{AtlasLumi2013} after applying beam, detector and data-quality requirements.

\section{The ATLAS detector}
\label{sec::ATLAS_detector}
A detailed description of the ATLAS detector can be found elsewhere~\cite{AtlasDetectorPaper}.
Muons are independently measured in the ID and in the MS. 

The ID measures tracks up to $|\eta|=2.5$ \footnote{ATLAS uses a right-handed coordinate system with its origin at the nominal interaction point (IP) in the centre of the detector and the $z$-axis along the beam pipe. The $x$-axis points from the IP to the centre of the LHC ring, and the $y$-axis points upward. Cylindrical coordinates $(r,\phi)$ are used in the transverse plane, $\phi$ being the azimuthal angle around the beam pipe. The pseudorapidity is defined in terms of the polar angle $\theta$ as $\eta=-\ln\tan(\theta/2)$.}  exploiting three types of detectors operated in an axial magnetic field
of $2$~T:  three layers of silicon pixel detectors closest to the interaction point, four layers of 
semiconductor microstrip detectors (SCT) surrounding the pixel detector, and a transition radiation straw-tube tracker (TRT) covering $|\eta|<2.0$ as the outermost part. The innermost pixel layer (known as the \emph{b-layer}) has a radius of $50.5$~mm in the barrel, whilst the outermost TRT tubes are at $r\approx 1$~m.
 
The electromagnetic and hadronic calorimeters surround the ID and cover
the pseudorapidity range $|\eta|<4.9$, far beyond the range over which muons are identified.
In the barrel and end-cap, in the region $|\eta|<3.2$ the electromagnetic calorimeter
consists of lead absorbers with liquid-argon (LAr) as active material.
The barrel hadronic tile calorimeter is a steel/scintillating-tile detector
and is extended by two end-caps with LAr as the active material and copper as absorber.
The total combined thickness of 11 interaction lengths ($\lambda$) includes
9.7\,$\lambda$ of active calorimeter and 1.3\,$\lambda$ of outer support.%
\begin{figure*}[t]
\begin{center}
    \includegraphics[width=0.8\linewidth]{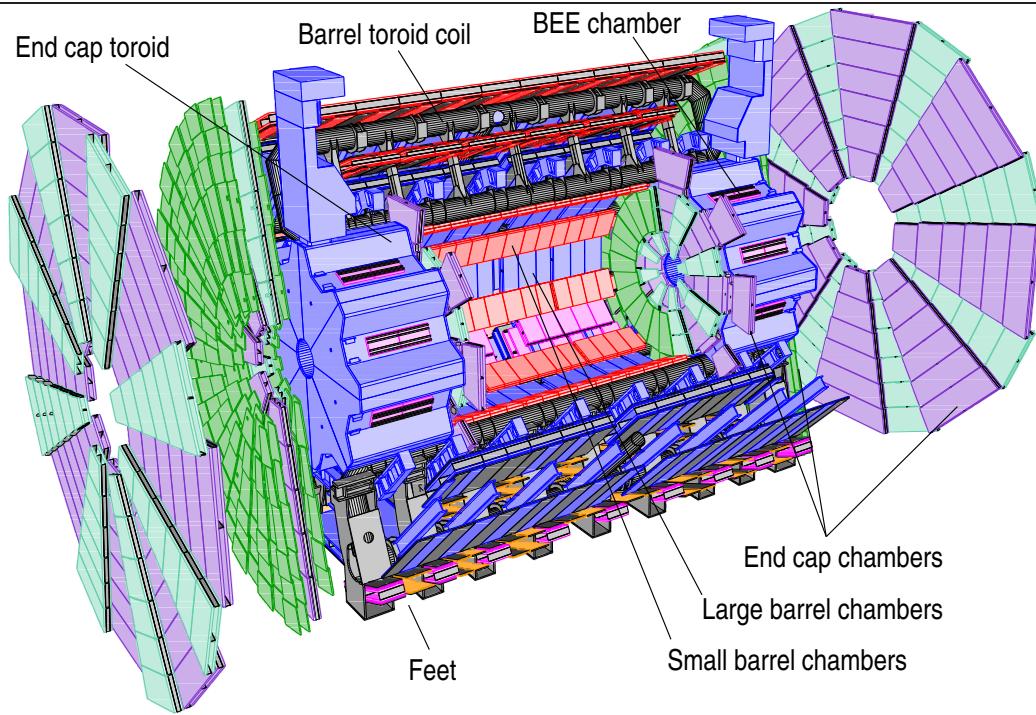}
    \caption{\label{det_schematic}Schematic drawing of the ATLAS muon spectrometer.}
\end{center}
\end{figure*}%

The magnetic field of the MS is produced by three large air-core superconducting toroidal magnet systems (two end-caps, where the average field integral is about 6~Tm, and one barrel, where the field integral is about 2.5~Tm). The field is continuously monitored by approximately 1800 Hall sensors distributed throughout the spectrometer volume. The deflection of the muon trajectory in this magnetic field
is measured via hits in three layers of precision monitored drift tube (MDT) chambers for
$|\eta|<2.0$ and two outer layers of MDT chambers in combination with one layer
of cathode strip chambers (CSCs) in the innermost end-cap wheels ($2.0\leq|\eta|<2.7$). Three layers of resistive plate chambers (RPCs) 
in the barrel ($|\eta|<1.05$) and three layers of thin gap chambers (TGCs) in the end-caps ($1.05<|\eta|<2.4$) are used by the
muon trigger (see below). The RPCs, TGCs and  CSCs  also measure the muon trajectory in the non-bending ($\phi$) plane of the spectrometer magnets. The following text frequently refers to chambers which make a measurement in the bending ($\eta$) plane as `precision chambers', since these have a much better spatial resolution (important for a good momentum resolution) than the chambers used for triggering. 

The chambers are monitored by an optical alignment system, designed to provide an accuracy of 30~$\mu$m in the barrel and 40~$\mu$m in the end-cap~\cite{MuonEndcapAlignment}.

The ATLAS detector has a three-level trigger system: level 1 (L1), level 2 (L2), and the event filter (EF). The MS
provides a L1 hardware muon trigger which is based on hit coincidences in different RPC and TGC detector layers within programmed geometrical windows which define the muon \pt. The L2 and EF muon triggers
perform a software confirmation of the L1 muon trigger using refined \pt~measurements from the precision chambers.

Figure \ref{det_schematic} shows a schematic drawing of the ATLAS MS. The barrel muon chambers are installed around the calorimeters in roughly cylindrical rings of approximately 5, 7 and 9~m radius. Large barrel chambers are mounted between the barrel toroid coil cryostats. Small barrel chambers are installed on the toroid coil cryostats. The barrel end-cap extra (BEE) chambers are mounted on the end-cap toroid cryostats. The end-cap chambers are arranged in disks with $z$-axis positions of approximately 7, 13 and 21~m from the centre of the detector, and which are orthogonal to the proton beams. 

\section{Muon reconstruction and identification in ATLAS \label{sec::muon_identification}}

Muon identification in ATLAS uses independent track reconstruction in the ID and MS, which are then combined.
Track reconstruction in the muon spectrometer is logically subdivided into the following stages: pre-processing of raw data to form drift-circles in the MDT’s or clusters in the CSC’s and the trigger chambers, pattern-finding and segment-making, segment-combining, and finally track-fitting. Track segments are defined as straight lines in a single MDT or CSC station. The search for segments is seeded by a reconstructed pattern of drift-circles and/or clusters.

Full-fledged track candidates are built from segments, typically starting from the outer and middle stations and extrapolating back through the magnetic field to the segments reconstructed in the inner stations (though other permutations are also explored). Each time a reasonable match is found, the segment is added to the track candidate. The final track-fitting procedure takes into
account all relevant effects (e.g. multiple scattering, field inhomogeneities, inter-chamber
misalignments, etc.). More details about the muon reconstruction can be found in Ref.~{\cite[p.165]{CSCBook}}.

A similar approach is followed by the ID track reconstruction where the pattern recognition uses space-points formed from the pixel and SCT clusters to generate track seeds. These seeds are then extended into the TRT and drift circles are associated. Finally the tracks are refitted with the information coming from all three detectors. More details about the ID track reconstruction can be found in Ref.~{\cite[p.19]{CSCBook}}.

The analyses presented here make use of three classes of reconstructed muons, as described below.
\begin{description}
\item[Stand-alone (SA) muon:] the muon trajectory is reconstructed only in  
the MS. The direction of flight and the impact parameter of the muon at the interaction point are determined by extrapolating the spectrometer track back to the point of closest approach to the beam line, taking into account the energy loss of the muon in the calorimeters.
\item[Combined (CB) muon:] track reconstruction is performed independently
in the ID and MS, and a combined track is formed from the successful combination
of a SA track with an ID track.
\item[Segment-tagged (ST) muon:] a track in the ID is identified
        as a muon if the track, extrapolated to the MS,
        is associated with at least one segment in 
        the precision muon chambers.
\end{description}

The main goal of this paper is the measurement of the reconstruction efficiencies and resolutions for combined (CB) and combined-plus-segment-tagged (CB+ST) muons, for which the use of the ID limits the acceptance to $|\eta|<2.5$. Stand-alone muons are employed to measure the muon reconstruction efficiency in the ID.

\bigskip
The CB muon candidates constitute the sample with the highest
purity.
The efficiency for their reconstruction is strongly affected
by acceptance losses in the MS, mainly in the two following regions:
\begin{itemize}
   \item at $\eta\sim0$, the MS is only partially equipped with muon 
         chambers in order to provide space for services of the ID
         and the calorimeters;
   \item in the region ($1.1<|\eta|<1.3$) between the barrel and the end-caps, 
         there are regions in $\phi$ where only one layer of chambers is traversed by
         muons in the MS, due to the fact that some chambers were not yet installed
         in that region during the 2010-2012 data-taking. Here no stand-alone momentum measurement is 
         available and the CB muon reconstruction efficiency is decreased.
\end{itemize}

The reconstruction algorithms for ST muons have higher efficiency than those for CB muons as they can recover muons which did not cross enough precision chambers to allow an independent momentum measurement in the MS.
They are also needed for the reconstruction of low-$p_\mathrm{T}$ muons which only reach the innermost layer of the muon chambers. Due to their lower purity and poorer momentum resolution, ST muons are only used in cases where no CB muon can be reconstructed.

In the early phase of the LHC operation, ATLAS used two entirely independent strategies for the reconstruction of both the CB and ST muons. These two approaches, known as \emph{chain~1} and \emph{chain~2} in the following, provide an invaluable cross-check on the performance of a very complex system, and allow ATLAS to ultimately take the best aspects of both. The chains have slightly different operating points, with chain~1 typically more robust against background, whilst chain~2 has a slightly higher efficiency.

In chain~1, the momentum of the muon is obtained from a statistical combination of the parameters of the tracks reconstructed by the ID and MS~\cite[p.166]{CSCBook}. SA muon tracks are required to have a sufficient number of hits in the precision and trigger chambers, to ensure a reliable momentum measurement. In chain~2, the combined muon momentum is the result of a simultaneous track fit to the hits in the ID and the MS. The requirements applied to the hit multiplicities in the MS are less stringent than in chain~1 because certain information, such as the trajectory in the plane transverse to the proton beams, is better provided by the ID in the simultaneous fit. In both chains, muon track segments can additionally be assigned to ID tracks to form ST muons, based on the compatibility of the segment with the extrapolated ID track. 

\begin{figure}[]
  \begin{center}
    \vspace*{-0.6ex}\hspace*{-2.0ex}
    \includegraphics[width=1.06\linewidth]{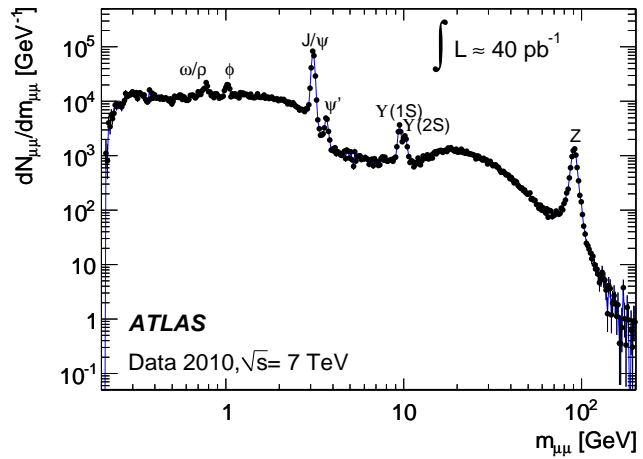}\hspace*{-0.2ex}
    \caption{\label{fig:dimuon_mass}Reconstructed invariant mass, $m_{\mu\mu}$, distribution of
     muon candidate pairs. The number of events is normalised by the bin width.
     The uncertainties are statistical only.}
\end{center}
\end{figure}
To illustrate the high purity of the ATLAS muon identification and
the size of the dimuon dataset, Fig.~\ref{fig:dimuon_mass} shows
the reconstructed invariant mass distribution of opposite-sign muon candidate pairs. The events
are selected by an unprescaled, 15 GeV \pt{} threshold single muon trigger,
which is reconfirmed offline by requiring at least one muon to have $\pt\!>\!15$\,GeV.
Both muons are required to be of CB type and to pass the ID track selection criteria 
of Sect.~\ref{sec:IDtrkselection}.
The distance of closest approach of the muon to the primary vertex is limited to 5\,mm
in the transverse plane and 200\,mm/$\sin\theta$ in the longitudinal direction.
The \jpsi, $\Upsilon$ and $Z$ peaks are clearly visible, and the muon reconstruction
has the capability to resolve close-by resonances, such as the \jpsi{} and $\psi^\prime$
as well as the $\Upsilon(1S)$ and $\Upsilon(2S)$.
The shoulder near $m_{\mu\mu}\!\approx\!15\,$GeV is caused by the kinematic selection.

\section{The tag-and-probe method \label{sec::tag_and_probe_method}}
As track reconstruction is performed independently in the ID and MS, the reconstruction efficiency for CB or ST muons is the product of the muon reconstruction efficiency in the ID, the reconstruction efficiency in the MS, and the matching efficiency between the ID and MS measurements (which includes the refit efficiency in the case of chain 2). It is therefore possible to study the full reconstruction efficiency by measuring these individual contributions. A tag-and-probe method is employed, which is sensitive to either the ID efficiency or the combined MS and matching efficiency.\footnote{Efficiencies determined with the tag-and-probe method, and with an alternative method based on Monte Carlo generator-level information, were found to agree to within statistical uncertainties~\cite[p.221]{CSCBook}, which also shows that any possible correlations between the tag and probe muons are negligible.}
This technique is applied to samples of dimuons from the \jpsi{} and $Z$ decays.

\begin{figure}[]
\begin{center}
\includegraphics[width=0.98\linewidth]{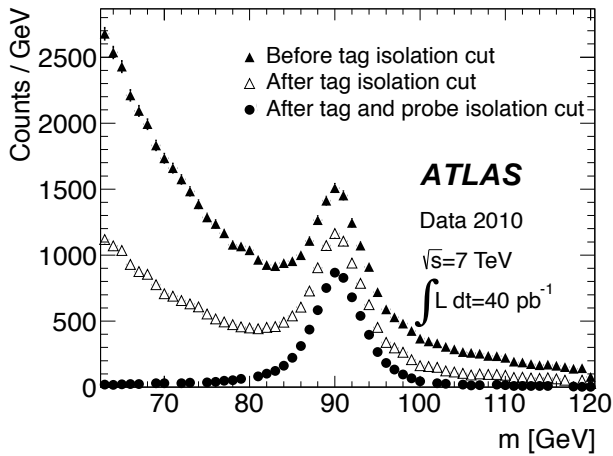}
    \caption{\label{fig2p}Invariant mass, $m$, distribution of
             pairs of tag muons (chain 2) and ID track probes for
             different sets of muon isolation requirements for the $Z$ boson analysis,
             as indicated in the legend.}
\end{center}
\end{figure}

\begin{figure}[]
\begin{center}
\includegraphics[width=0.98\linewidth]{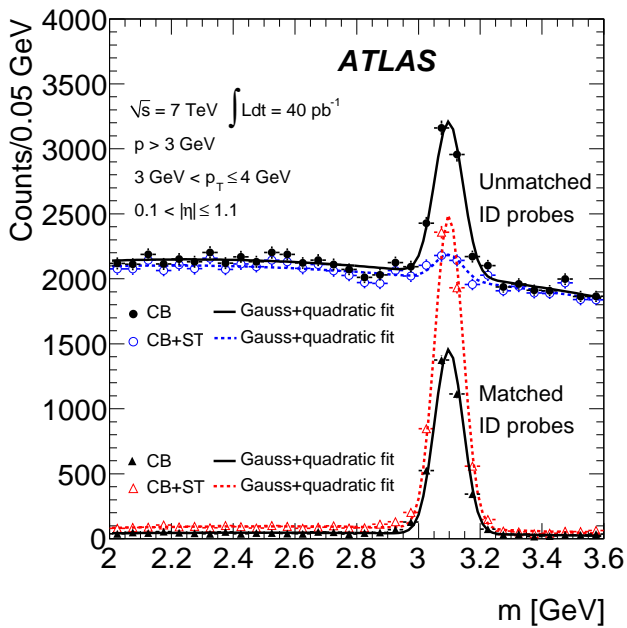}
    \caption{\label{fig2}
    Distribution of the invariant mass, $m$, of the unmatched (upper distributions) and matched (lower distributions) tag-and-probe pairs for CB and CB+ST muons of chain 2, for the \jpsi{} analysis with a probe muon selection as described in the legend.
    Also shown are the results of the fit using a Gaussian signal and a quadratic background contribution.}
\end{center}
\end{figure}

For $Z\to\mu^+\mu^-$ decays, events are selected by requiring two oppositely charged isolated tracks with a dimuon invariant mass near the mass of the $Z$ boson. One of the tracks is required to be a CB muon candidate, and to have triggered the readout of the event (see \mySect~\ref{sec::tag_and_probe_selection}). This muon is called the {\itshape tag}. The other track, the so-called {\itshape probe}, is required to be a SA muon if the ID efficiency is to be measured. If the MS reconstruction and matching efficiency is to be measured the probe must be an ID track. The ID reconstruction efficiency is defined as the fraction of SA probes which can be ascribed to an inner detector track. The combined MS and matching efficiency is the fraction of ID probes which can be associated to a CB or ST muon. 

The invariant mass spectra of $Z$ boson tag-and-probe pairs, shown in \myFig~\ref{fig2p}, illustrate how muon isolation requirements (see \mySects~\ref{sec::tag_and_probe_selection} and \ref{sec::iso_eff}) almost entirely remove contributions from background processes, resulting in a relatively pure sample of muon tag-and-probe pairs. Monte Carlo studies show that the contribution from other sources is below 0.1\% when MS probes are used and below 0.7\% when ID probes are used. These backgrounds arise from  $Z\to\tau^+\tau^-$,
$W^\pm\!\to\!\mu^{\pm\,}{}^{\scriptscriptstyle(\!}_{\phantom{(\!}}
\bar{\nu}^{\scriptscriptstyle)}_{\raisebox{0.3ex}{$\scriptstyle\mu$}\phantom{)}}$
$W^\pm\!\to\!\tau^{\pm\,}{}^{\scriptscriptstyle(\!}_{\phantom{(\!}}
\bar{\nu}^{\scriptscriptstyle)}_{\raisebox{0.3ex}{$\scriptstyle\tau$}\phantom{)}}$
$b\bar{b}$, $c\bar{c}$, and $t\bar{t}$.
The presence of backgrounds in the data leads to an apparent decrease in the muon efficiency in the range $\pt\lesssim30$~GeV, for both reconstruction chains. This is taken into account by comparing the measured efficiencies to efficiencies predicted using simulated samples which include these background contributions. 

To investigate the reconstruction efficiency at lower transverse momenta, dimuon pairs from $J/\psi\to\mu^+\mu^-$ decays
are used in the same way as those from $Z\to\mu^+\mu^-$ decays. Because $J/\psi$ mesons are produced inside jets, isolation requirements cannot be used to select a pure sample. In this case, the invariant mass distribution of the tag-and-probe pairs is fitted using the sum of a quadratic background term and a Gaussian signal term~\cite{JPsiObservation}. This is illustrated in \myFig~\ref{fig2} for probe muons selected in the range $0.1<|\eta|<1.1$ and $3\mathrm{\,GeV}<p_{\mathrm{T}}<4$\,GeV. The invariant mass spectra are shown for tag-and-probe pairs in which the probes are matched to reconstructed muons (see \mySect~\ref{sec::tag_and_probe_selection_matching}) and for unmatched tag-and-probe pairs. The muon reconstruction efficiency is then extracted from a simultaneous fit to the distributions obtained from the matched and unmatched tag-and-probe pairs.

\section{Monte Carlo samples and expectations\label{sec::MC}}

The measurements presented in this paper are compared with predictions of Monte Carlo (MC) simulations. For the efficiency measurements in the region $\pt>20$~GeV, five million $Z\to\mu^+\mu^-$ events were simulated with PYTHIA 6.4~\cite{Pythia}, passed through the full simulation of the ATLAS detector~\cite{:2010wqa}, based on GEANT4~\cite{Geant42003,GEANT42006}, and reconstructed with the same reconstruction programs as the experimental data.

During the 2010 data taking, the average number of $pp$ interactions per
bunch crossing was about 1.5. This ``pile-up'' is modelled by overlaying
simulated minimum bias events on the original hard-scattering event.
It is found to have a negligible impact for these measurements.
The following background samples were used:
$Z\!\to\!\tau^+\tau^-$,
$W^\pm\!\to\!\mu^{\pm\,}{}^{\scriptscriptstyle(\!}_{\phantom{(\!}}
\bar{\nu}^{\scriptscriptstyle)}_{\raisebox{0.3ex}{$\scriptstyle\mu$}\phantom{)}}$
$W^\pm\!\to\!\tau^{\pm\,}{}^{\scriptscriptstyle(\!}_{\phantom{(\!}}
\bar{\nu}^{\scriptscriptstyle)}_{\raisebox{0.3ex}{$\scriptstyle\tau$}\phantom{)}}$
$b\bar{b}$, $c\bar{c}$, and $t\bar{t}$.  More details can be found in Ref.~{\cite{WZCrossSectionPaper}}.

The reconstruction efficiency at low $\pt$ was studied with a simulated
sample of five million prompt $\jpsi$ events
generated with PYTHIA using the PYTHIA implementation of the colour-octet model.
In order to increase the number of events at the higher end of the low-$\pT$ region, this sample was supplemented with a  sample of one million $pp\!\to\!b\bar{b}$ events also generated with PYTHIA, in which at least one $\jpsi$ decaying into muons of $\pT > 2.5$~GeV
was required in the $b$-quark decay chain. 

\begin{figure}[]
\begin{center}
\includegraphics[width=0.46\textwidth]{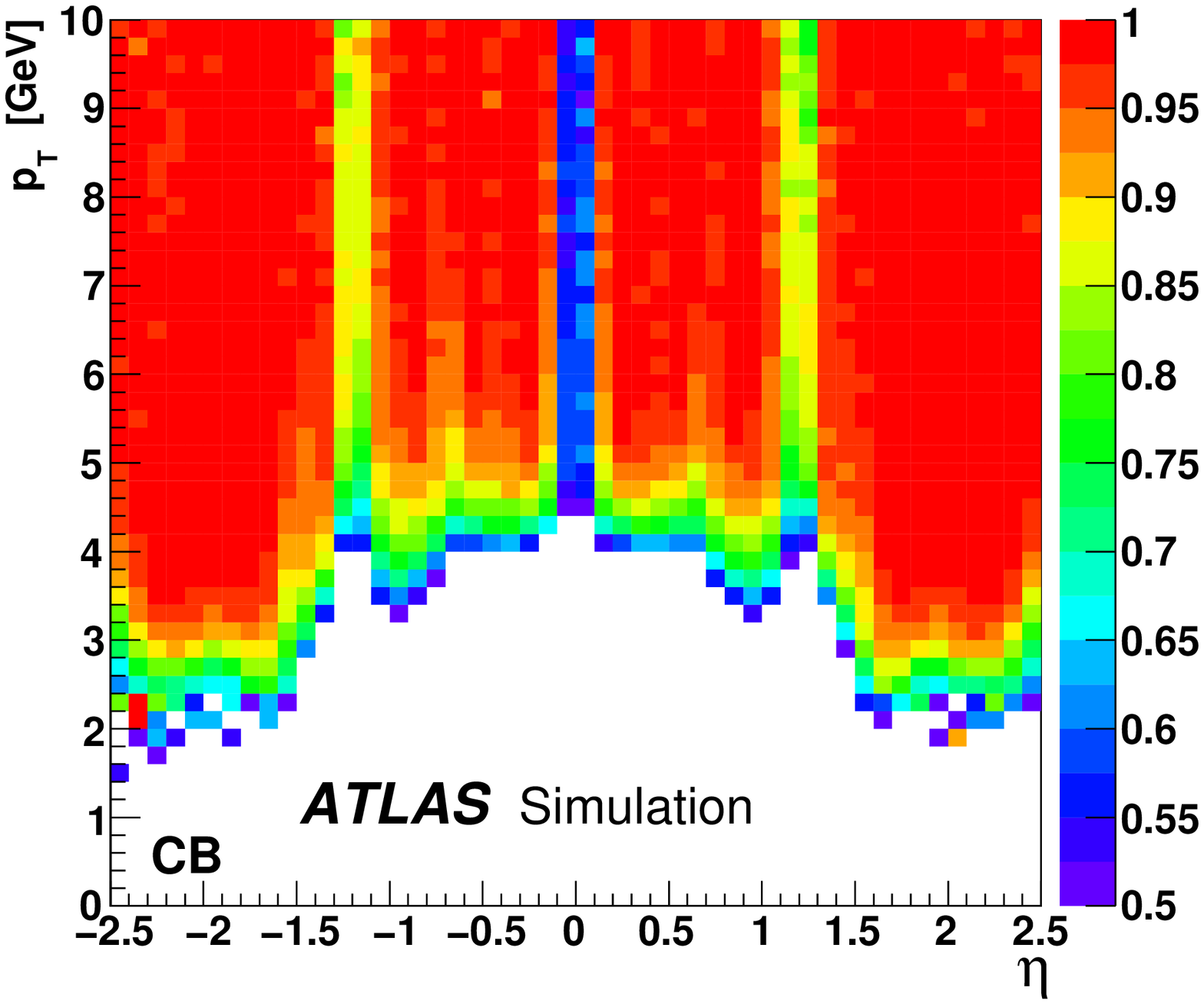}
\includegraphics[width=0.46\textwidth]{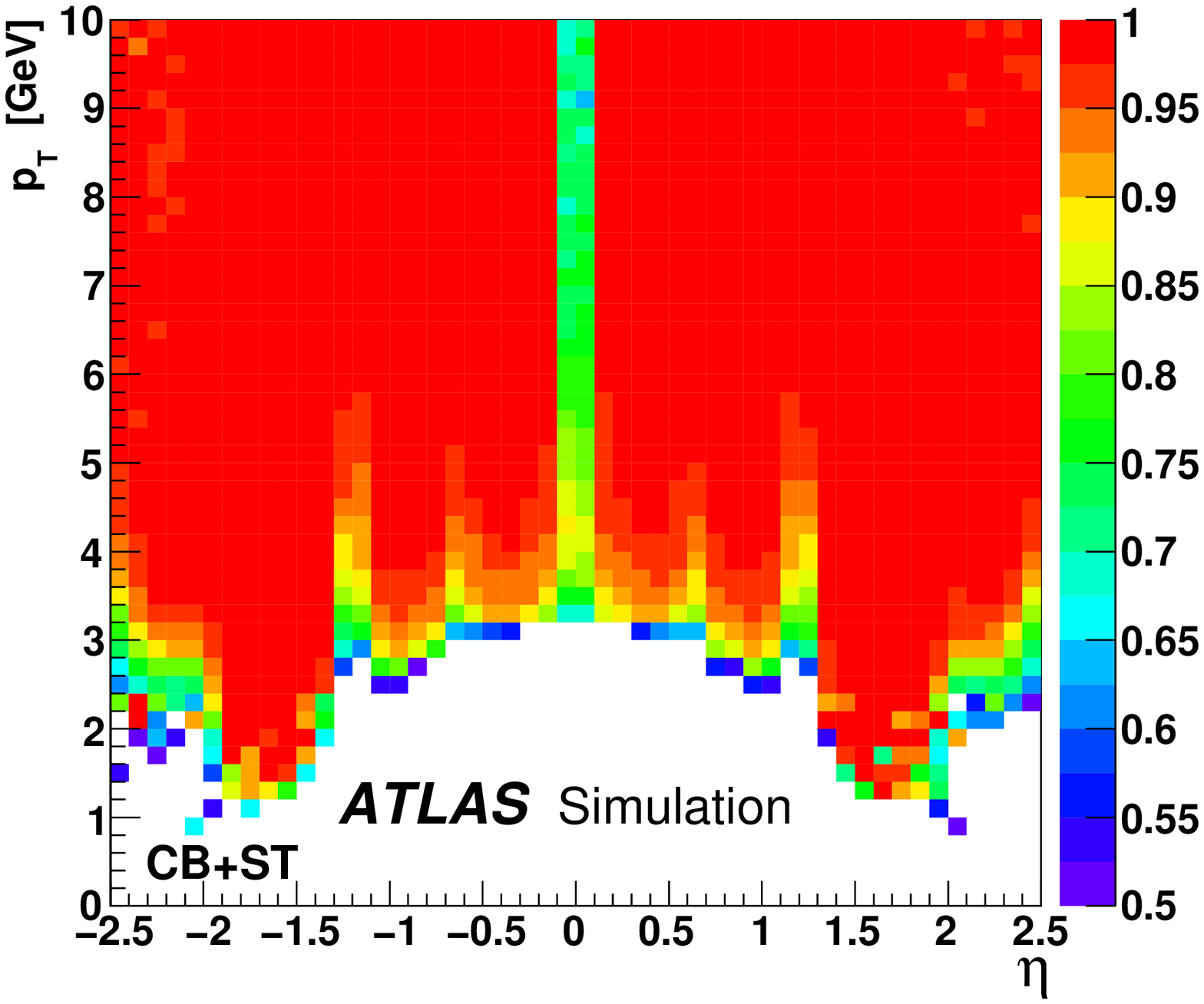}
\caption{The chain~1 muon reconstruction efficiency from simulated $\jpsi$ decays for 
CB  (top) and  CB+ST (bottom) muons as a function
of $\eta$ and $\pt$ for efficiency values above 50\%.}
\label{fig:sim}
\end{center}
\end{figure}

The reconstruction efficiencies obtained from the analysis of the $\jpsi$
Monte Carlo samples are shown in \myFig~\ref{fig:sim}, as a function of
$\pt$ and $\eta$, for CB and CB+ST muons from chain~1.
The most discernible features are the areas of lower efficiency at fixed $\eta$ that result
from the un-instrumented (`crack') region in the MS at $\eta \sim 0$ and from the barrel/end-cap transition regions  where the chamber configuration ($1.1<|\eta|<1.3$)
and the magnetic field ($1.1<|\eta|<1.7$) are rather non-uniform.
Also visible is the impact of the energy loss in the calorimeter on the efficiency, for muons with $\pT$ of less than 2--5~GeV (depending on the $\eta$ region), which are absorbed in the calorimeter.
For $|\eta|<2.0$, the CB+ST muon reconstruction starts to be efficient at $\pt$ values lower than in the reconstruction of pure CB muons, since it includes muons reaching only the inner layer of
MDT chambers.  For $|\eta|>2.0$ the CB and CB+ST
efficiencies are very similar for \Staco, because cases with only one segment in the CSC chambers, corresponding to the inner layer of
precision chambers in this region, are not considered for ST muons. Chain 2 does make use of these segments, and shows an improved CB+ST efficiency in this region (see \mySect~\ref{sec::Z_results_CBST}).
These detector features motivate the binning used for the determination of the $\pt$ dependence of the reconstruction efficiency at low $\pt$.

For the $J/\psi\!\to\!\mu^+\mu^-$ analysis the measured efficiencies are separated into five pseudorapidity intervals according to the different MS regions:\medskip

\begin{tabular}{@{\hspace*{-0ex}}rl}
$|\eta|<0.1$ &the $\eta=0$ crack region;\\
$0.1<|\eta|<1.1$ & the barrel region;\\
$1.1<|\eta|<1.3$ &the transition region \\
                  &between barrel and end-cap;\\
$1.3<|\eta|<2.0$ &the end-cap region;\\
$2.0<|\eta|<2.5$ &the forward region.\\
\end{tabular}

\smallskip

Muons from $Z\!\to\!\mu^+\mu^-$ decays were required to have $\pt>20$~GeV. In contrast to the case of lower-$\pt$ muons from $J/\psi$ decays, the $\phi$ deflections of these muons by the magnetic fields in the detector are so small that one can use the muon directions of flight at the $pp$ interaction point to associate them with specific ($\eta,\phi$) regions of the MS. Ten different regions are defined,
corresponding to ten different physical regions in the MS~\cite{AtlasDetectorPaper}. In each of these, the muon traverses a particular set of detector layers and encounters a different quality of detector alignment, a different amount of material or a different magnetic field configuration. The ten regions are described below (see also \myFig~\ref{det_schematic}).
\begin{itemize}
 \item \emph{Barrel large}: the regions containing large barrel chambers only, which are mounted between the barrel toroid coils.
 \item \emph{Barrel small}: the regions containing small barrel chambers only, which are mounted on the barrel toroid coils.
 \item \emph{Barrel overlap}: the regions where small and large barrel chambers have slight overlaps in acceptance.
 \item \emph{Feet}: the detector is supported by `feet' on its bottom half, which results in a loss of acceptance due to missing chambers, making muon reconstruction more challenging.
 \item \emph{Transition}: the region $1.1<|\eta|<1.3$, between the barrel and the end-cap wheels.
 \item \emph{End-cap small}: the small end-cap sectors, consisting of MDT chambers.
 \item \emph{End-cap large}: the large end-cap sectors, consisting of MDT chambers and which (in contrast to the \emph{Barrel large} regions) contain the toroid coils.
 \item \emph{BEE}: the regions containing barrel end-cap extra chambers, which are mounted on the end-cap toroid cryostats.
 \item \emph{CSC small}: the  end-cap sectors consisting of small CSC chambers.
 \item \emph{CSC large}: the  end-cap sectors consisting of large CSC chambers.
\end{itemize}

\section{Selection of tag-and-probe pairs \label{sec::tag_and_probe_selection}}

\subsection{Event selection}
The events used for the efficiency measurements were selected online with a single-muon trigger. For the studies with $J/\psi\!\to\!\mu^+\mu^-$ decays, a combined muon is required, with minimum $\pt$ thresholds of 4, 6, 10, or 13~GeV (as it was necessary to increase the thresholds during the year, in order to keep the trigger rate within limits). For the studies with $Z\!\to\!\mu^+\mu^-$ decays, events have to pass the lowest $\pt$ threshold muon trigger that was unprescaled. The thresholds of the selected triggers range from 10~GeV to 13~GeV, well below the transverse momentum threshold of the tag muon in the analysis. To suppress non-collision background events, a reconstructed collision vertex with at least three associated ID tracks is required.

\subsection{Inner detector track selection}
\label{sec:IDtrkselection}
Tracks in the ID are required to satisfy requirements on the number of
hits in the silicon detectors for qualifying as a muon candidate.
They must have at least two pixel hits, including at least
one in the b-layer, and at least six SCT hits.
In order to reduce inefficiencies due to known inoperative
sensors,\footnote{The fraction of inoperative sensors was
$\approx 3\%$ for the pixel detector and $<1\%$ for the SCT.}
the latter are counted as hits for tracks crossing them.
Within  $|\eta| < 1.9$, a good-quality extension of the muon trajectory into the TRT is enforced by requirements on the numbers of associated good TRT hits and TRT outliers.  
The TRT outliers appear in two forms in the track reconstruction: as straw tubes
with a signal from tracks other than the one in consideration,
or as a set of TRT measurements in the extrapolation of a track which fail
to form a smooth trajectory together with the pixel and SCT measurements.
The latter case is typical of a hadron decay-in-flight, and can be rejected
by requiring that the outlier fraction (the ratio of outliers to total TRT hits)
is less than 90\%.
In the region $|\eta| <1.9$ the sum of the numbers of TRT hits and outliers is required to be greater than five, with an outlier fraction less than 90\%. 
At higher $|\eta|$ the requirement on the total number of TRT hits and outliers is
not applied, but tracks which do pass it are also required to  pass the cut on the outlier fraction.
These quality cuts suppress fake tracks and discriminate against muons from $\pi / K $ decays. 

\subsection{Tag selection  \label{sec::tag_selection}}
For each of the two reconstruction chains, tag muons are defined as CB muons from the interaction vertex. Different selection cuts are applied for the measurements using $J/\psi\!\to\!\mu^+\mu^-$ and $Z\!\to\!\mu^+\mu^-$ decays to account for the different kinematics and final-state topologies. For the studies with $J/\psi\!\to\!\mu^+\mu^-$ a tag muon has to pass the following requirements:
\begin{itemize}
    \item   the tag muon triggered the readout of the event;
    \item   $\pT> 4$~GeV, $|\eta| < 2.5$;
    \item   the distance of closest approach of the muon to the primary vertex, in the
            transverse plane, has transverse coordinate $|d_0|<0.3$~mm, and longitudinal coordinate
            $|z_0|<1.5$~mm, and  significances\break
            $|d_0|/\sigma(d_0)<3$,
            $|z_0|/\sigma(z_0)<3$, respectively.
\end{itemize}
For the studies with $Z\to\mu^+\mu^-$ decays an additional quantity is used, namely track isolation

\begin{equation} 
  \label{sec::tandpselection::trackisolationeqn}
$$\TisolRfour = \sum{ \emph{p}_{\rm{T}}(\Delta R <0.4)} / \emph{p}_{\mathrm{T}}(\mathrm{tag}),$$ 
\end{equation}
where the sum extends
over all tracks with $\pt>1$~GeV (excluding the track on which the tag was based), within a cone of $\Delta R\equiv\sqrt{(\Delta\eta)^2+(\Delta\phi)^2} = 0.4$ around the tag. 
A tag muon must  pass the following requirements:
\begin{itemize}
    \item   the tag muon triggered the readout of the event (restricting the tag muon to the trigger acceptance, $|\eta|<2.4$);
    \item   $\pt>20$~GeV;\smallskip
    \item   $\TisolRfour<0.2$. 
\end{itemize}

\subsection{Probe selection}
Probes are either SA muons or ID tracks, depending on which efficiency measurement is being performed. They have to satisfy the following criteria for studies using $J/\psi\to\mu^+\mu^-$ decays:
\begin{itemize}
    \item   an ID track fulfilling the hit requirements described in \mySect~\ref{sec:IDtrkselection} (SA muons are not used, as the ID efficiency is not measured using these decays); 
    \item   reconstructed momentum, $p>3\GeV$, $|\eta|<2.5$;
    \item   the tag and the probe are oppositely charged;
    \item   the tag and the probe must be associated with the same vertex;
    \item   $\Delta R<3.5$ between the tag and probe.
    \item   the invariant mass of the tag-and-probe pair is within the range of $2<m<3.6$~GeV
\end{itemize}
Different cuts are applied in case of $Z\to\mu^+\mu^-$ decays:
\begin{itemize}
   \item an ID track fulfilling the hit requirements or a SA muon with at least one $\phi$ measurement;
   \item $\pt>20$~GeV, $|\eta|<2.5$;
   \item the tag and the probe are oppositely charged;
   \item the tag and the probe are associated with the same vertex; 
   \item azimuthal separation of the tag and the probe,\break
         $\Delta\phi>2.0$;\smallskip
   \item $\TisolRfour<0.2$;\smallskip
   \item the invariant mass of the tag-and-probe pair is within 10 GeV of $m_Z$.
\end{itemize}

\subsection{Matching of probes to ID tracks and muons\label{sec::tag_and_probe_selection_matching}}
After selecting all tag-and-probe pairs, an attempt is made to match probe tracks
to the objects for which the efficiency is to be measured, i.e.~SA probe
tracks to ID tracks in the case of the ID efficiency, or ID tracks to CB or
CB+ST muons in the case where the reconstruction efficiencies for these two
classes of muons are investigated.
A match between an ID probe and a reconstructed muon is considered successful if they have the same charge and are close in ($\eta,\phi$) space: $\Delta R\le0.01$. Similarly, a match
between an SA probe and an ID track is considered successful if $\Delta R\le0.05$.

\section{Low-\pt~ reconstruction efficiency measured with $J/\psi\to\mu^+\mu^-$ decays \label{sec::Jpsi_results}}

\begin{figure*}[p]
\begin{center}
    \begin{tabular}{@{\hspace*{0.2ex}}cc}
    \includegraphics[width=0.48\textwidth]{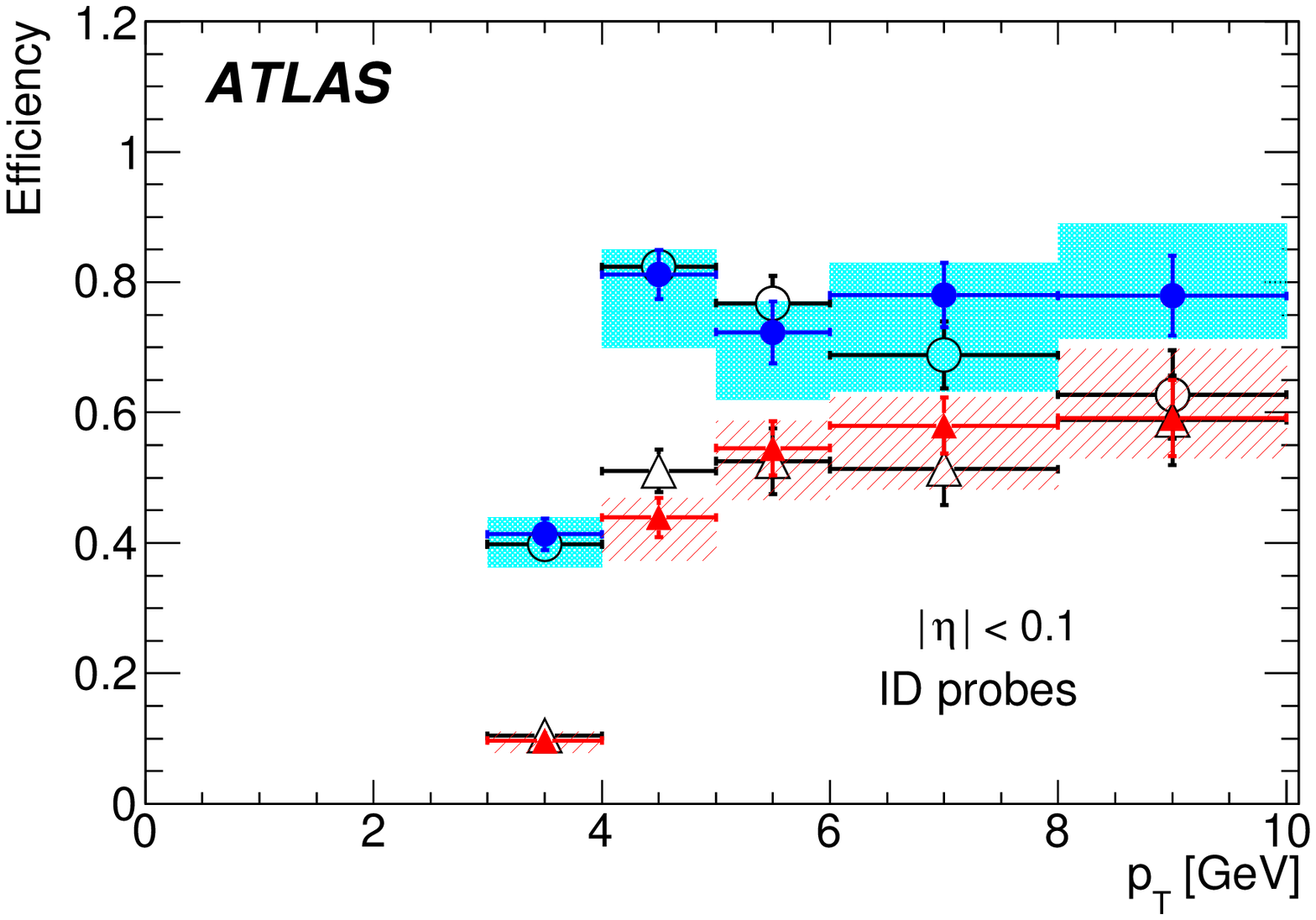}&
    \includegraphics[width=0.48\textwidth]{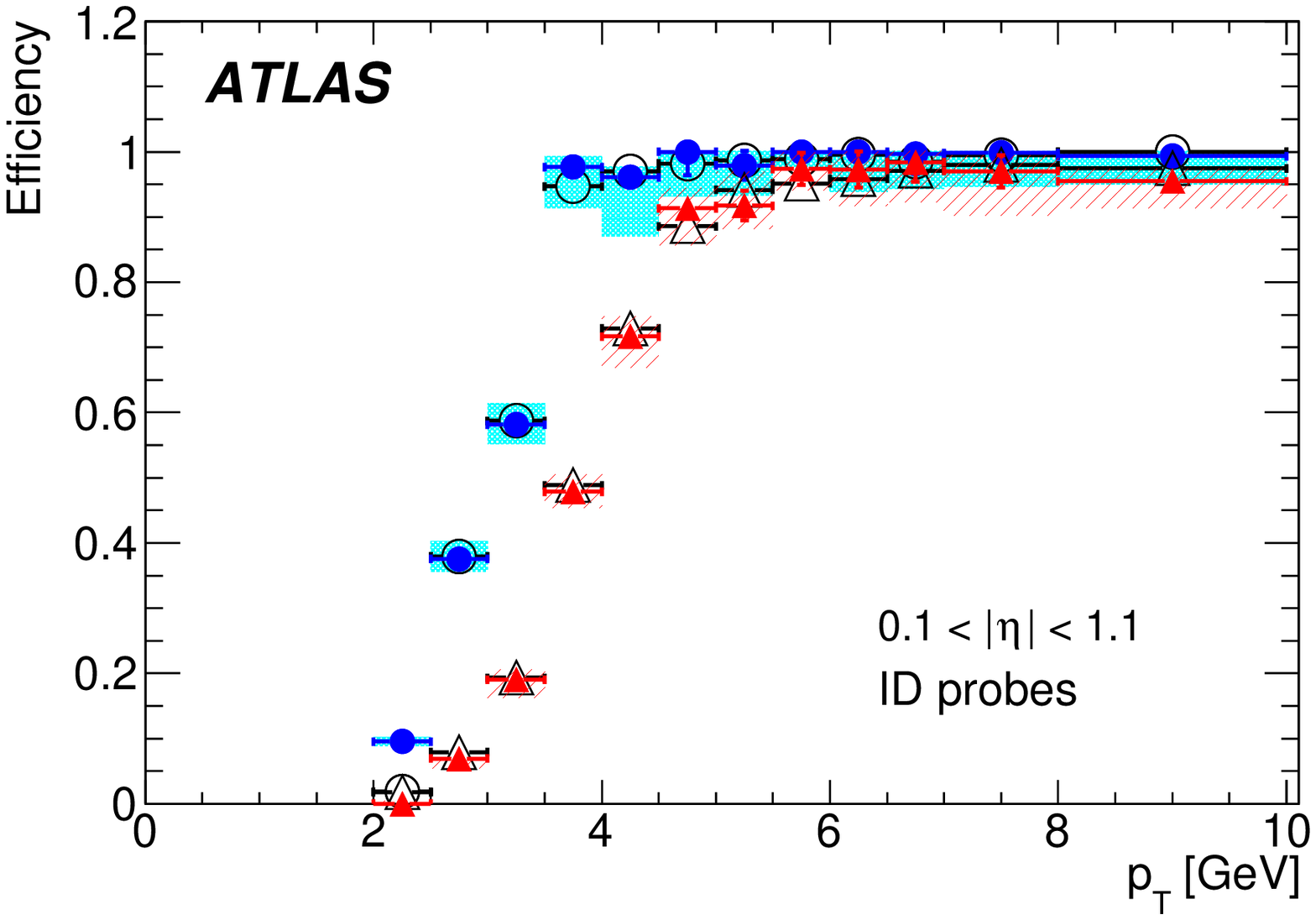}\\
    \includegraphics[width=0.48\textwidth]{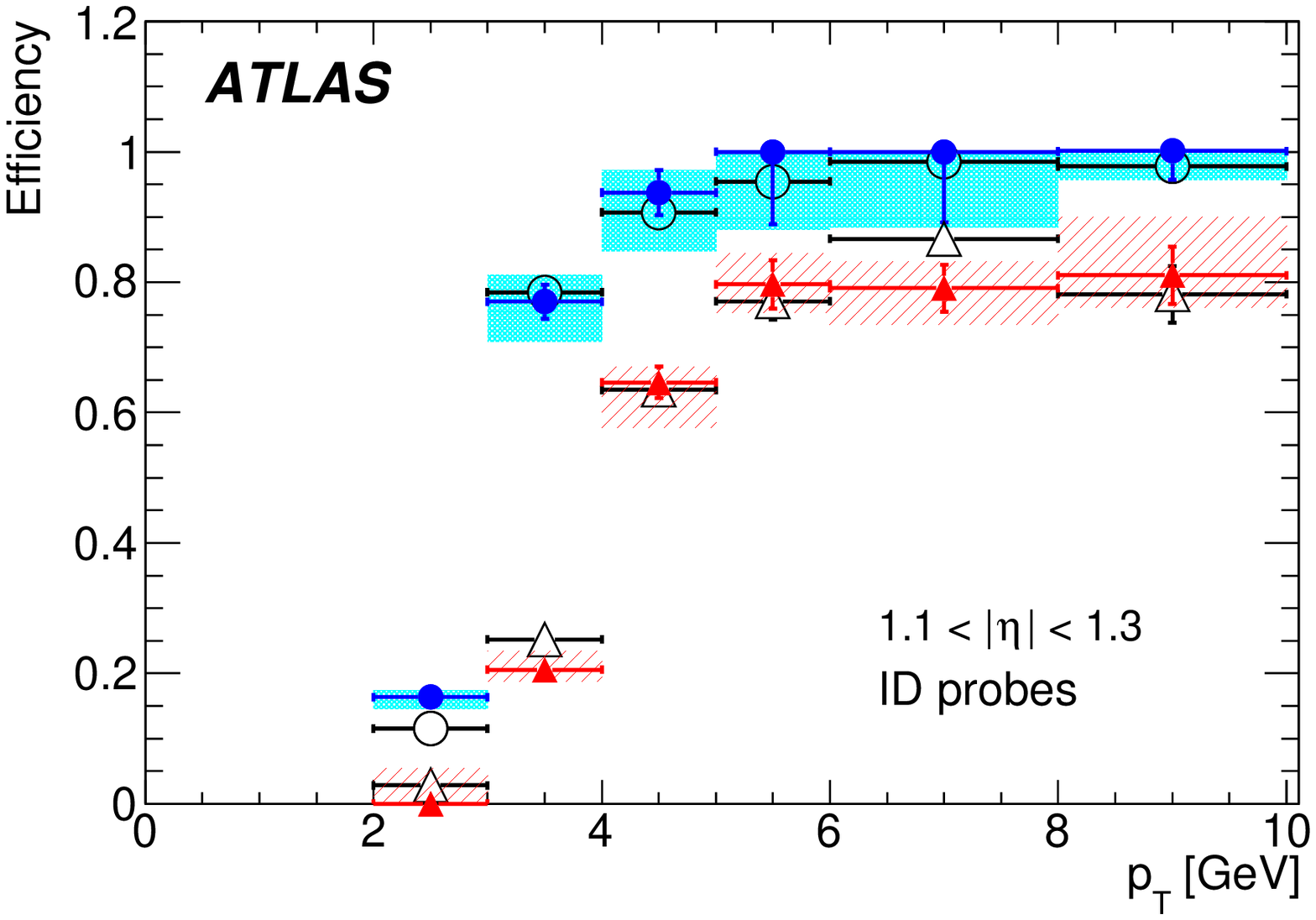}&
    \includegraphics[width=0.48\textwidth]{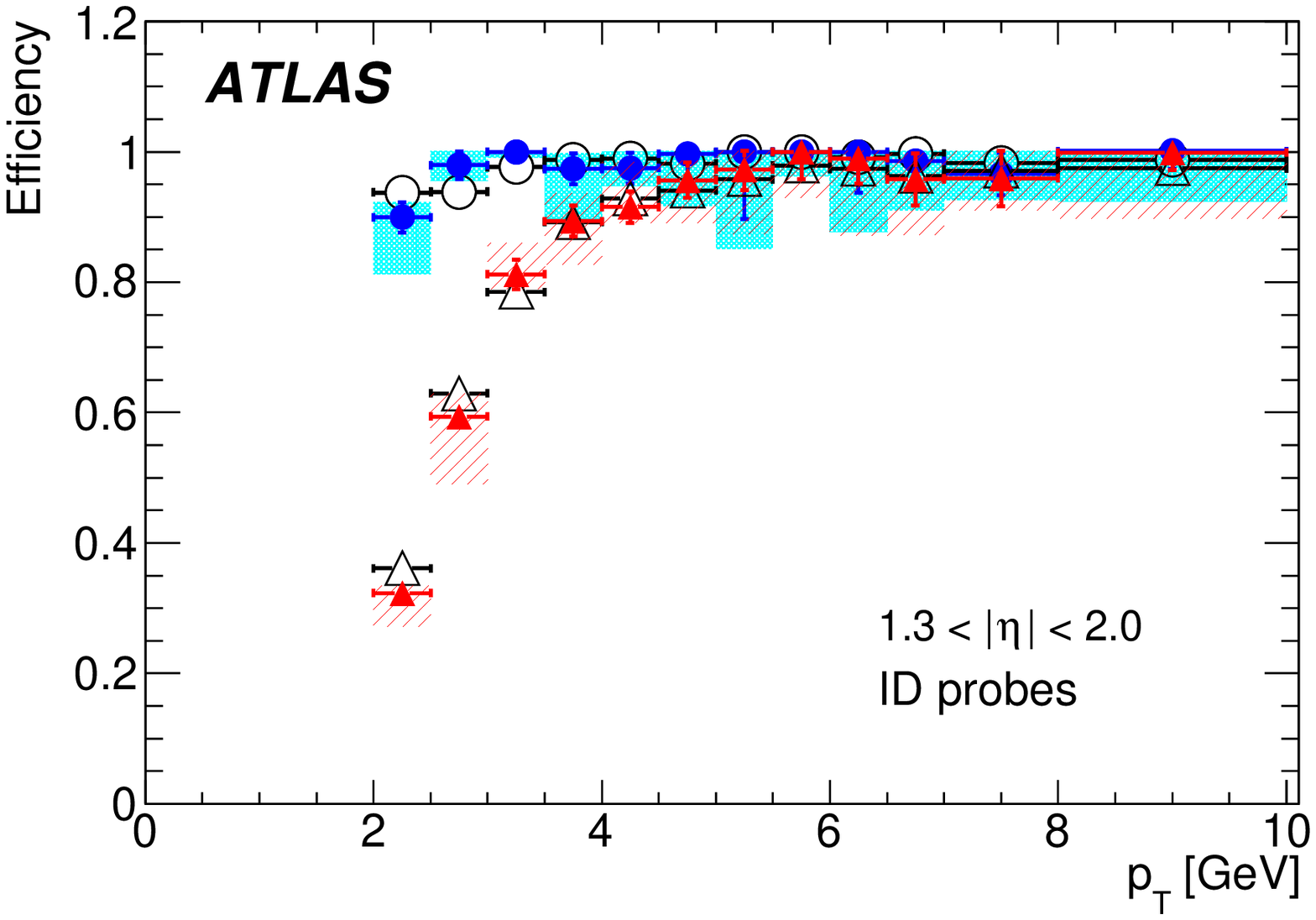}\\ 
    \includegraphics[width=0.48\textwidth]{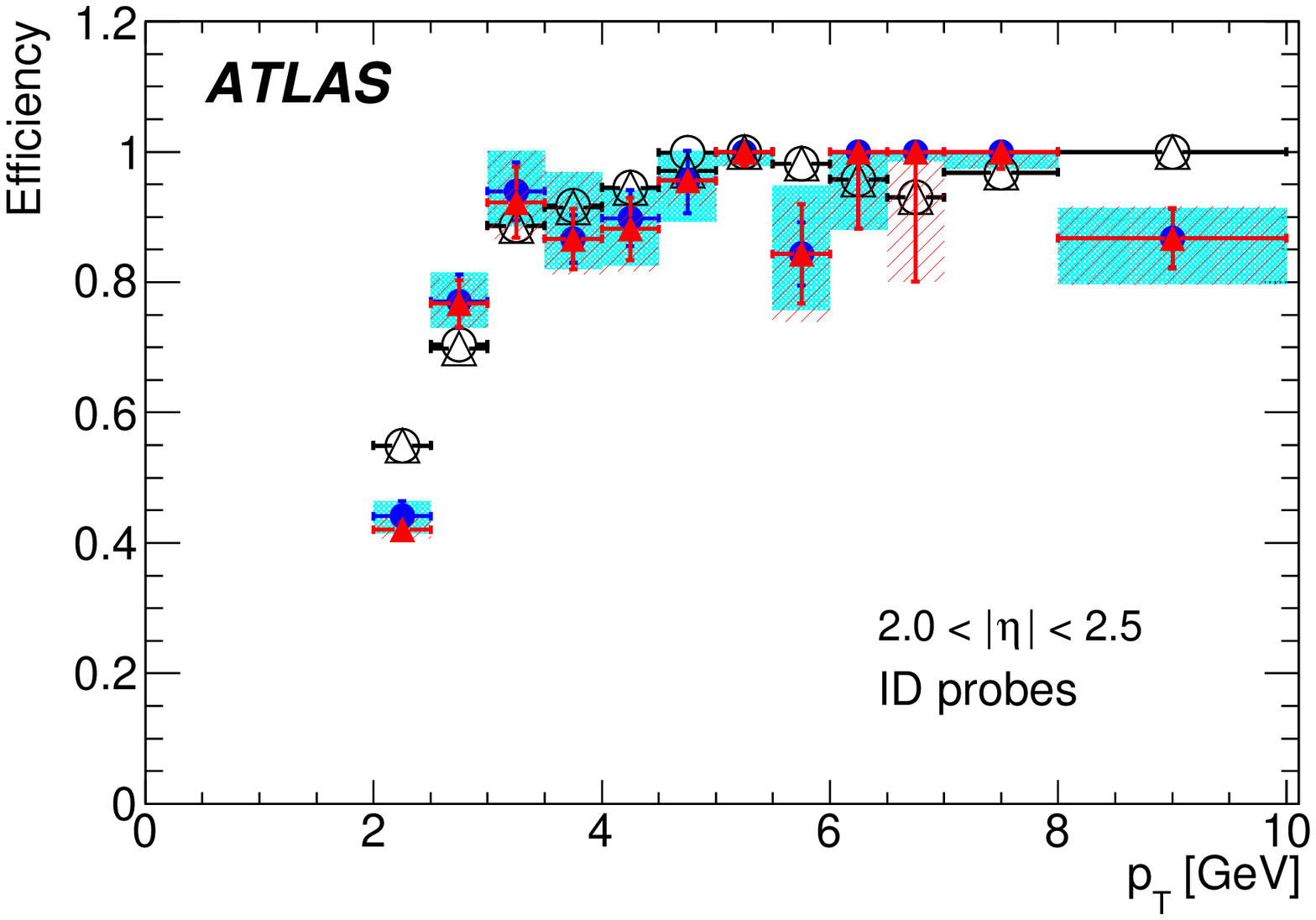}& 
    \includegraphics[width=0.48\textwidth]{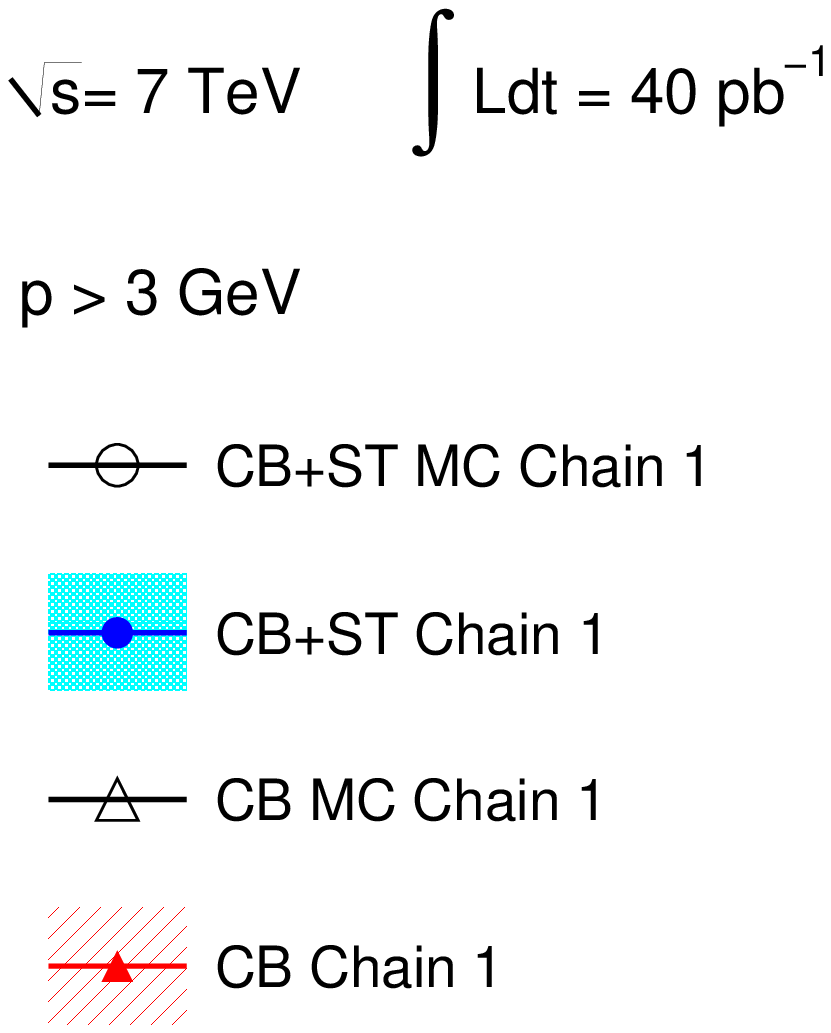}\\ 
    \end{tabular}
    \caption{\label{fig::Jpsi_results_chain_1} Efficiency for chain~1 CB and CB+ST muons
    with momentum $p>3$\,GeV (from $J/\psi$ decays), as a function of \pt,
    for five bins in $|\eta|$ as described in the legend, for data and MC events.
    The error bars represent the statistical uncertainties while the bands around
    the data points represent the statistical and systematic uncertainties added in quadrature.}
\end{center}
\end{figure*}

\begin{figure*}[p]
\begin{center}
    \begin{tabular}{@{\hspace*{0.2ex}}cc}
    \includegraphics[width=0.48\textwidth]{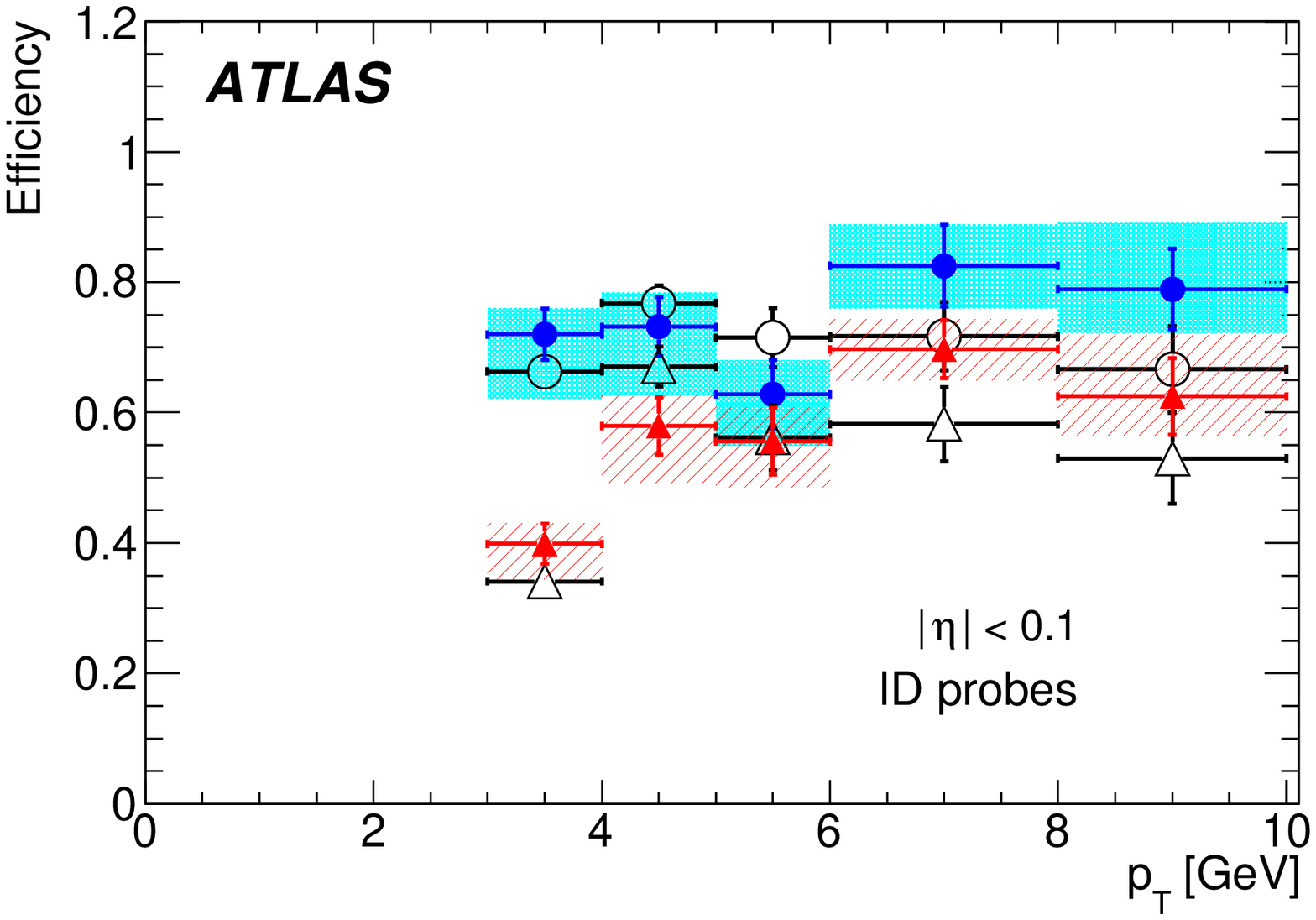}&
    \includegraphics[width=0.48\textwidth]{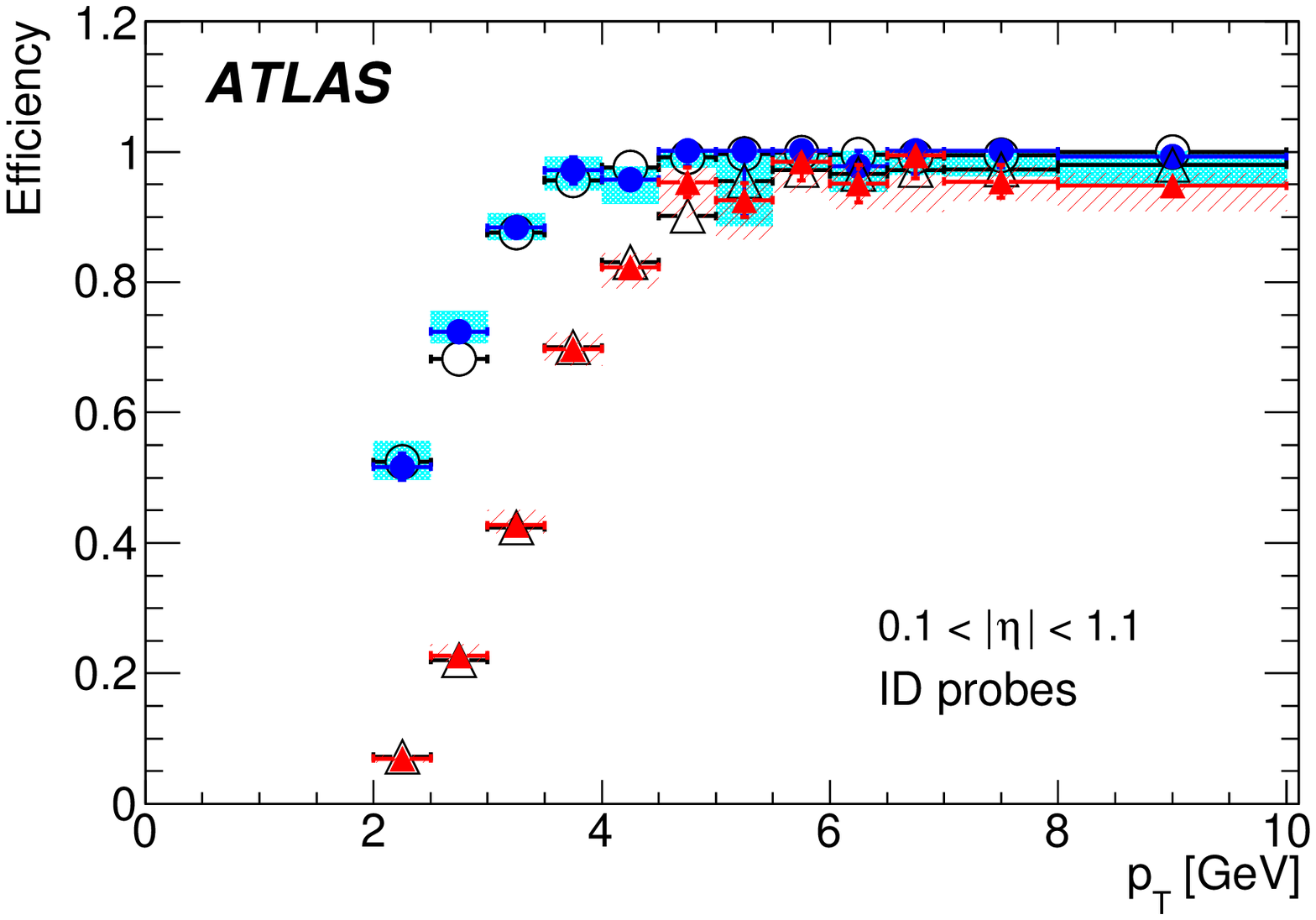}\\
    \includegraphics[width=0.48\textwidth]{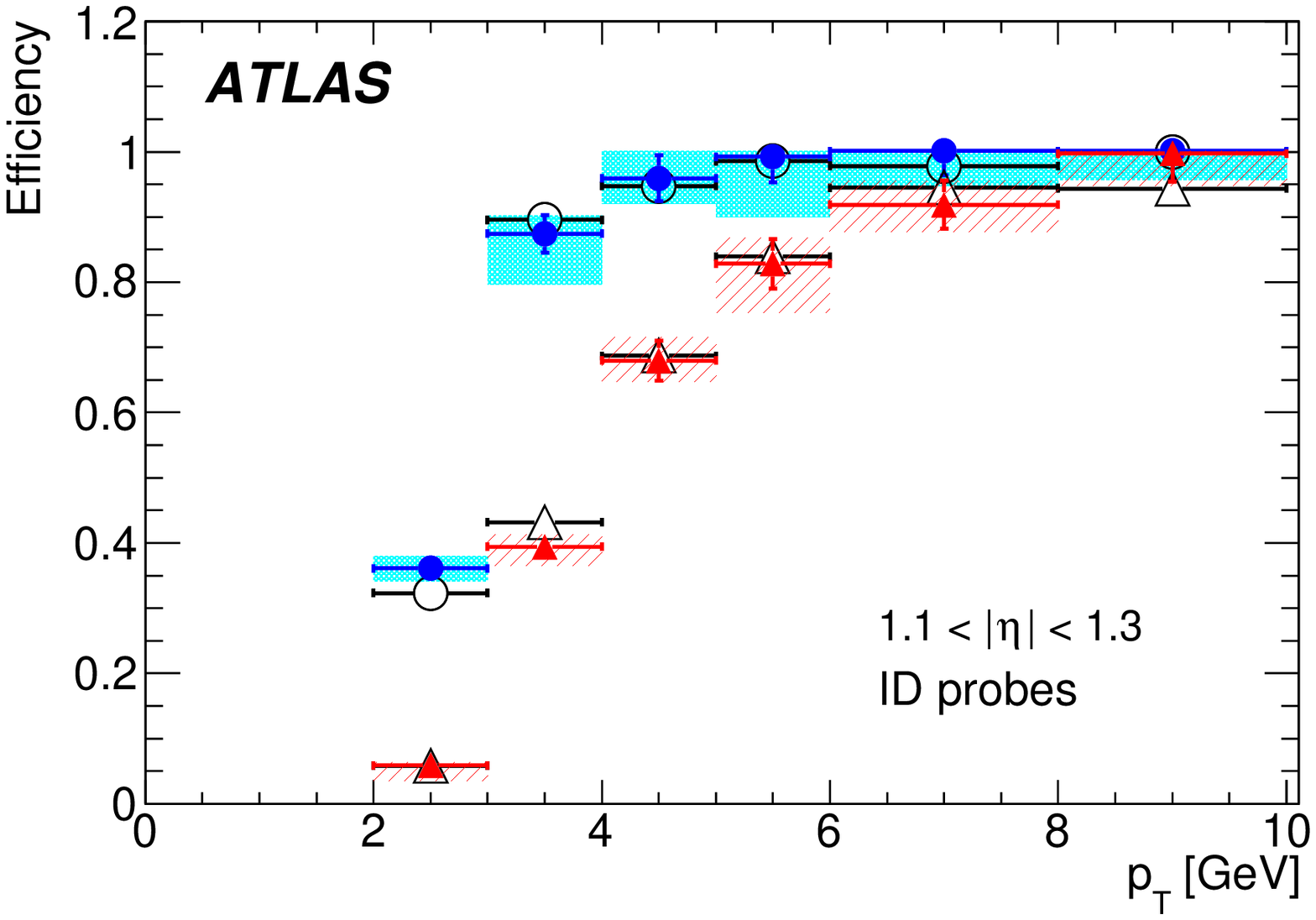}&
    \includegraphics[width=0.48\textwidth]{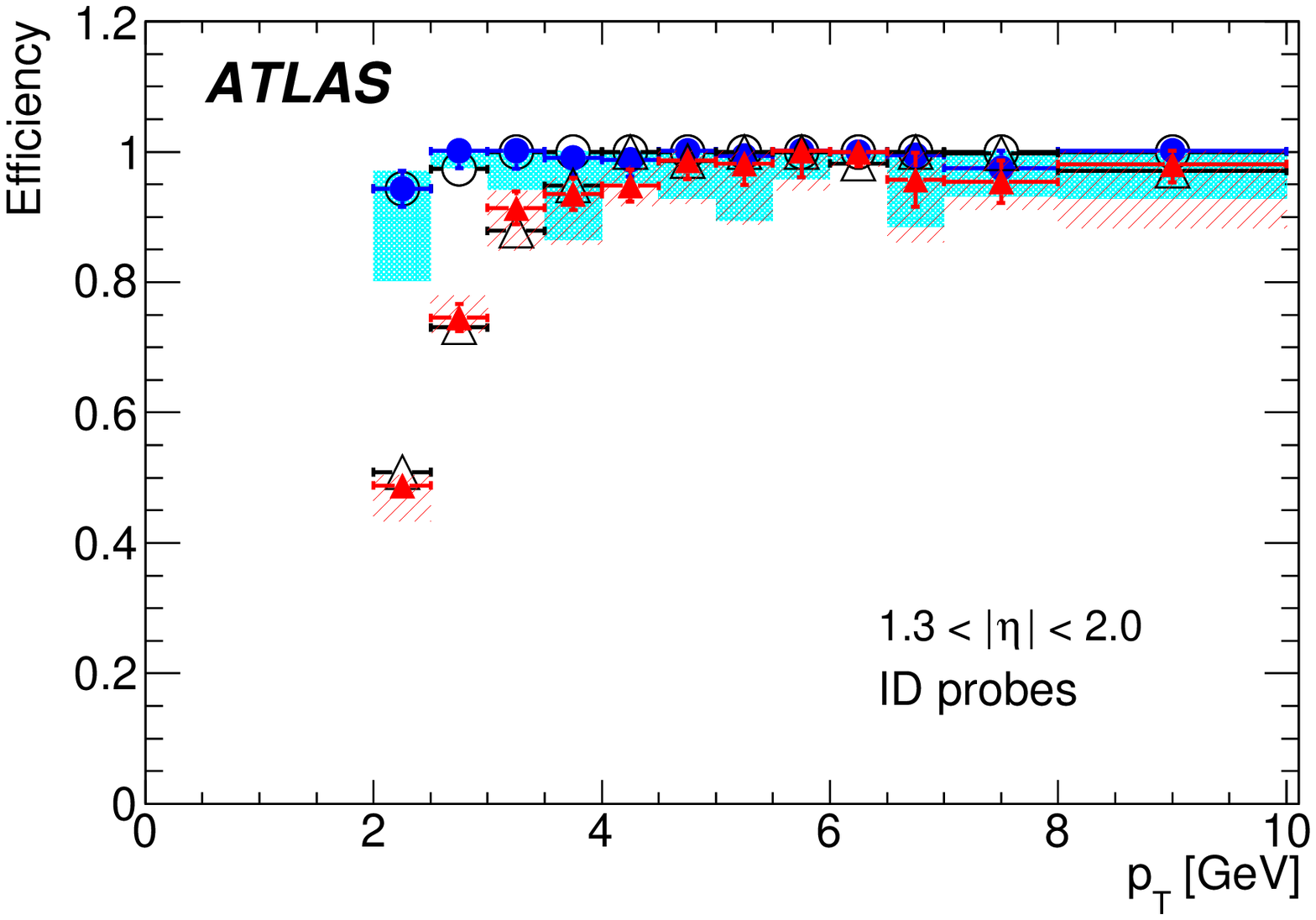}\\ 
    \includegraphics[width=0.48\textwidth]{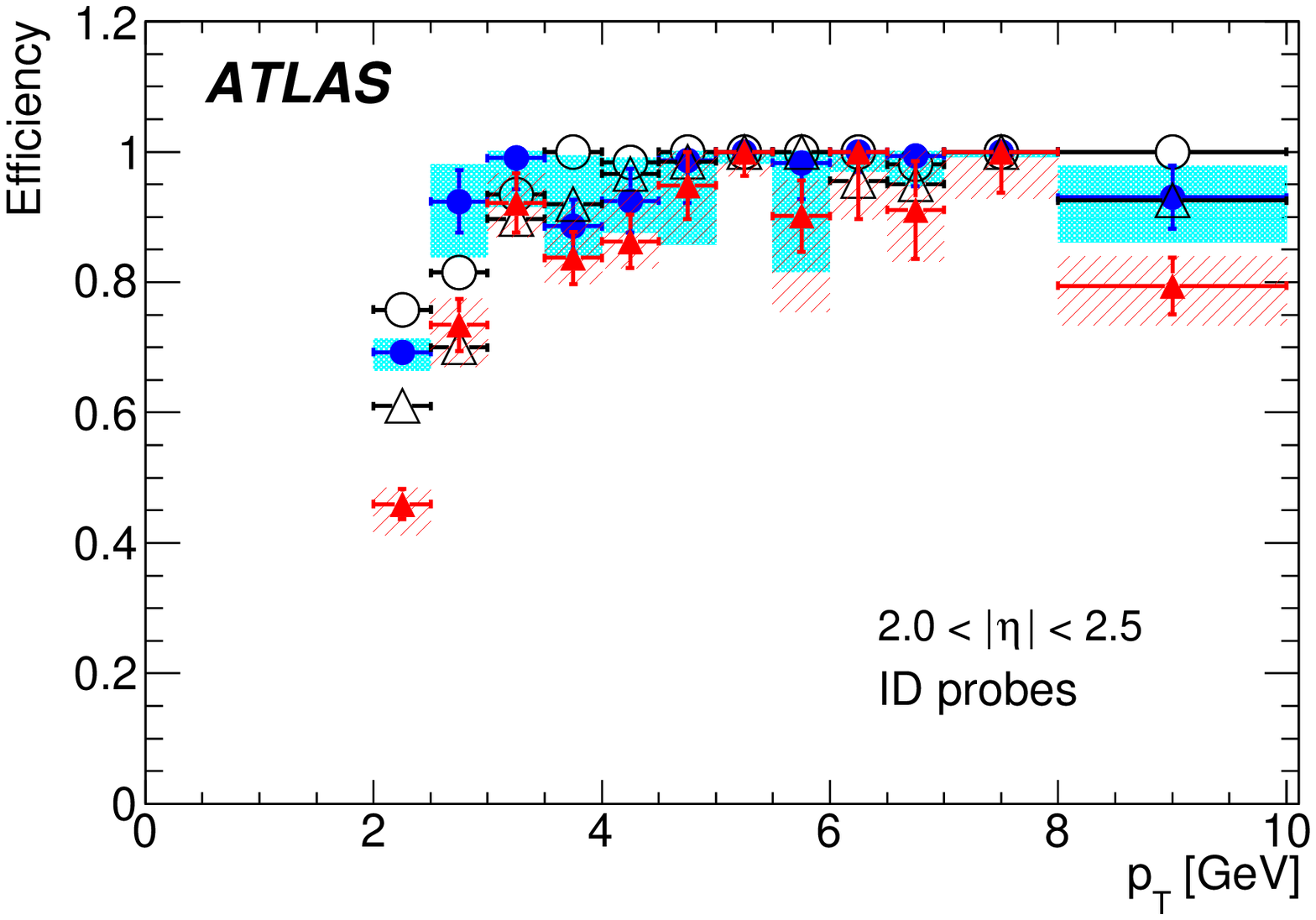}& 
    \includegraphics[width=0.48\textwidth]{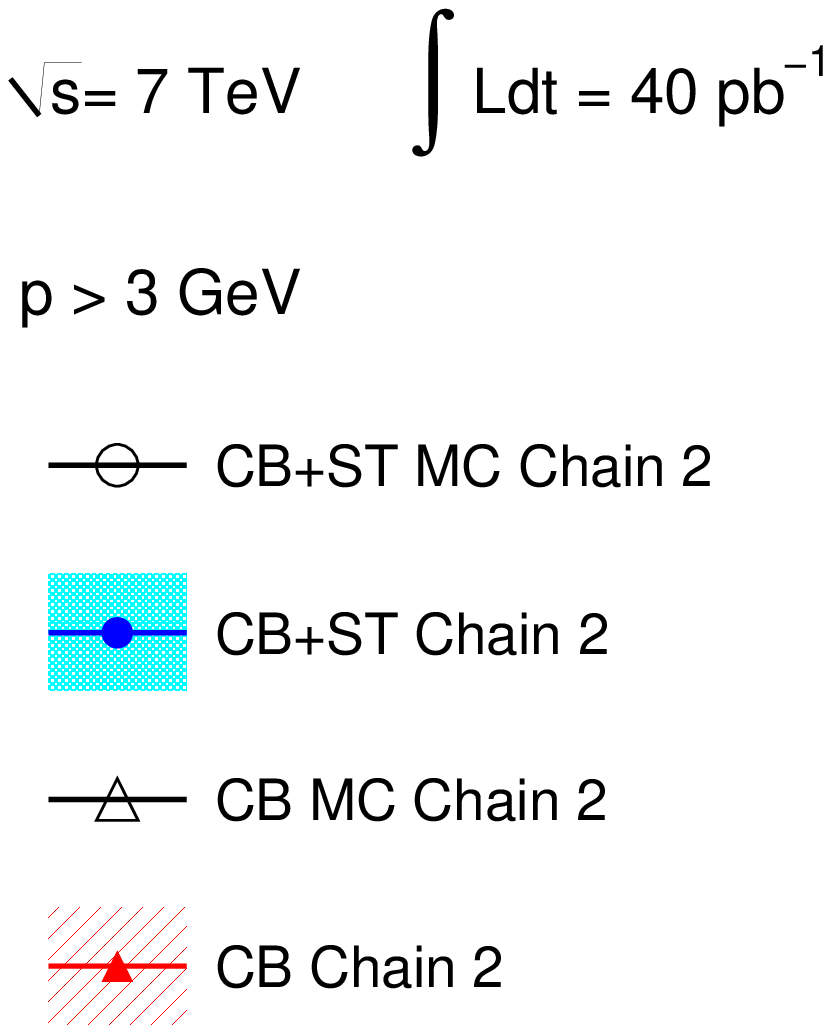}\\ 

    \end{tabular}
    \caption{\label{fig::Jpsi_results_chain_2} Efficiency for chain~2 CB and CB+ST
    muons with momentum $p>3$\,GeV (from $J/\psi$ decays), as a function of
    \pt, for five bins in $|\eta|$ as described in the legend, for data and MC events.
    The error bars represent the statistical uncertainties while the band around the
    data points represents the statistical and systematic uncertainties added in
    quadrature.}
\end{center}
\end{figure*}

Figures \ref{fig::Jpsi_results_chain_1} and \ref{fig::Jpsi_results_chain_2} show the reconstruction efficiencies for chain 1 and chain 2 with respect to ID tracks with
momentum $p>3$~GeV, as a function of the probe \pt, for the five bins in probe $|\eta|$ described in \mySect~\ref{sec::MC}. Also shown are the Monte Carlo predictions, which agree with data within the statistical and systematic uncertainties of the measurements.%

A number of checks were performed to study the dependence of the results on analysis details and assumptions.
\begin{enumerate}
    \item   Signal shapes: the means and the widths of the two (matched and unmatched) Gaussians in the fit were allowed to vary independently.
    \item   Background shape: a linear background function was used in the fit, instead of the quadratic parameterisation; in this case the fit 
    was performed in the reduced mass range of 2.7--3.5~GeV (instead of 2.0--3.6~GeV).
    \item   Alternative fit: an independent fit to the matched and the total (matched + unmatched) distributions, rather than to 
    matched and unmatched, was used and the efficiency estimated as the ratio of the signal normalisations in the two distributions. 
    While this option does not provide for an easy propagation of the uncertainty from the background subtraction and does not directly 
    account for the correlations between the two samples, it profits from a higher stability of the two simpler fits, whereas the default 
    method needs some care in the choice of the initial conditions, in particular in cases of very high efficiency or small overall sample size. 
\end{enumerate}
The largest positive and negative variations obtained from any of the three
checks were taken as systematic uncertainties and added in quadrature to the
statistical uncertainty to obtain the total upper and lower uncertainties.
The statistical uncertainties were found to be at the level of a few percent.

\clearpage

\section{Intermediate- and high-\pt~reconstruction efficiencies measured with
$Z\to\mu^+\mu^-$ decays \label{sec::Z_results}}

For higher momentum muons, with \pt$>20$~GeV, $Z$ decays are used to measure the reconstruction efficiencies.

\subsection{Inner detector reconstruction and identification efficiency}
\begin{figure}[]
\begin{center}
    \vspace*{-0.6ex}\hspace*{-0.8ex}
    \includegraphics[width=1.03\linewidth]{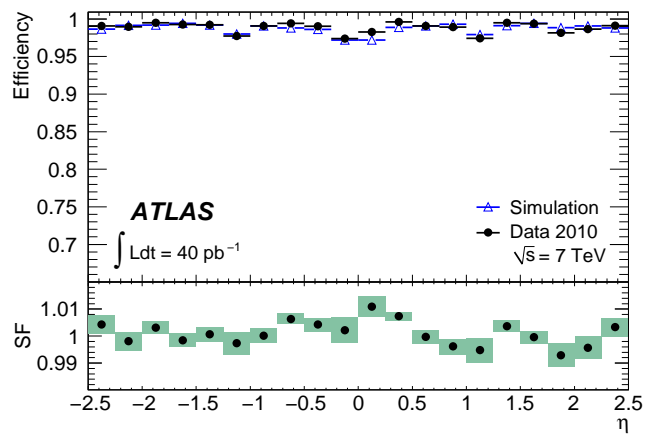}\hspace*{-0.2ex}
    \caption{\label{fig:ID_eff}Measured ID reconstruction and identification efficiency for muons
    (from $Z$ decays), as a function of $\eta$, for data and Monte Carlo simulation.
    The scale factors (SF), defined as the ratio of the measured efficiency to the
    predicted efficiency, are shown in the lower panel of the plot. The uncertainties are statistical. The systematic uncertainty is discussed in \mySect~\ref{sec::Z_results_systematics}.}
\end{center}
\end{figure}

Figure~\ref{fig:ID_eff} shows the reconstruction and identification efficiency in the ID as a function of $\eta$, for data and simulation, as determined using SA probes. The simulation includes all considered backgrounds.
The scale factors (SF), defined as the ratio of the data efficiency to the Monte Carlo efficiency, are displayed in the lower panel (the smallness of the background correction, as described in \mySect~\ref{sec::tag_and_probe_method}, means that its effect on the SF is negligible).

As discussed earlier, the efficiency for the combined reconstruction varies with the detector region, and with $\pt$ in the range below 6~GeV. In contrast, the ID reconstruction efficiency is independent of $\phi$ and $\pt$~\cite{AtlasDetectorPaper}, and shows only a slight dependence on $\eta$. 

The slightly lower efficiencies at $\eta\sim0$ and $|\eta|\sim1$ are caused by the ID hit requirements for muon identification described in \mySect~\ref{sec:IDtrkselection}: at $\eta\sim0$, ID tracks pass through an inactive region near the middle of the TRT barrel where straws produce no TRT hits; at $|\eta|\sim1$, there is a small region in the transition between the barrel and the end-caps of the ID in which muons cross fewer than six SCT sensors~\cite{AtlasDetectorPaper}.
The measured ID muon reconstruction and identification efficiencies agree with the Monte Carlo
predictions within 1\%, and, for the most part, within the statistical uncertainties.
The average ID efficiency is 0.991$\pm$0.001 with the small loss being due to the hit requirements imposed on the ID muon tracks. These results are independent of the choice of the algorithm chain for the stand-alone muon.

\subsection{Reconstruction efficiencies for CB muons}
\begin{figure*}[]
  \begin{center}
      \vspace*{-0.6ex}
      \begin{tabular}{@{\hspace*{-0.3ex}}cc}
          \includegraphics[width=0.51\textwidth]{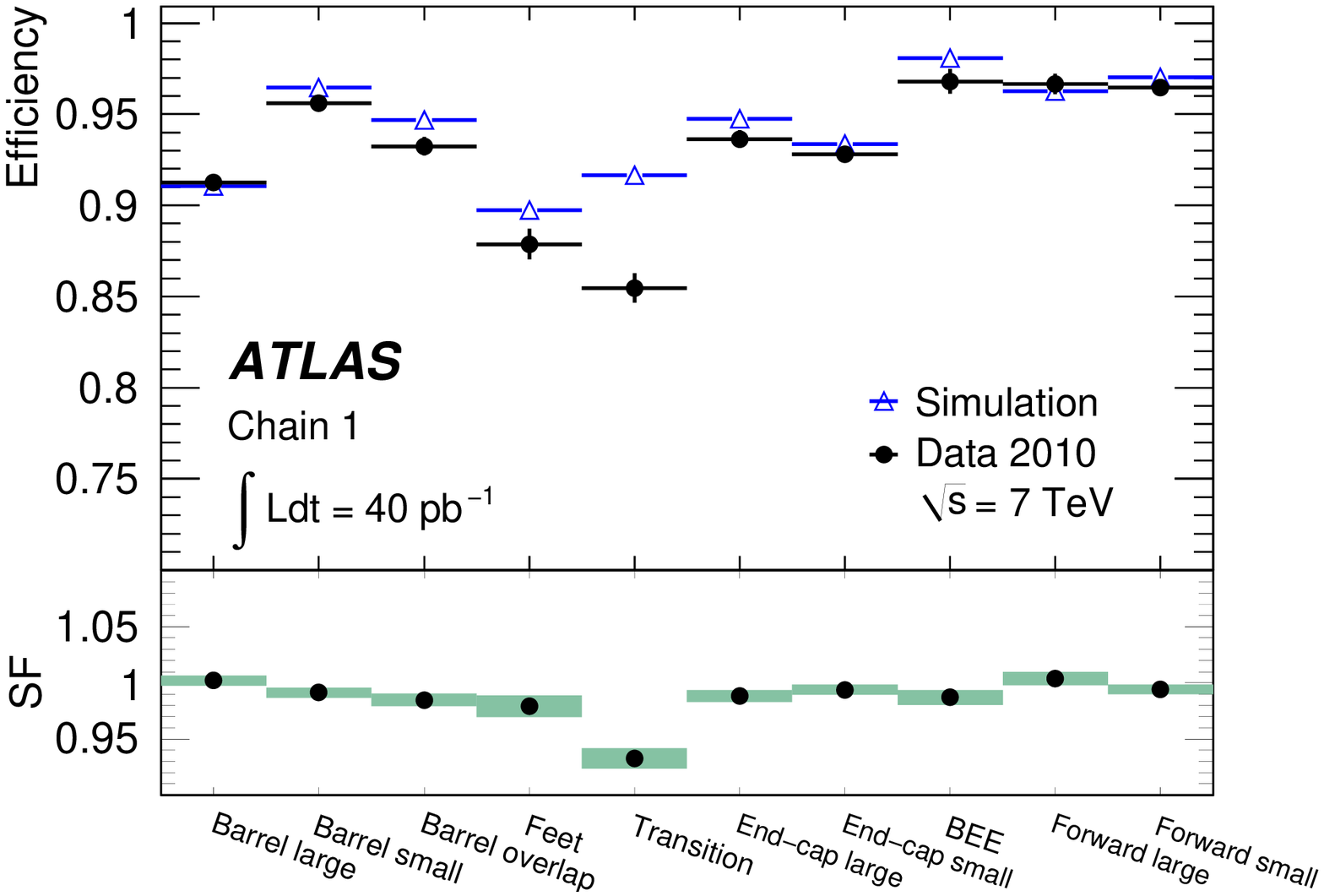}\hspace*{-2ex}&
          \includegraphics[width=0.51\textwidth]{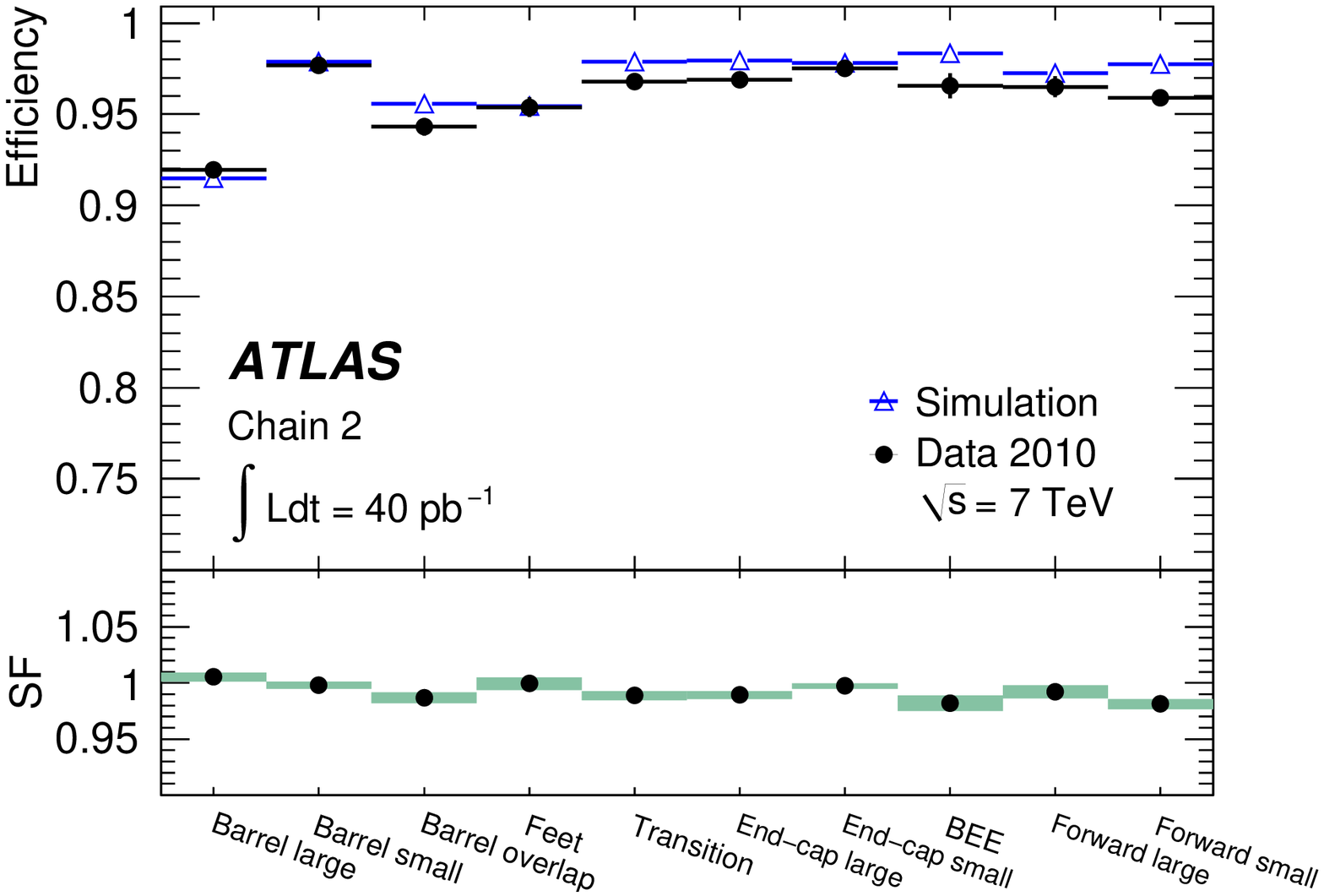}\hspace*{-2ex}\\
          \includegraphics[width=0.48\textwidth]{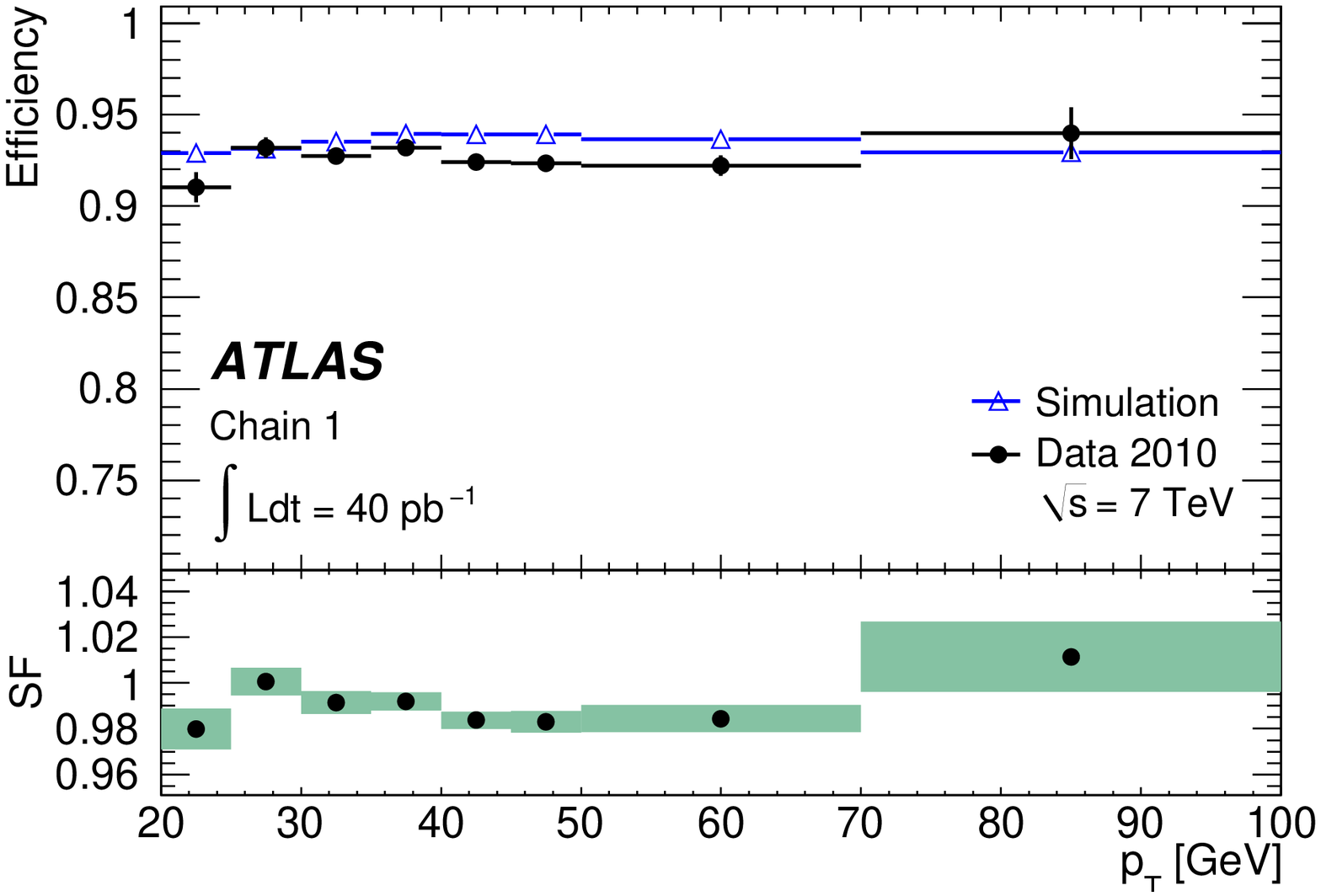}&
          \includegraphics[width=0.48\textwidth]{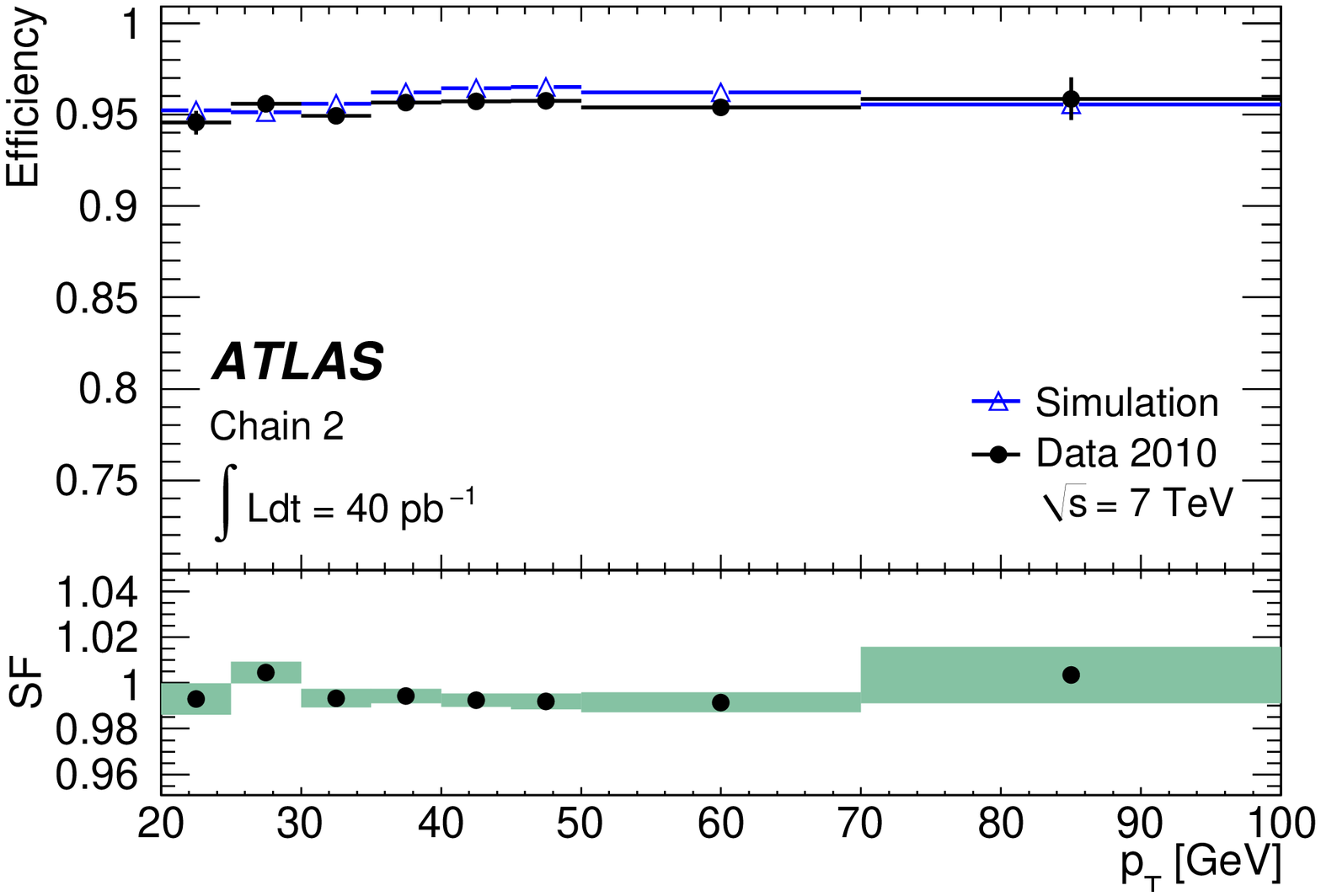}\\[-2pt]
          \includegraphics[width=0.48\textwidth]{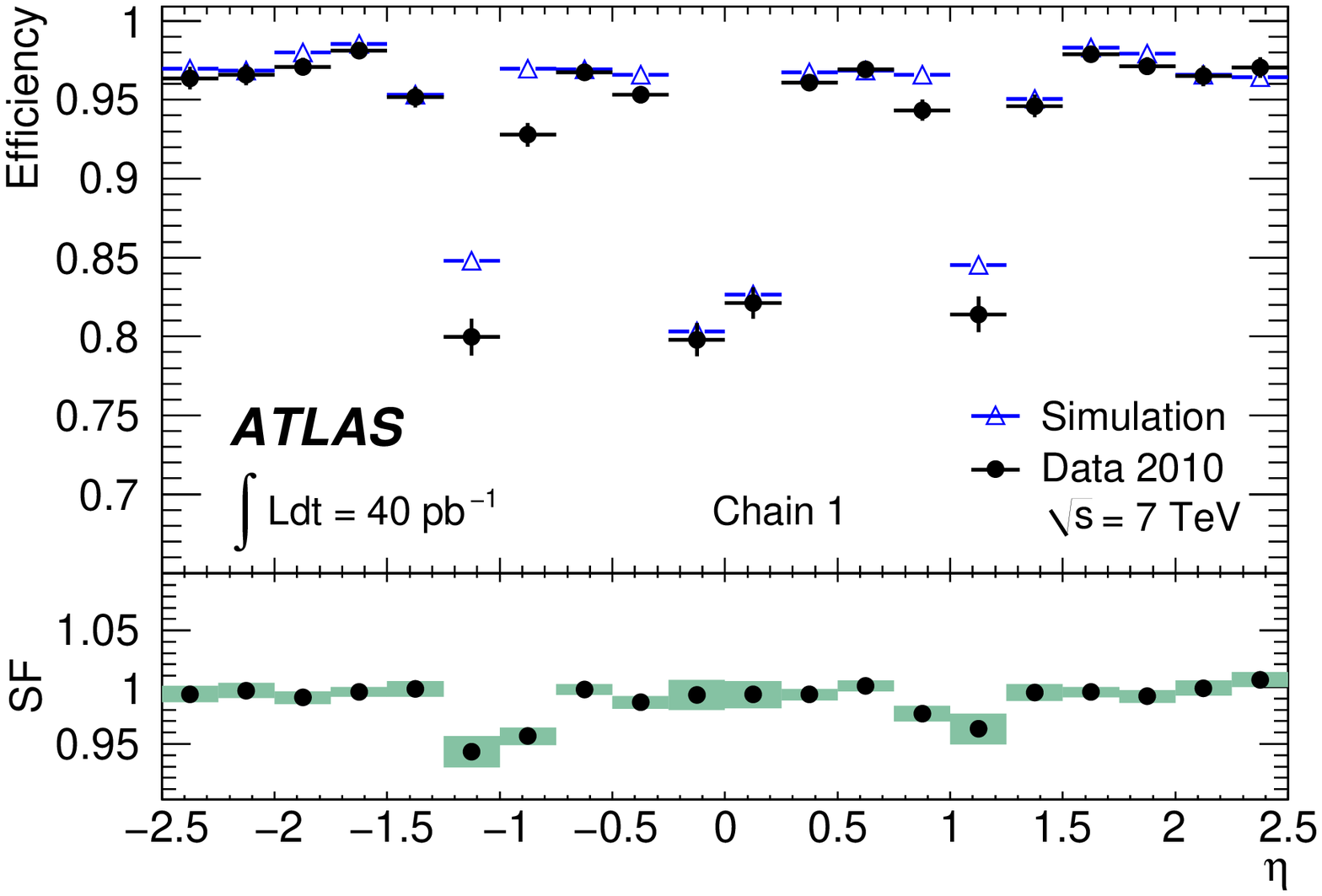}&
          \includegraphics[width=0.48\textwidth]{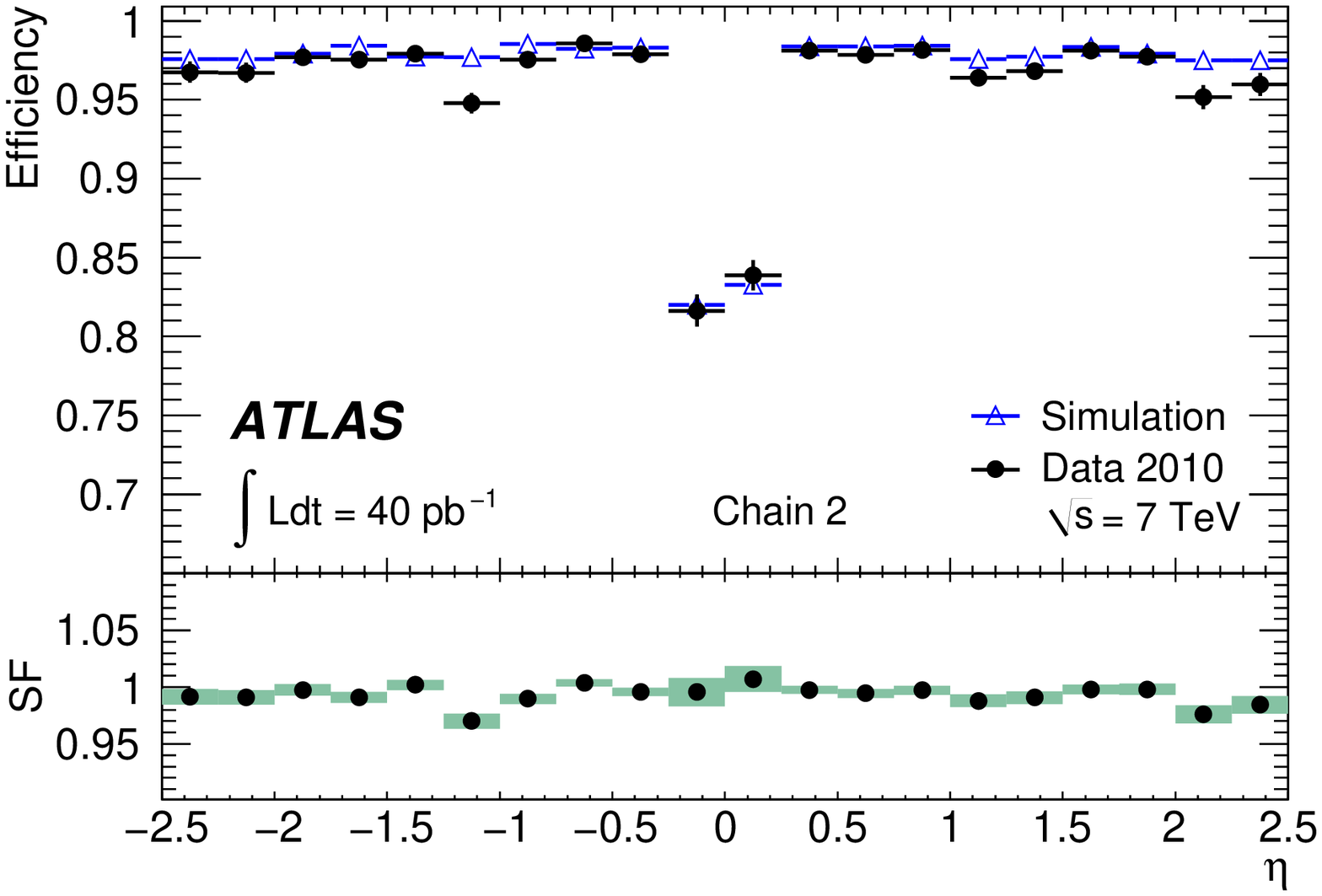}\\
      \end{tabular}
  \end{center}
  \caption{Reconstruction efficiencies (relative to the ID efficiency)
    and scale factors for CB muons (from $Z$ decays) as a function of detector
    region, muon $\pt$ and muon $\eta$ as indicated in the figure.
    The efficiencies for the two reconstruction chains, obtained from data
    (without background correction) 
    and Monte Carlo simulation (including
    backgrounds) are shown in the upper part of each figure.
    The corresponding scale factors are shown in the lower panels.
    The uncertainties are statistical only. The systematic uncertainties
    are discussed in \mySect~\ref{sec::Z_results_systematics}.}
  \label{fig:EffFullStat}
\end{figure*}

Figure~\ref{fig:EffFullStat} shows the reconstruction efficiency (relative to the ID reconstruction efficiency) for CB muons
as a function of the detector region, $\pt$ and $\eta$,  for data and simulation (with all
considered backgrounds included). The scale factors are displayed in the lower panel of each plot. 

The mean value of the $\eta$-dependent scale factor is
$0.989\pm0.003$ for \Staco{} and 
$0.995\pm0.002$ for \Muid{},
where the errors are statistical. 
The 1\% deviation from unity in the overall efficiency scale factor of \Staco is caused mainly by the data/MC disagreement in the transition region ($\mathrm{SF}=0.94$). The lower data efficiency in the transition region is attributed to the limited accuracy of the magnetic field map used in the reconstruction of the ATLAS data in this region, which leads to a small mis-measurement of the stand-alone muon momentum. This in turn may affect the combination of the MS and ID tracks, as their momenta may not be compatible. The transition region efficiency drop can be recovered, and the overall efficiency significantly increased by including ST muons, which are tagged by only one muon layer, as described in detail below.
The scale factors determined in bins of $p_\mathrm{T}$ agree, within 1.5 standard deviations, with the average scale factor for the
algorithm in question. 

\begin{figure*}[]
    \begin{center}
       \vspace*{-0.6ex}
        \begin{tabular}{@{\hspace*{-0.3ex}}cc}
            \includegraphics[width=0.50\textwidth]{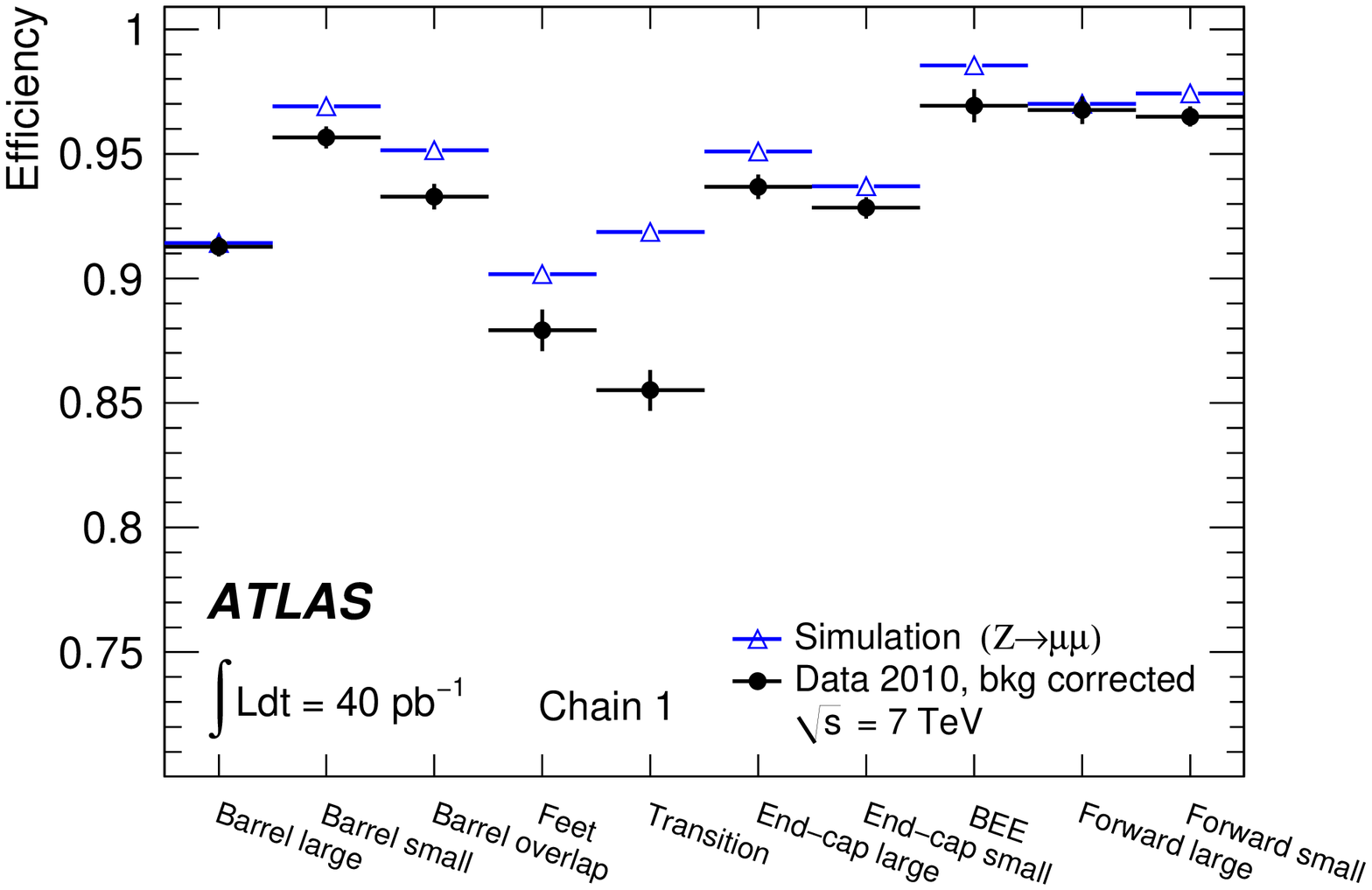}\hspace*{-2ex}&
            \includegraphics[width=0.50\textwidth]{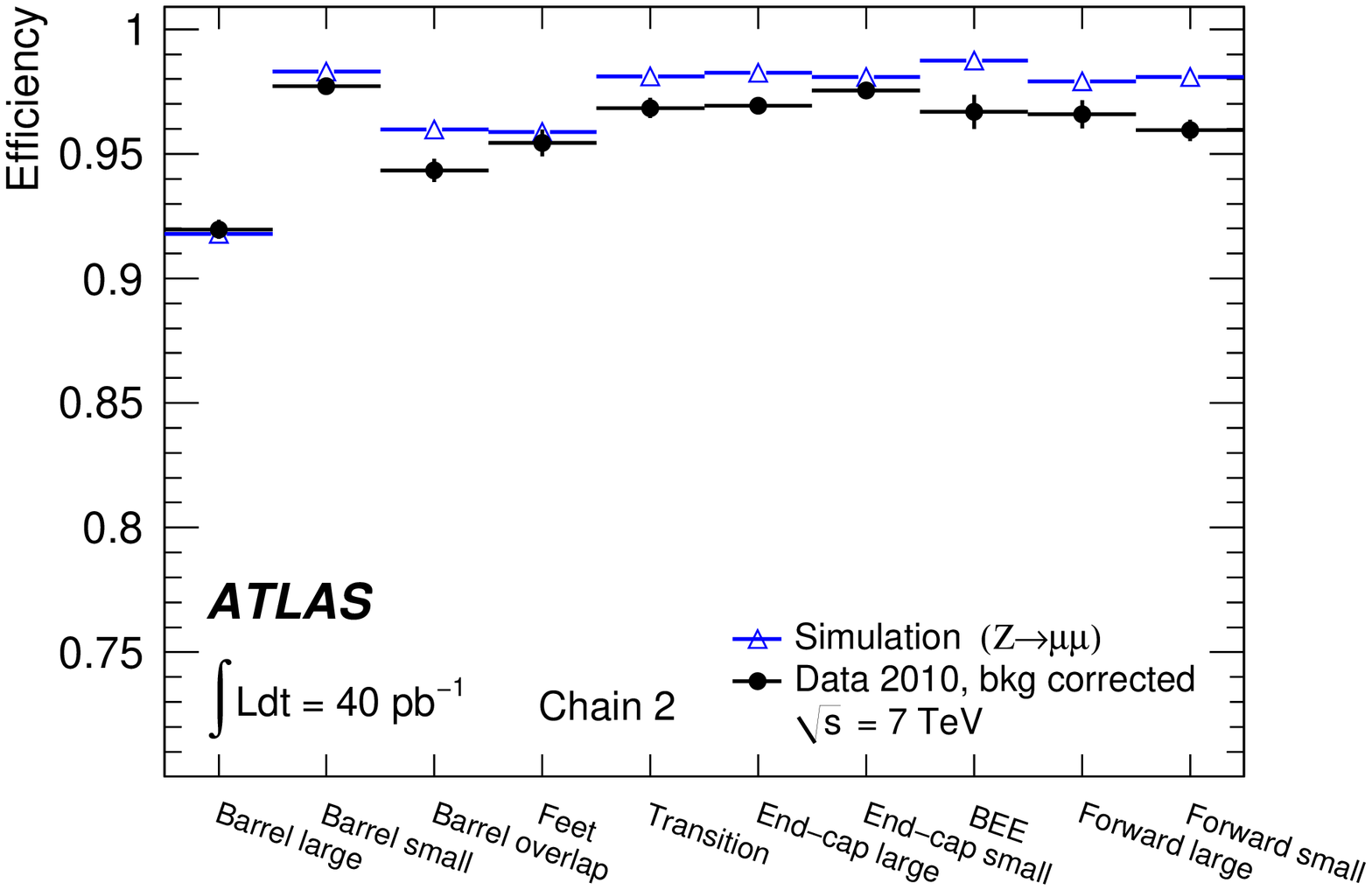}\hspace*{-2ex}\\
            \includegraphics[width=0.47\textwidth]{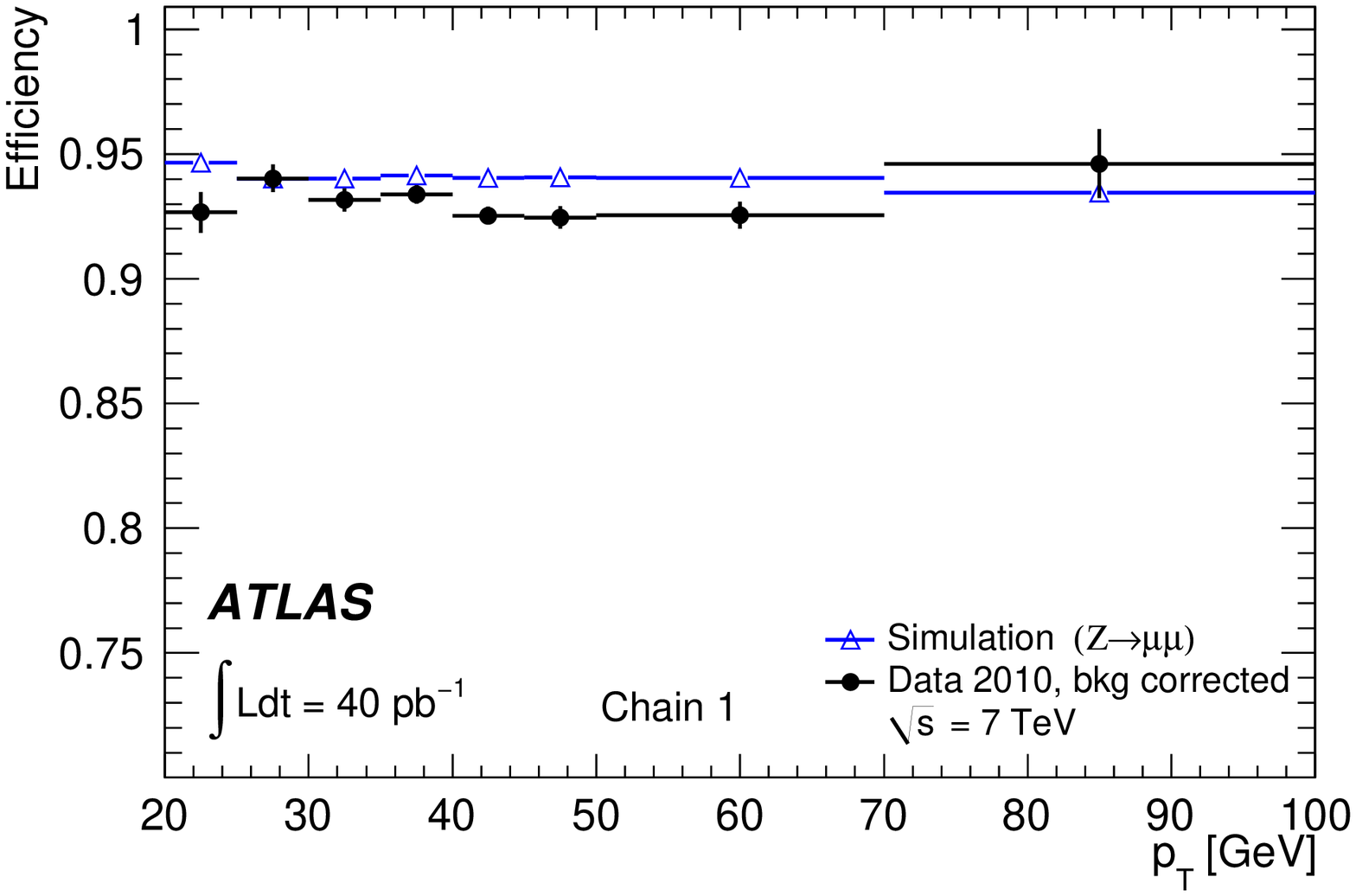} &
            \includegraphics[width=0.47\textwidth]{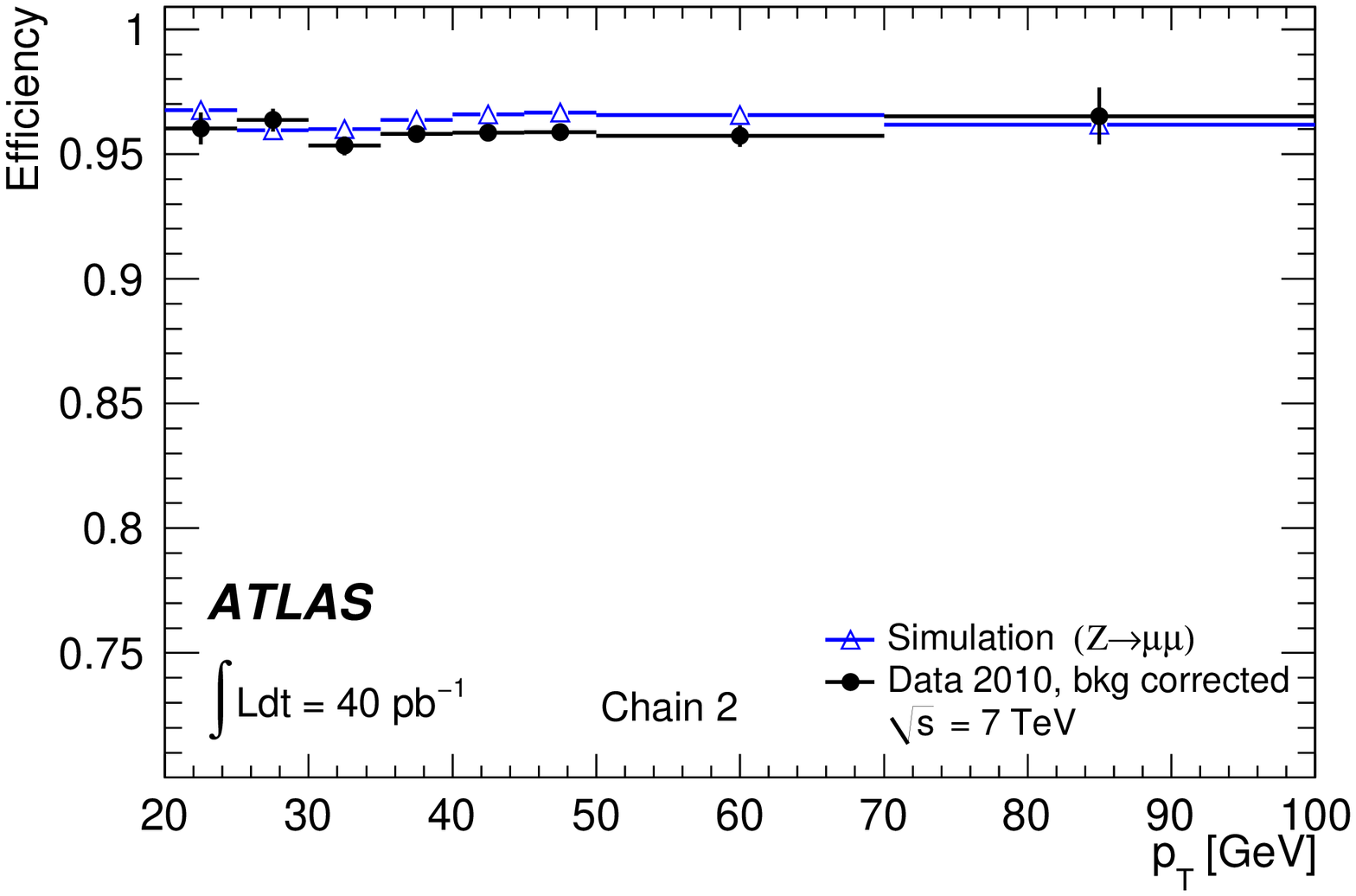} \\[-2pt]
            \includegraphics[width=0.47\textwidth]{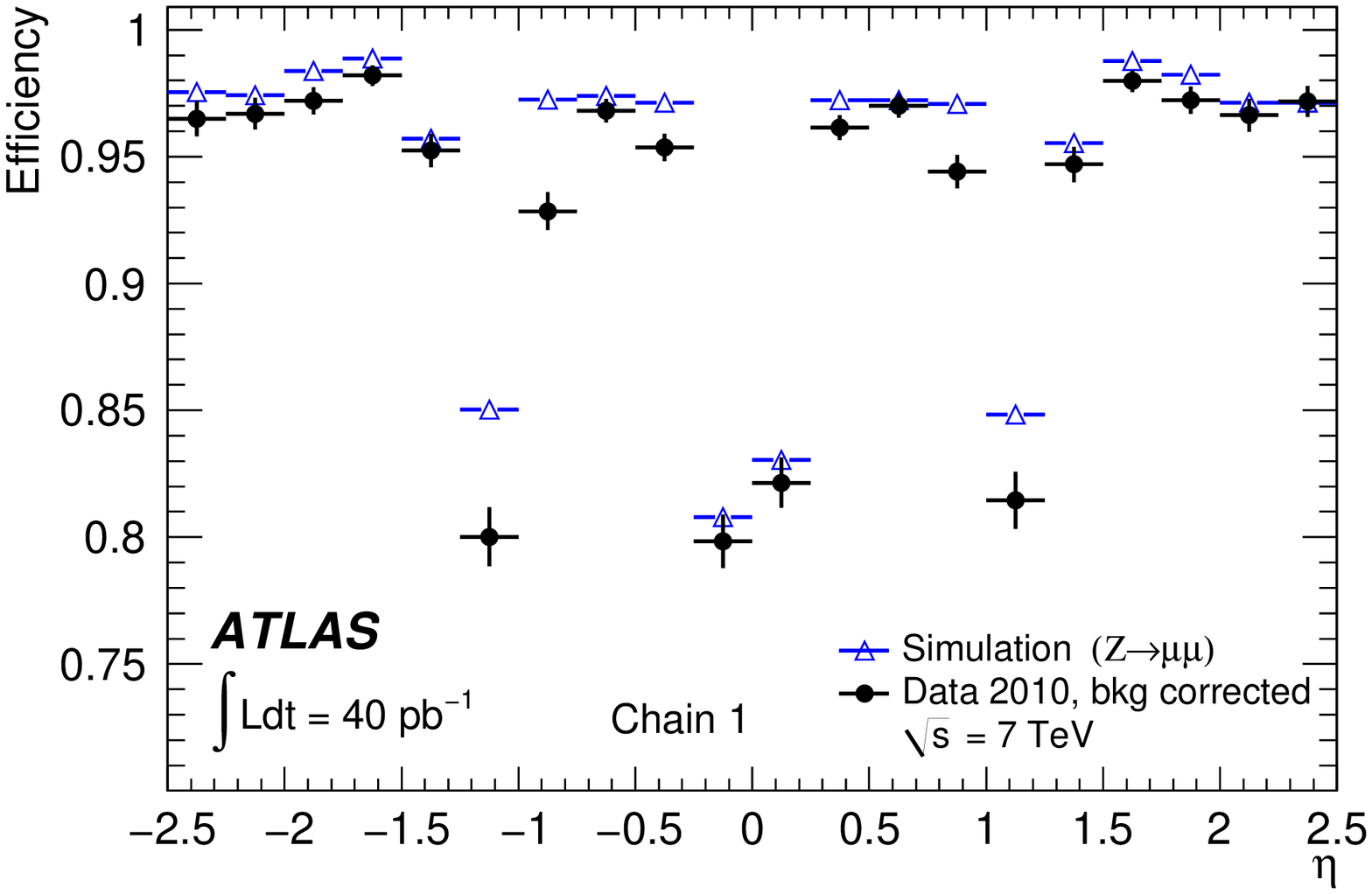} & 
            \includegraphics[width=0.47\textwidth]{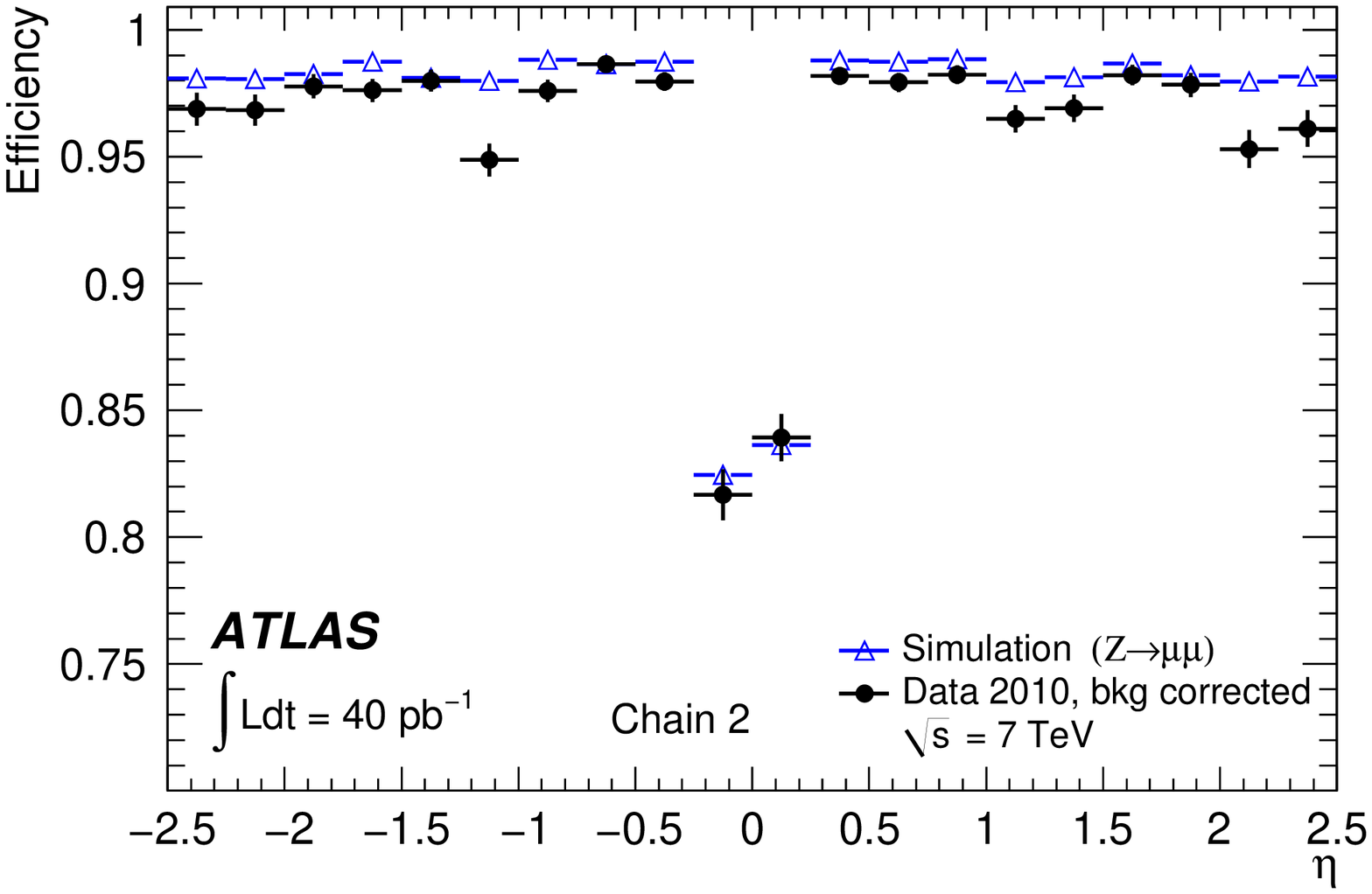} \\
    \end{tabular}
    \end{center}
    
    \caption{Background-corrected efficiencies for CB muons (from $Z$ decays)
             as a function of detector region, muon $\pt$ and muon $\eta$ as
             indicated in the figure, obtained from data and Monte Carlo
             simulation for the two reconstruction chains. The uncertainties
             are statistical only. The systematic uncertainties are discussed
             in \mySect~\ref{sec::Z_results_systematics}.}
    \label{fig:BgCorrectedEff}
\end{figure*}

The background-corrected efficiencies for CB muons
are shown in \myFig~\ref{fig:BgCorrectedEff}.
The background is estimated from Monte Carlo simulation,
as described in \mySect~\ref{sec::tag_and_probe_method}, and is subtracted bin by bin. The average CB muon reconstruction efficiency is $0.928\pm0.002$ for \Staco{} and \mbox{$0.958\pm0.001$} for \Muid{}. 
The difference in efficiency between the two chains arises mainly from the more stringent requirements on the reconstructed MS tracks in chain 1. The ratios between data and MC efficiencies are almost identical to the SFs already discussed for \myFig~\ref{fig:EffFullStat} as a consequence of the smallness of the background correction. 

\subsection{Reconstruction efficiencies for CB+ST muons}
\label{sec::Z_results_CBST}
\begin{figure*}[]
    \begin{center}
        \vspace*{-0.6ex}
        \begin{tabular}{@{\hspace*{-0.3ex}}cc}
            \includegraphics[width=0.50\textwidth]{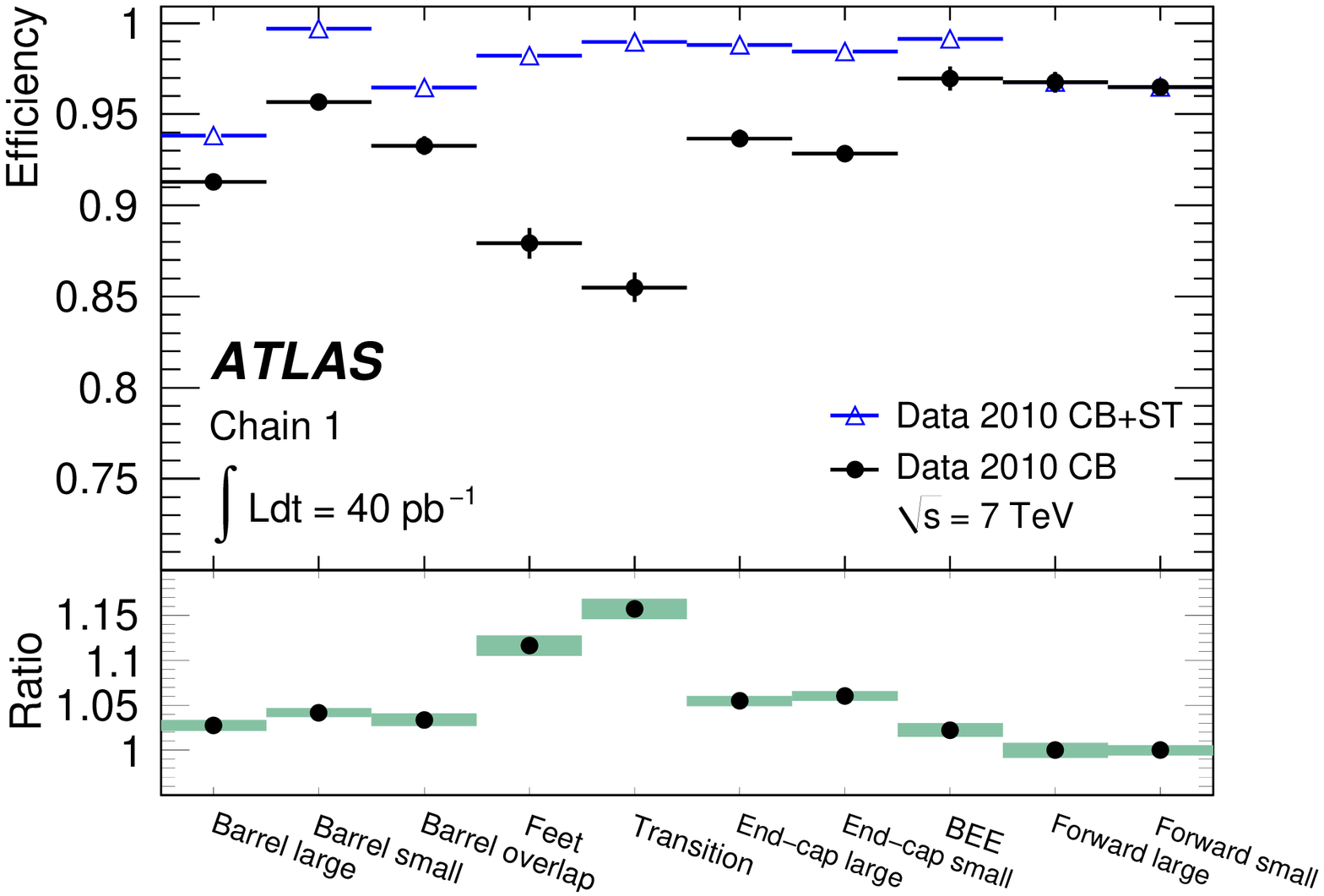}\hspace*{-2ex}&
            \includegraphics[width=0.50\textwidth]{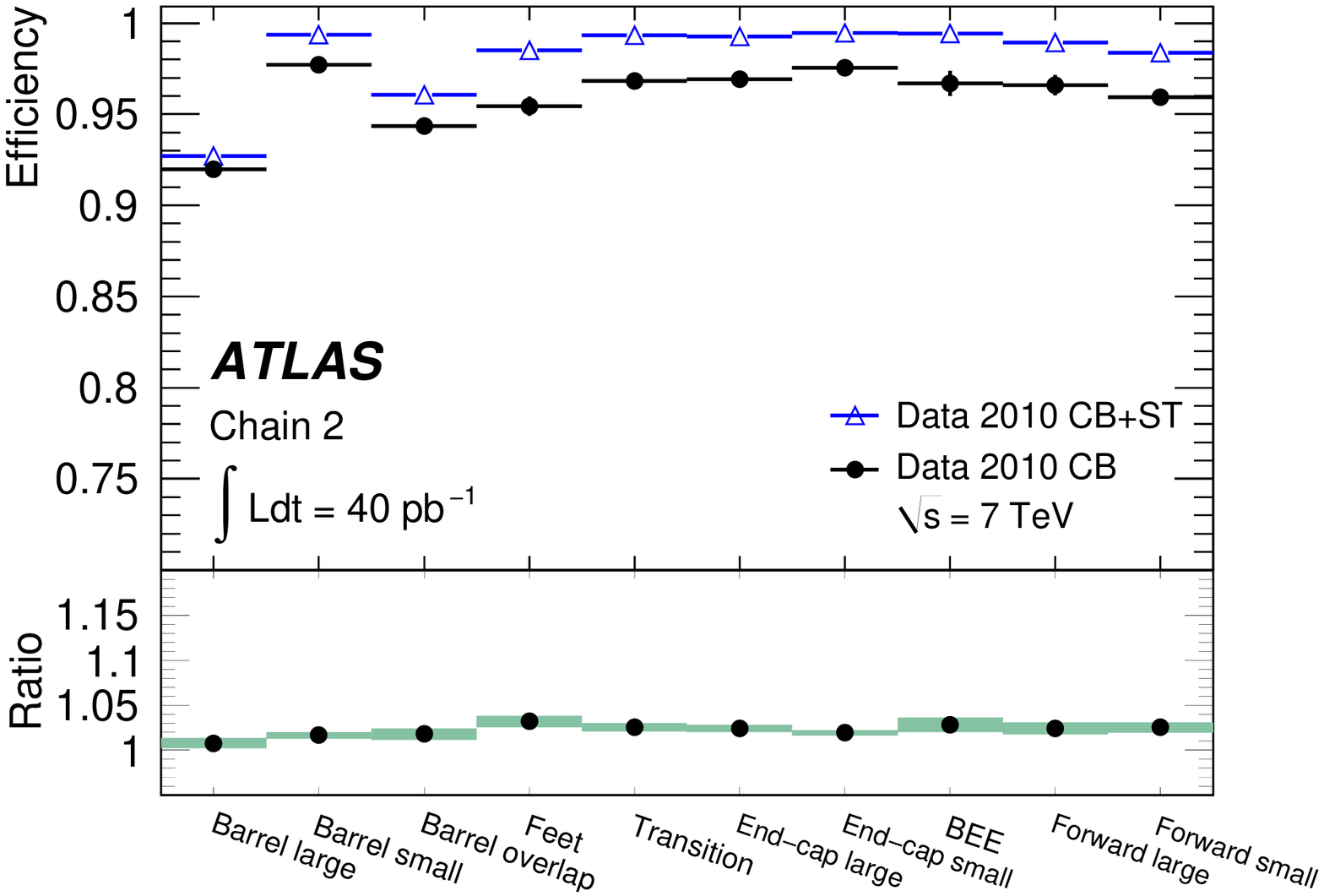}\hspace*{-2ex}\\
            \includegraphics[width=0.47\textwidth]{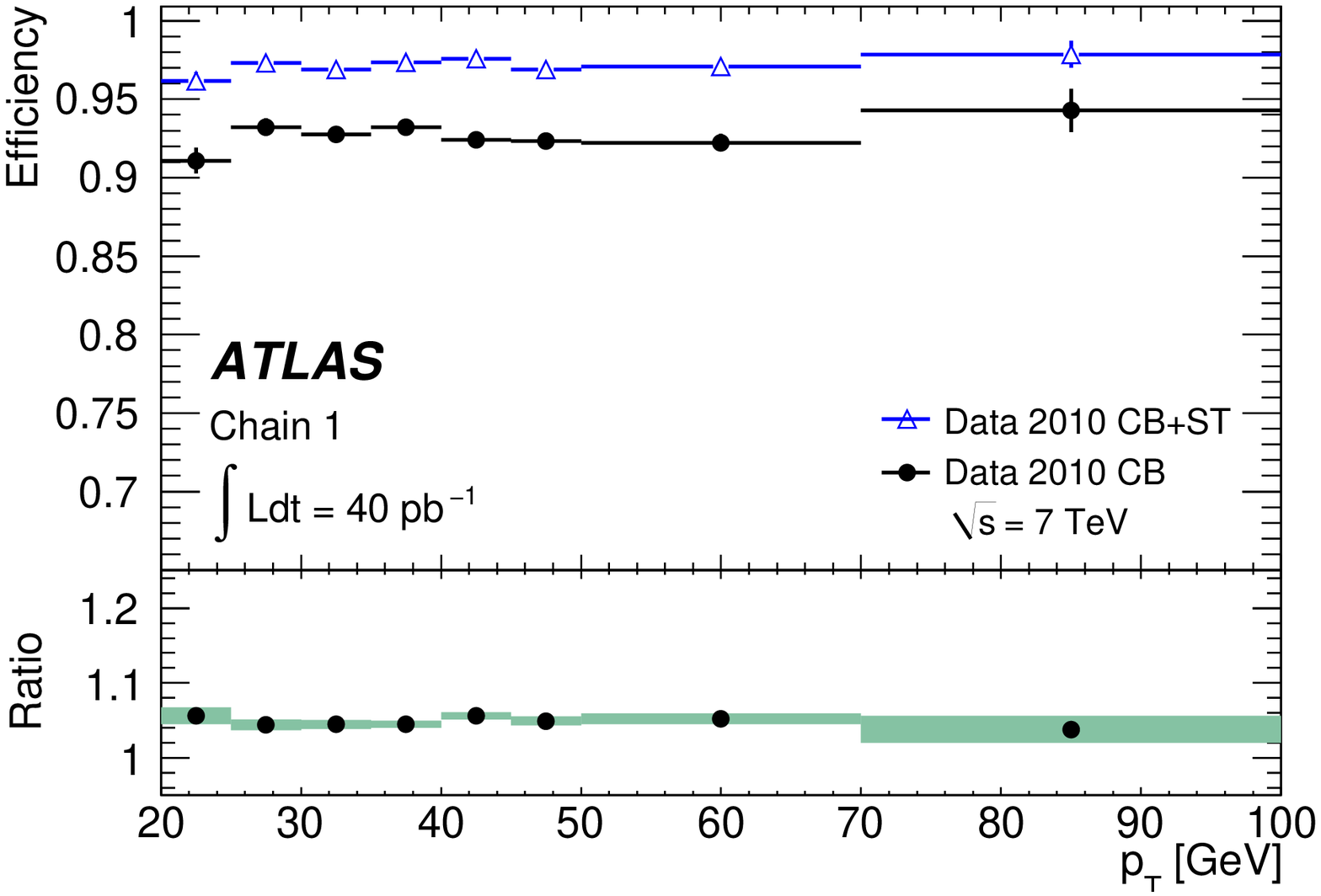}&
            \includegraphics[width=0.47\textwidth]{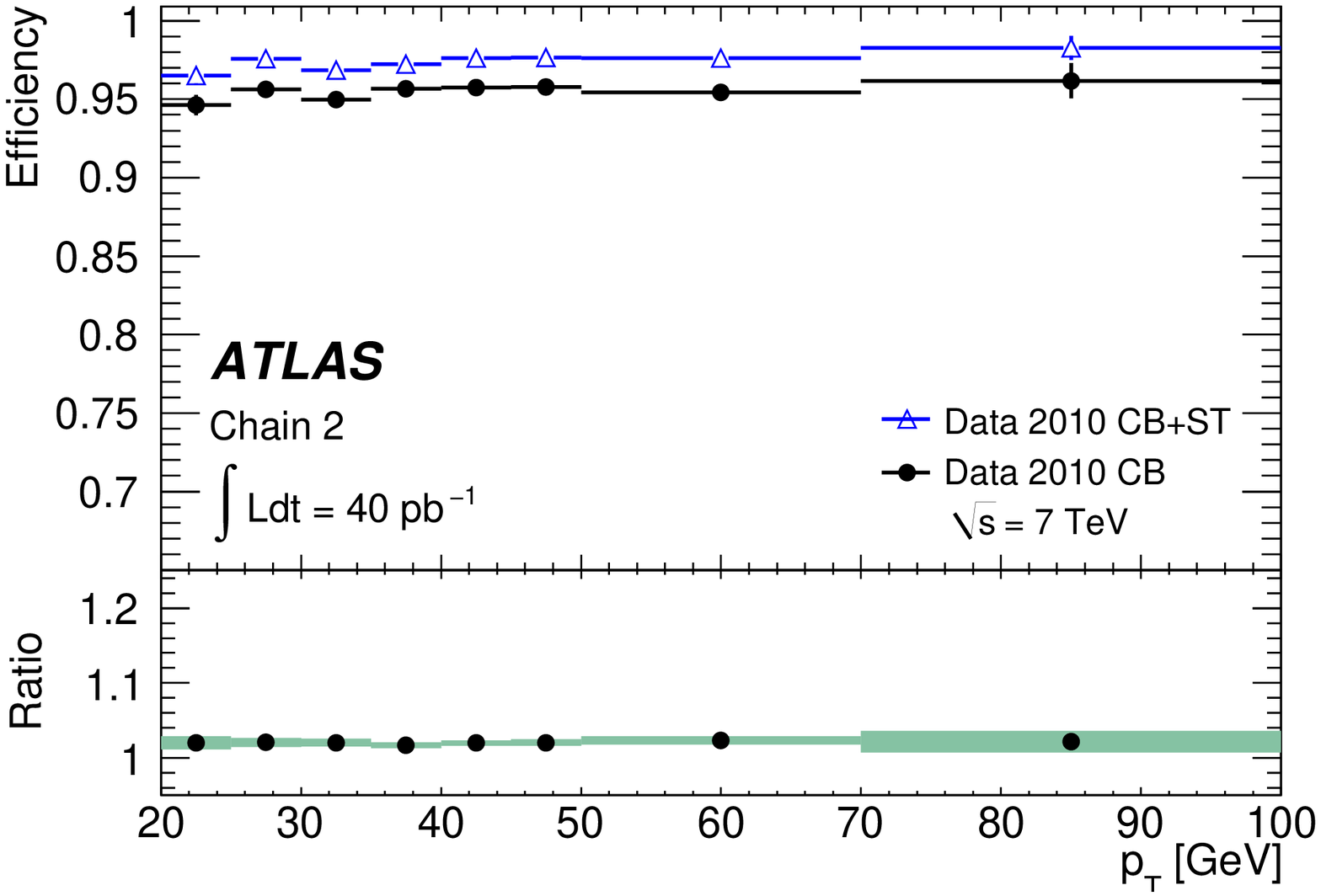}\\[-2pt]
            \includegraphics[width=0.47\textwidth]{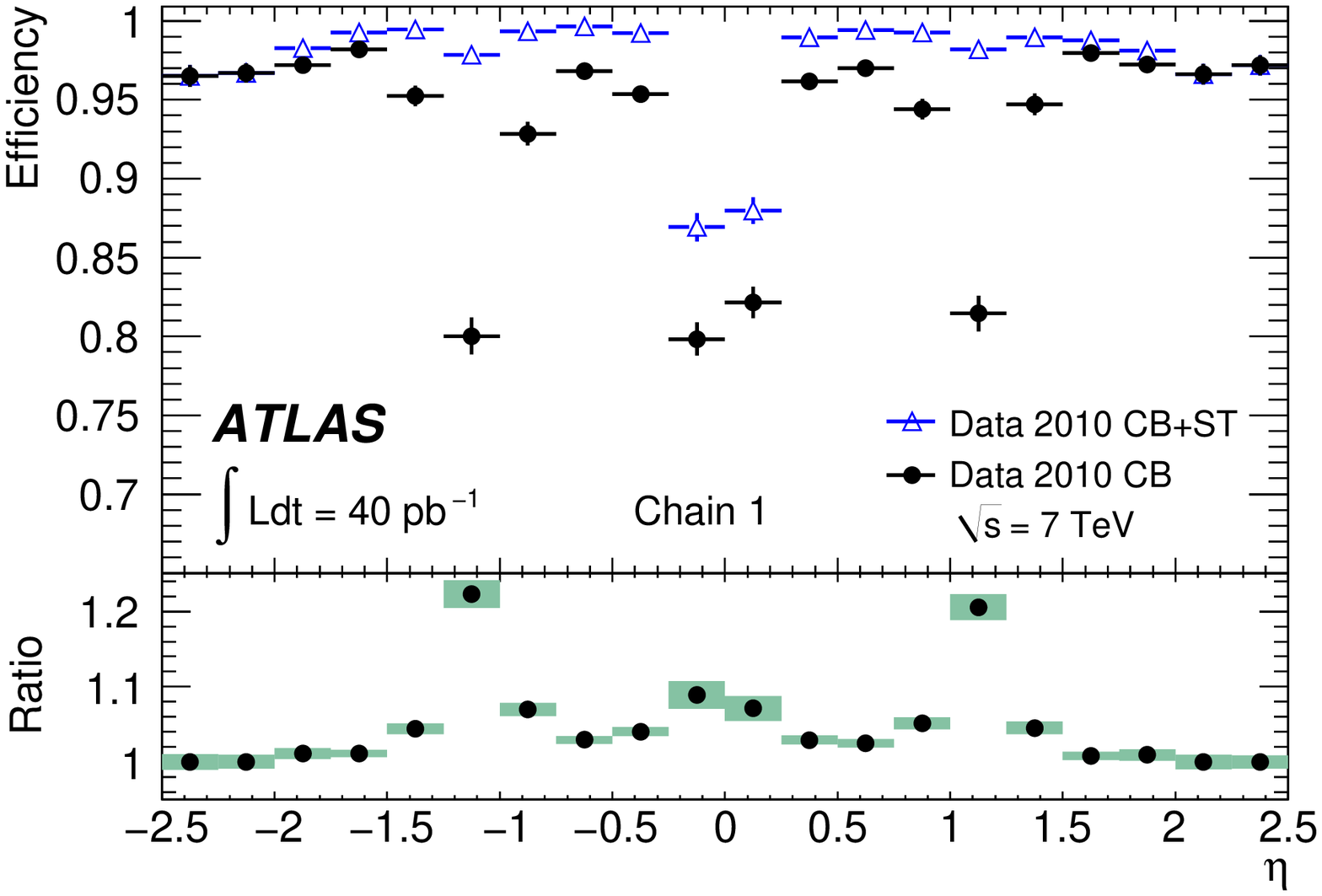}& 
            \includegraphics[width=0.47\textwidth]{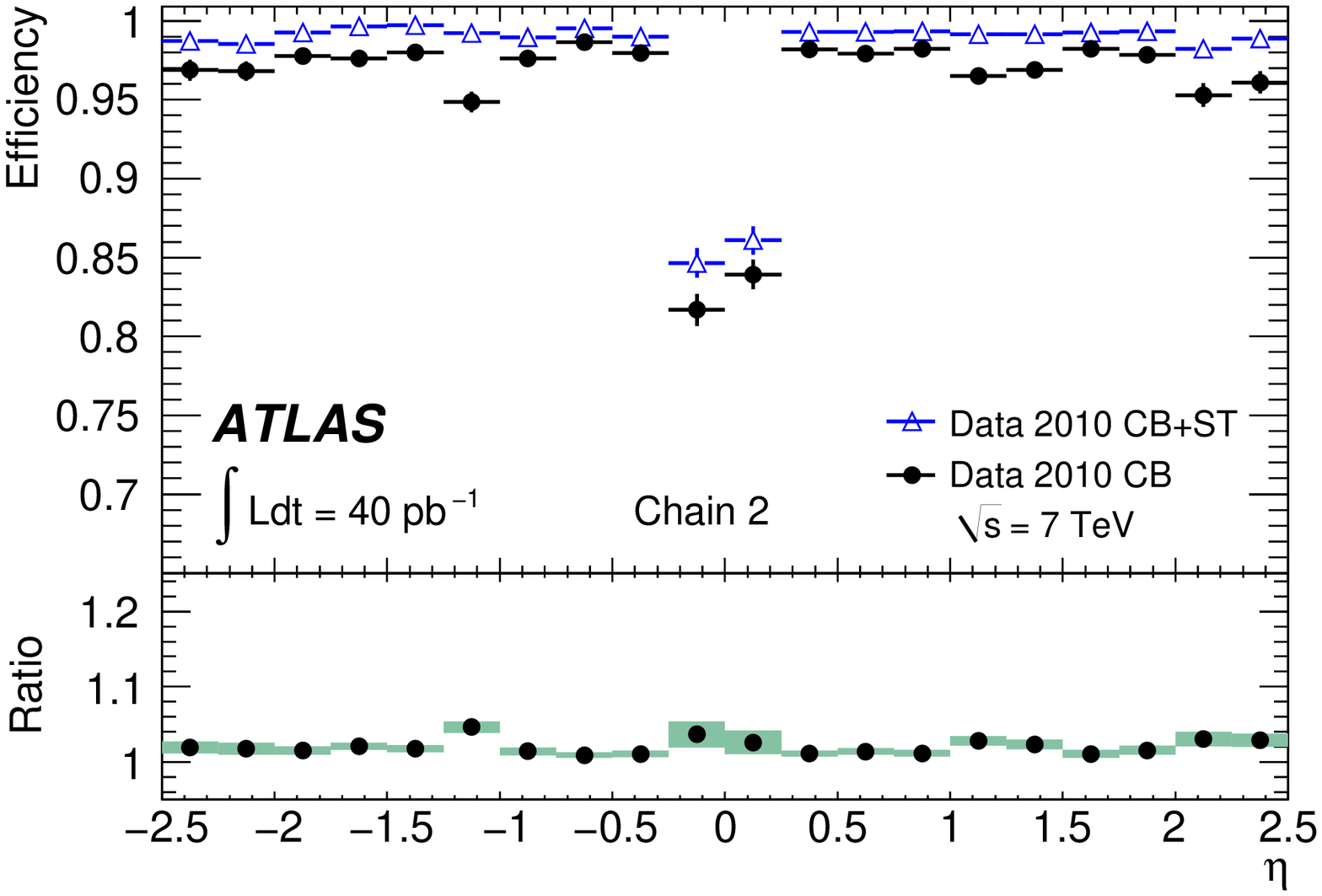}\vspace*{-0.5ex}
        \end{tabular}
    \end{center}
    \caption{Efficiencies for CB+ST muons (from $Z$ decays) in comparison
    to those for CB muons only, for the two reconstruction chains and
    as a function of detector
    region, muon $\pt$ and muon $\eta$ as indicated in the figure.
    The relative gain is shown in the lower panel of each figure.
    The uncertainties are statistical only. The systematic uncertainties
    are discussed in \mySect~\ref{sec::Z_results_systematics}.}
    \label{fig:tagEff-combined}
\end{figure*}%
\begin{figure*}[]
  \begin{center}
    \vspace*{-0.6ex}
    \begin{tabular}{@{\hspace*{-0.3ex}}cc}
        \includegraphics[width=.495\textwidth]{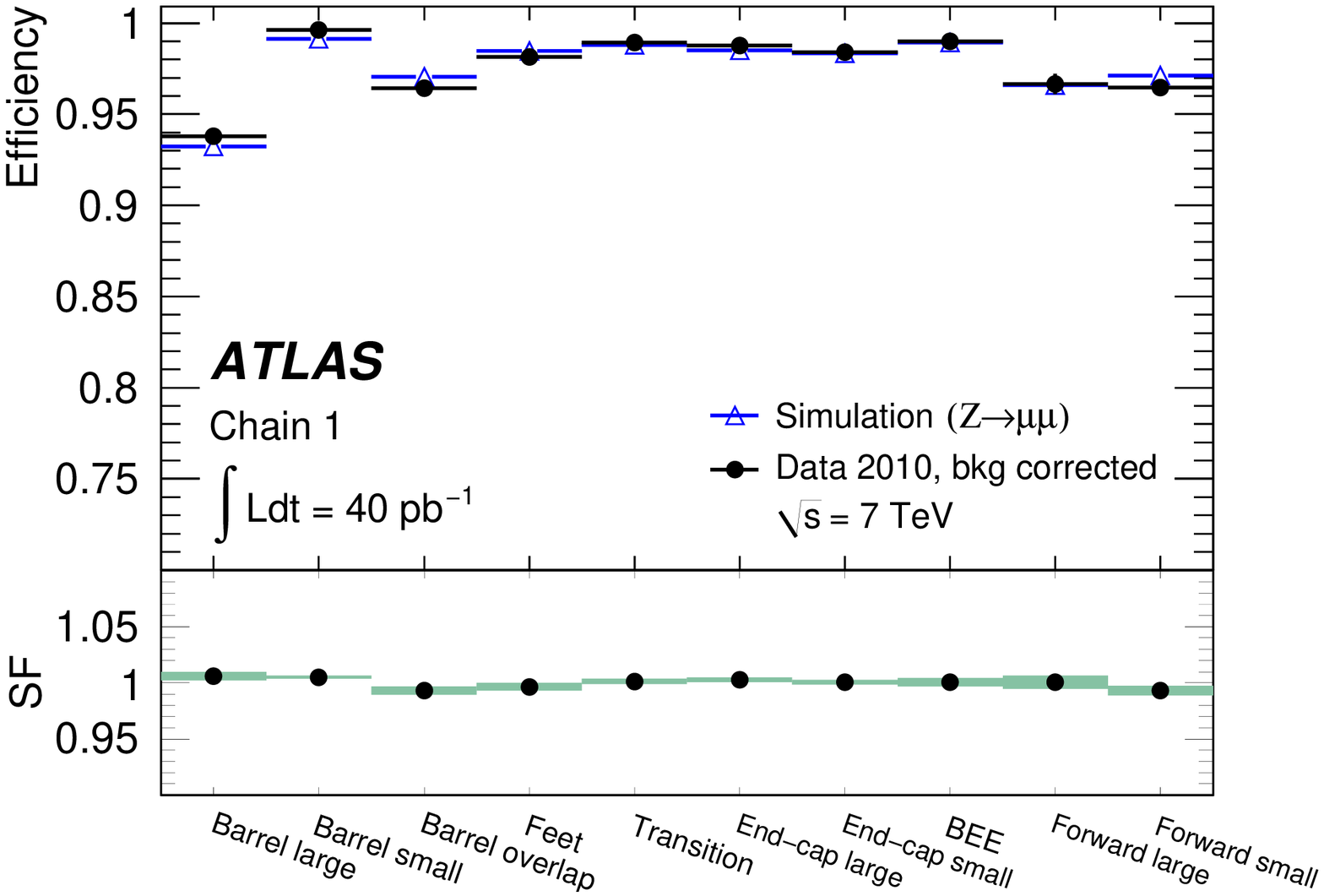}\hspace*{-2ex} &
        \includegraphics[width=.495\textwidth]{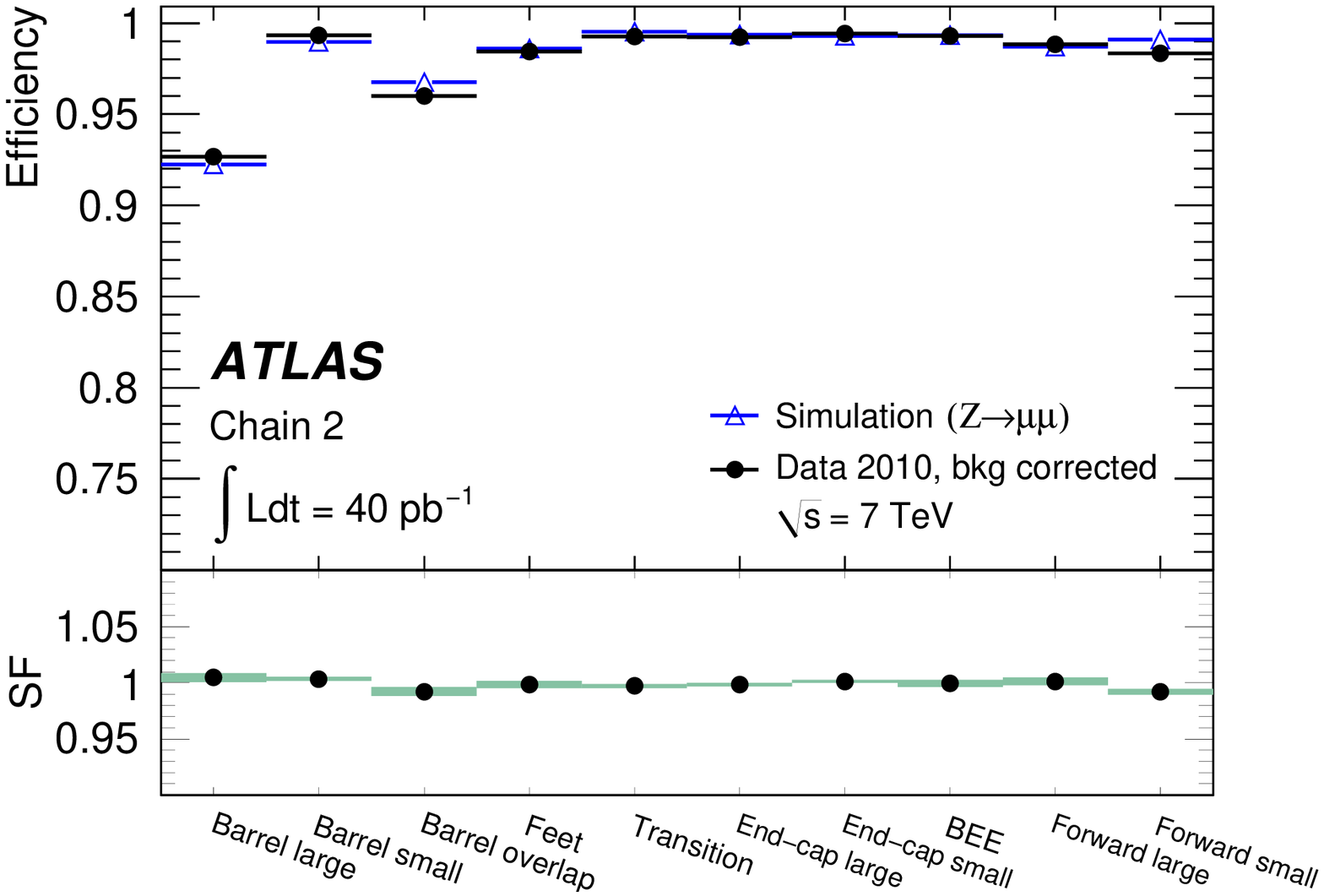}\hspace*{-2ex} \\
        \includegraphics[width=.465\textwidth]{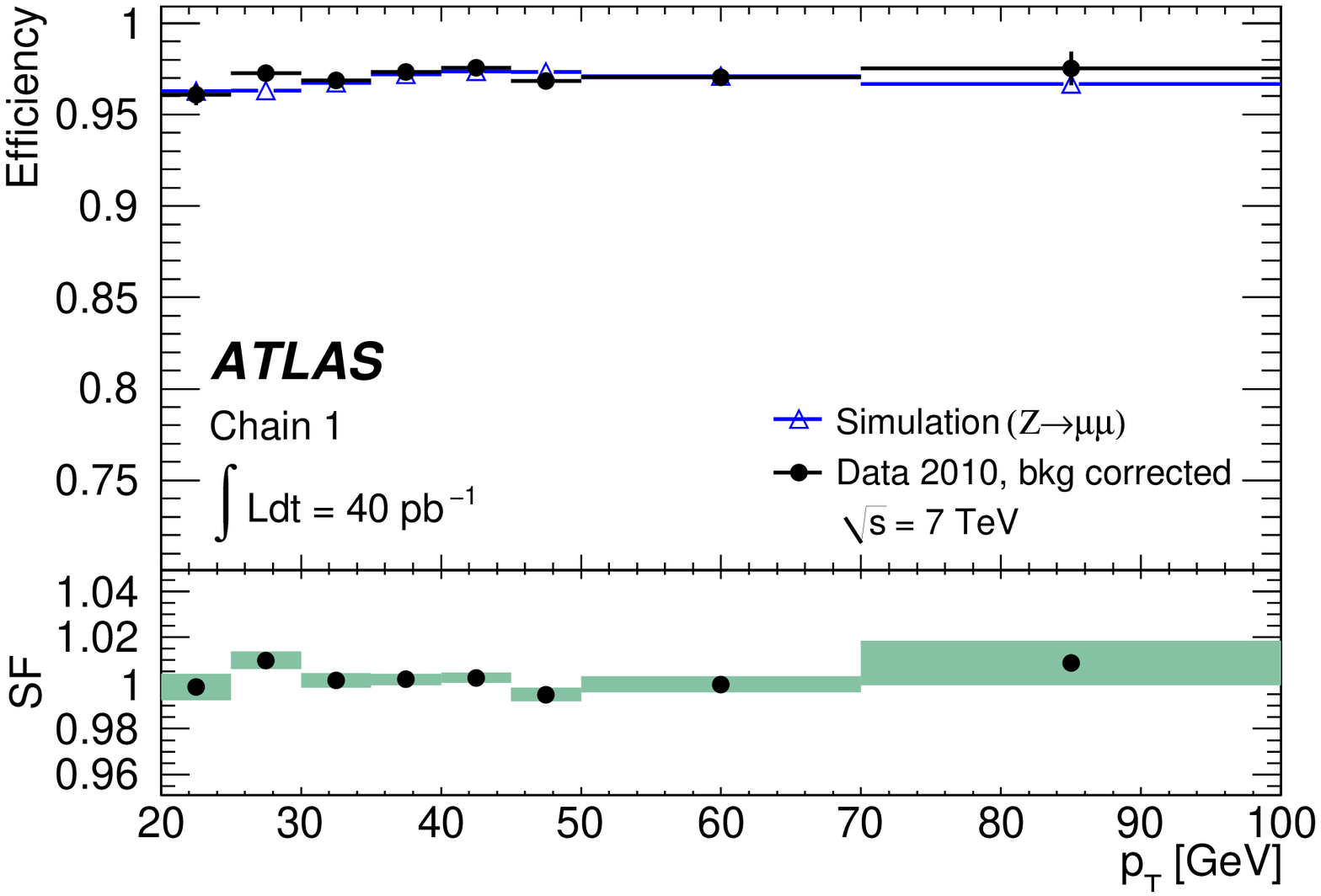}  &
        \includegraphics[width=.465\textwidth]{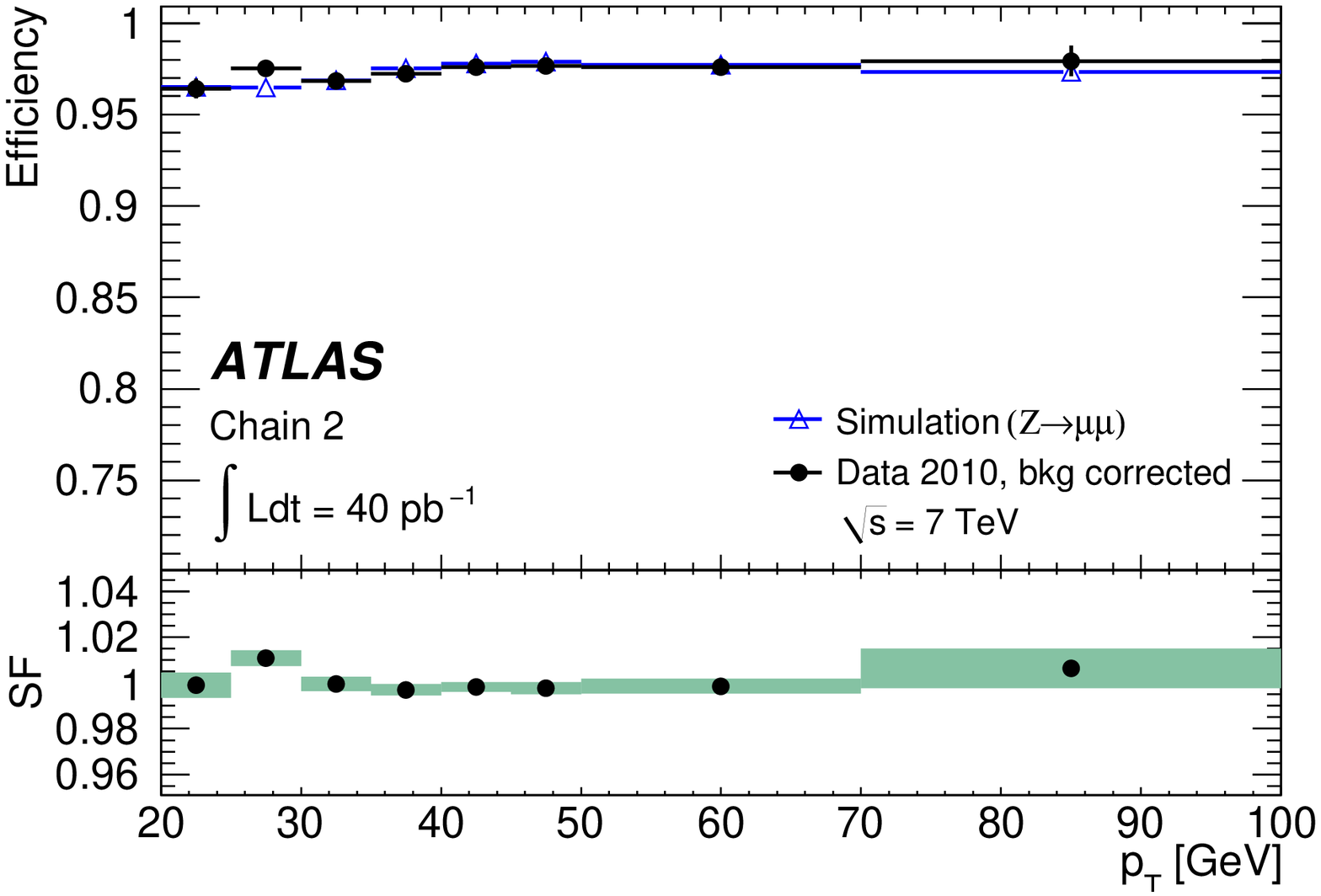}   \\[-2pt]
        \includegraphics[width=.465\textwidth]{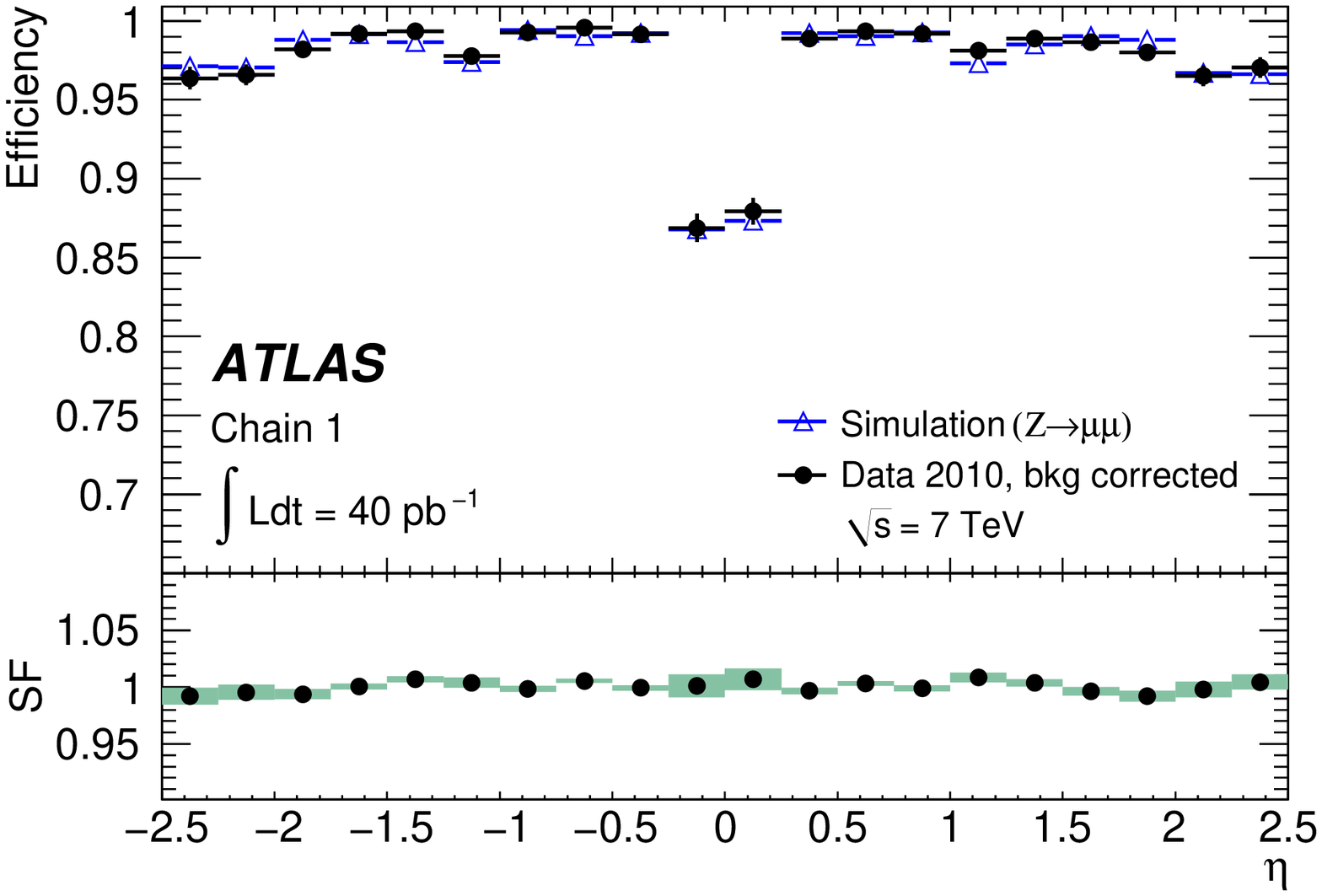} & 
        \includegraphics[width=.465\textwidth]{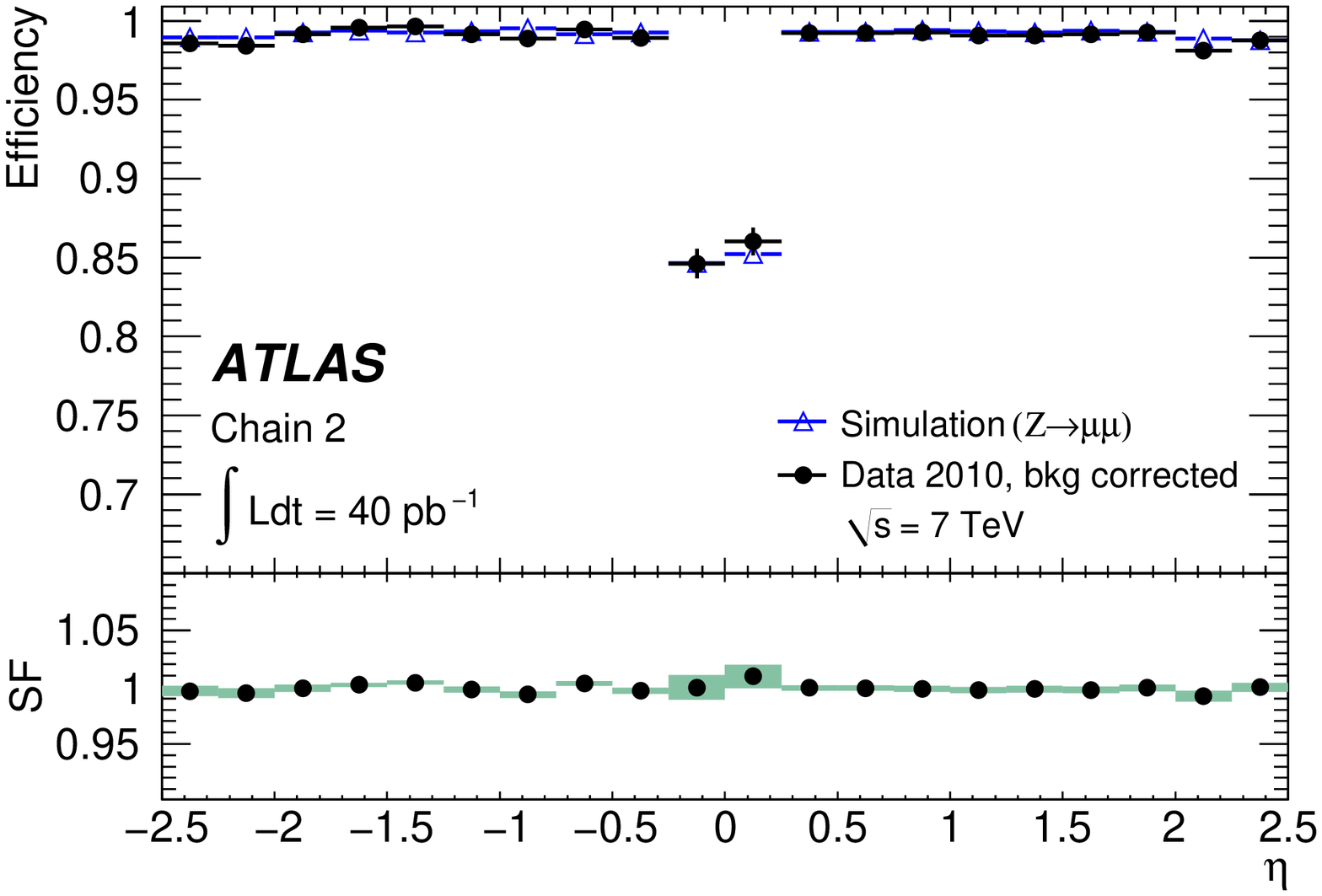}
    \end{tabular}\vspace*{-0.5ex}%
  \end{center}
  \caption{Efficiencies for CB+ST muons (from $Z$ decays), for the two reconstruction chains
    as a function of detector region, muon $\pt$ and muon $\eta$ as indicated in the figure.
    The efficiencies are obtained from data with background correction and
    from Monte Carlo simulation of the signal. 
    The corresponding scale factors are shown in the lower panel of each plot. The uncertainties are statistical only. The systematic uncertainty is discussed in \mySect~\ref{sec::Z_results_systematics}.}
  \label{fig:tagEffMC-Data}
\end{figure*}%
The degree to which segment tagging can recover some muons, in particular in detector regions with only partial muon coverage, is studied by measuring the efficiency for CB+ST muons.
The same tag-and-probe method is used with the only difference being that
the probe is matched to a CB or ST muon. 
Figure~\ref{fig:tagEff-combined} shows the measured CB+ST muon efficiencies as functions of the detector region, $\pt$ and $\eta$, in comparison with the corresponding CB muon efficiencies. The gains in efficiency when using ST muons in addition to the CB muons are presented in the lower panels of the plots.
These are largest in the ATLAS Feet ($13\%$)
and Transition ($15\%$) regions of the detector for $\Staco$.
For $\Muid$ the largest gain is $3\%$ in the Feet and BEE regions.
Figure~\ref{fig:tagEff-combined} also shows that the two chains have similar overall efficiencies for CB+ST muons, 0.970$\pm$0.001 for \Staco{} and 0.980$\pm$0.001 for \Muid{}.

In \myFig~\ref{fig:tagEffMC-Data}, the efficiency for CB+ST muons
measured from data is compared to the Monte Carlo expectations and scale factors are presented. Remarkable agreement between the measured and predicted efficiencies is achieved. The scale factors for CB+ST muons are
1.003$\pm$0.002 for \Staco{} and
1.001$\pm$0.002 for \Muid{}.

\subsection{Systematic uncertainties \label{sec::Z_results_systematics}}
Uncertainties on the background contributions and on the resolution of the detector are considered as sources of systematic uncertainties. The uncertainty due to the description of the finite detector resolution is estimated by varying the selection cuts when determining the efficiencies from MC-simulated data. For CB muons, the cuts on the mass window around $m_Z$ and the cut on the transverse momentum of the tag are each varied within $\pm1\sigma$ of the $m_{\mumu}$ and \pT{} resolutions. Other cuts are varied by $\pm10\%$. The resulting changes in the scale factors are quoted as systematic uncertainties. The normalisation of the background contribution inside the mass window is varied by $\pm10\%$ and the resulting differences in the scale factors are considered as additional systematic uncertainties. The individual uncertainties are considered to be uncorrelated and are added in quadrature to estimate the total systematic uncertainty. For values which result from an upwards and downwards variation, the larger value is used. The largest contribution arises from the level of background contamination, which depends primarily on the choice of the mass window and the normalisation of the backgrounds. Another important contribution is due to the variation of the probe isolation criteria.
The overall systematic uncertainty on the CB muon efficiency is $0.2\%$ for both chains. 

As the same tag-and-probe selection is used for the measurements of the CB+ST muon efficiencies, the same systematic uncertainties are expected for the corresponding scale factors. The systematic uncertainties on the ID muon efficiency scale factors are substantially smaller, principally due to the high purity of the MS probe muons.

\begin{figure*}[]
    \begin{center}
        \begin{tabular}{@{\hspace*{-0.3ex}}cc}
            \includegraphics[width=0.48\textwidth]{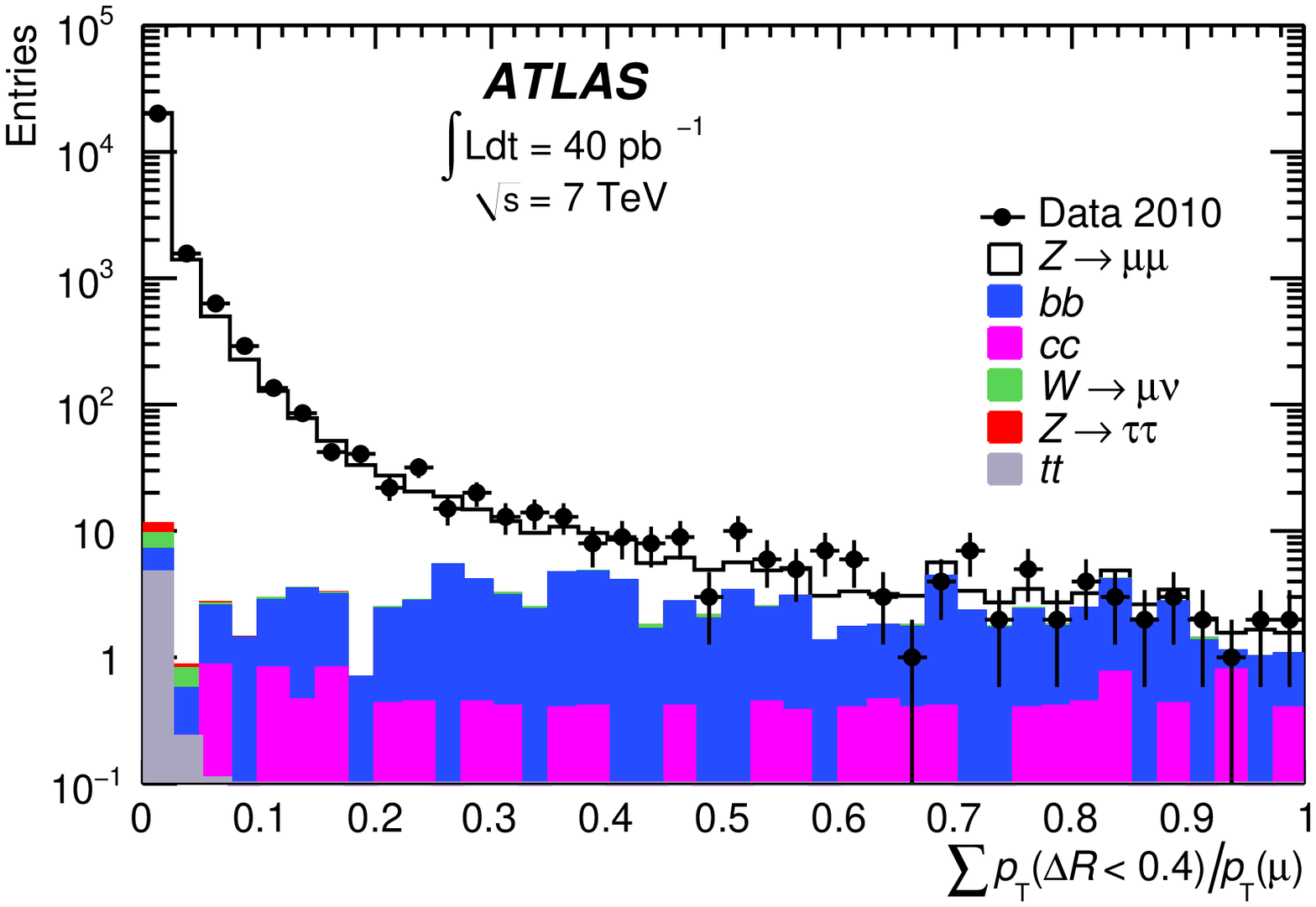} &
            \includegraphics[width=0.48\textwidth]{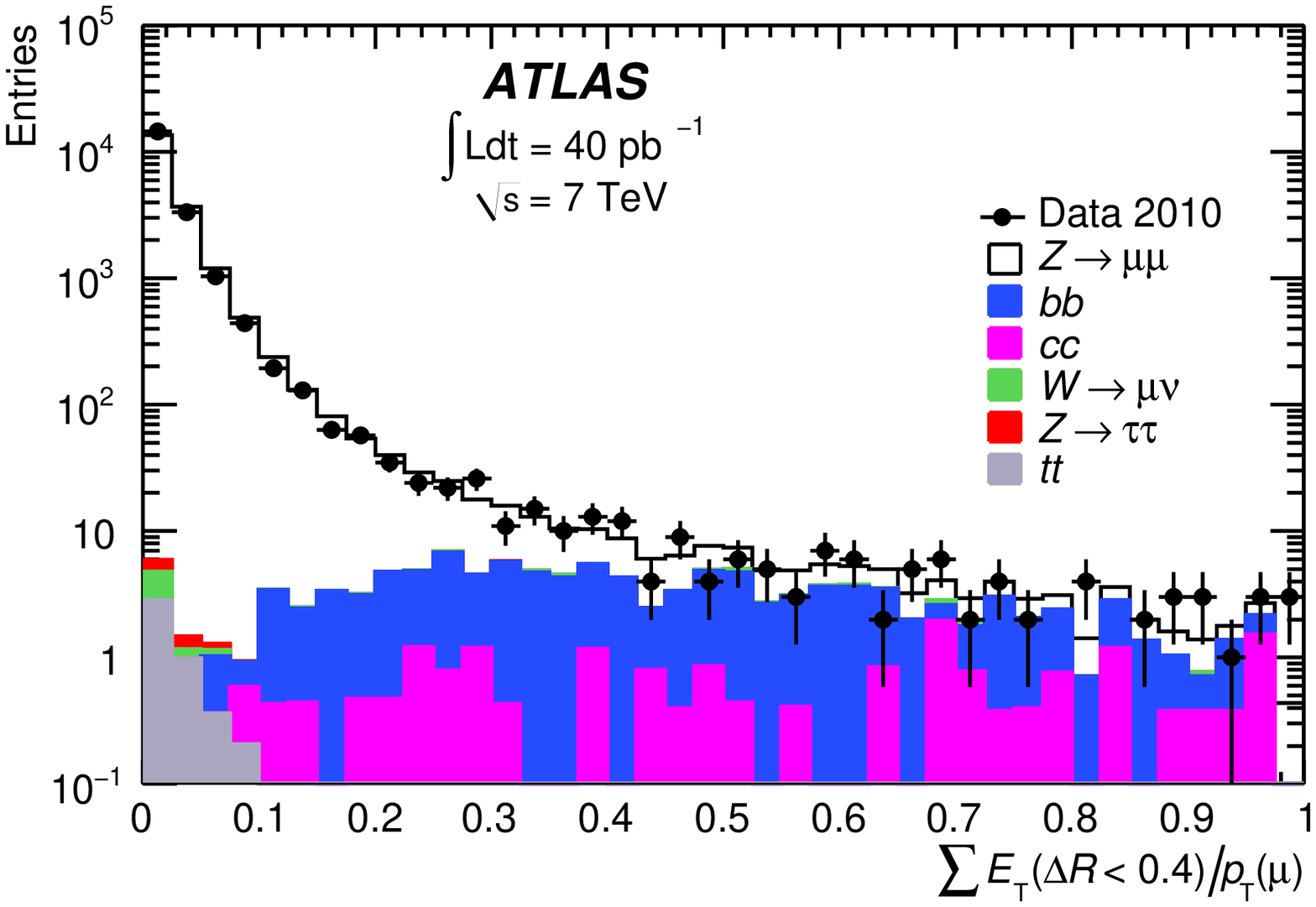} \\
            \includegraphics[width=0.48\textwidth]{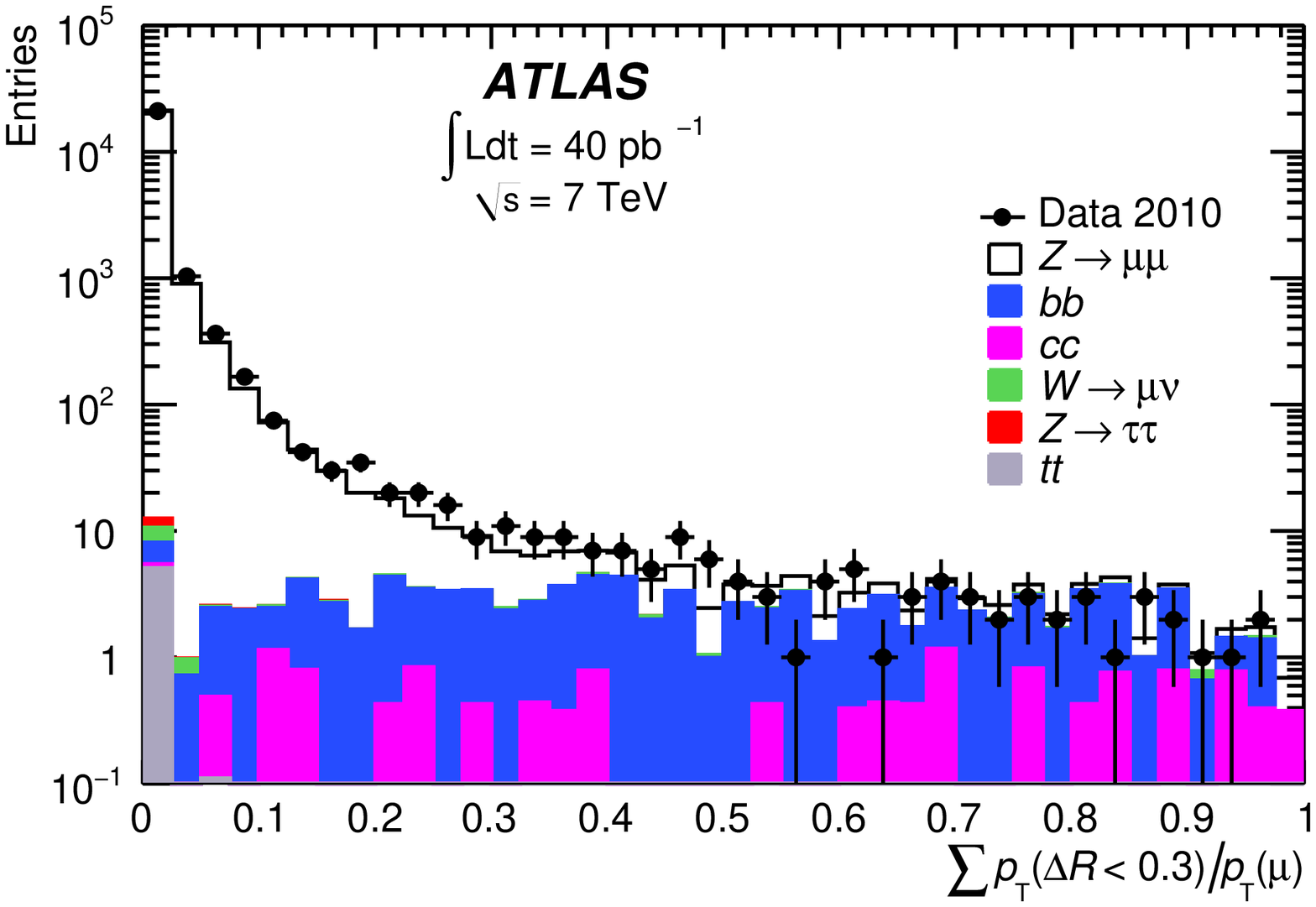}    &
            \includegraphics[width=0.48\textwidth]{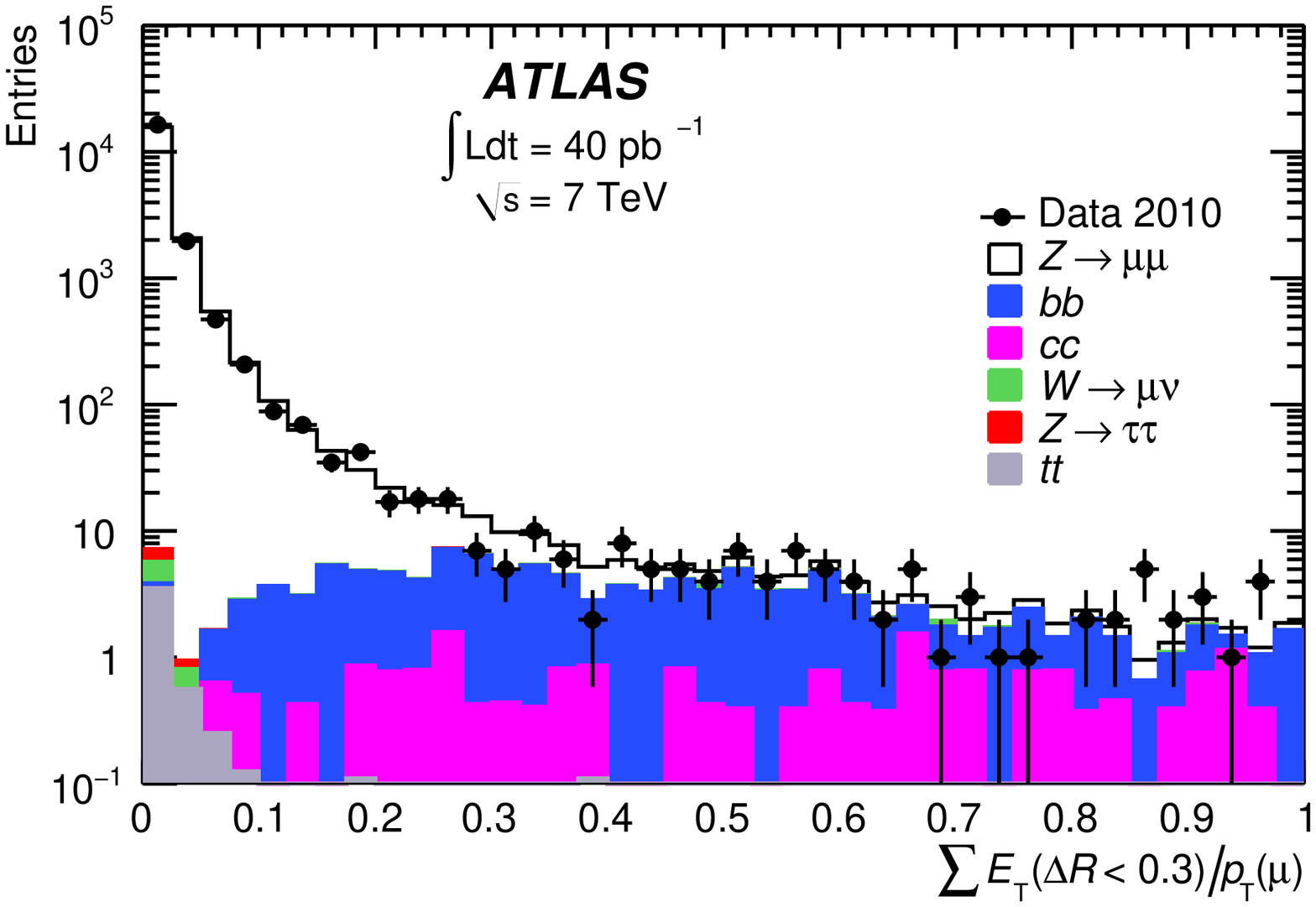}    \\
        \end{tabular}                                                       
    \end{center}
    \caption{Comparison of the measured track isolation (left) and calorimeter isolation (right) distributions of the probe muon (from $Z$ decays) with the Monte Carlo predictions, for two different cone sizes using the isolation variables defined in the text. The upper and lower plots correspond to $\Delta R=0.4$ and $\Delta R=0.3$, respectively. The simulation includes the effects of pile-up, as described in the text. The uncertainties are statistical.}
    \label{fig:isodist}
\end{figure*}%
\section{Measurement of the muon isolation efficiency}
\label{sec::iso_eff}
Muon isolation is a powerful tool for a high-purity event selection in many physics analyses, and is also used for rejecting muons from hadron decays in the $Z$ decay tag-and-probe analyses presented here. It is therefore desirable to quantify the reliability of the Monte Carlo prediction of the isolation efficiency (simulated using PYTHIA).\footnote{The effects of pile-up are taken into account in the simulation as described in \mySect~\ref{sec::MC}.} This is studied using the same event selection that was used for the reconstruction efficiency measurements, up to and including the selection of the tag muon (the specific chain used is not shown, since the performance is comparable for both). In this case, the probe muon is defined as a CB muon with $\pt>20$~GeV that fulfils the ID hit requirements described in \mySect~\ref{sec::tag_and_probe_selection}. We consider the following isolation variables:

\begin{itemize}
    \item track isolation\footnote{The track isolation, $\TisolRfour$, was defined in \mySect~\ref{sec::tag_selection}.} -- the summed {\pT} of tracks (excluding that of the muon) in cones of size $\Delta R=0.3$ and $\Delta R=0.4$ around the muon, divided by the {\pT} of the muon;
    \item calorimeter isolation -- the transverse energy ($E_\mathrm{T}$) deposition in the calorimeter  in cones of size $\Delta R=0.3$ and $\Delta R=0.4$ around the muon (with the muon's energy loss subtracted~\cite[p.194]{CSCBook}), divided by the {\pT} of the muon. 
\end{itemize}
The tag-and-probe selections, as described in \mySect~\ref{sec::tag_and_probe_selection}, only make use of ${\TisolRfour<0.2}$. However, the choice of isolation criteria depends on the analysis and this section presents the comparisons of 
data and Monte Carlo simulations for the following combinations of isolation variables:
\begin{itemize}
\item $\TisolRfour<0.2$ and $E_\mathrm{T}^{\Delta R<0.4} / \pt(\mu)<0.2$; \medskip
\item $\TisolRfour<0.1$ and $E_\mathrm{T}^{\Delta R<0.4} / \pt(\mu)<0.1$; \medskip
\item $\TisolRthree<0.1$ and $E_\mathrm{T}^{\Delta R<0.3} / \pt(\mu)<0.1$.
\end{itemize}

Figure~\ref{fig:isodist} compares the distributions of the measured isolation variables for the probe muons with the Monte Carlo predictions. The experimental and simulated distributions agree well, leading to a reliable prediction as a function of $\pt$, of the isolation efficiency, which is defined as the fraction of probe muons passing a given set of isolation cuts.

\begin{figure*}[]
  \begin{center}
      \begin{tabular}{@{\hspace*{-0.3ex}}cc}
        \includegraphics[width=0.48\textwidth]{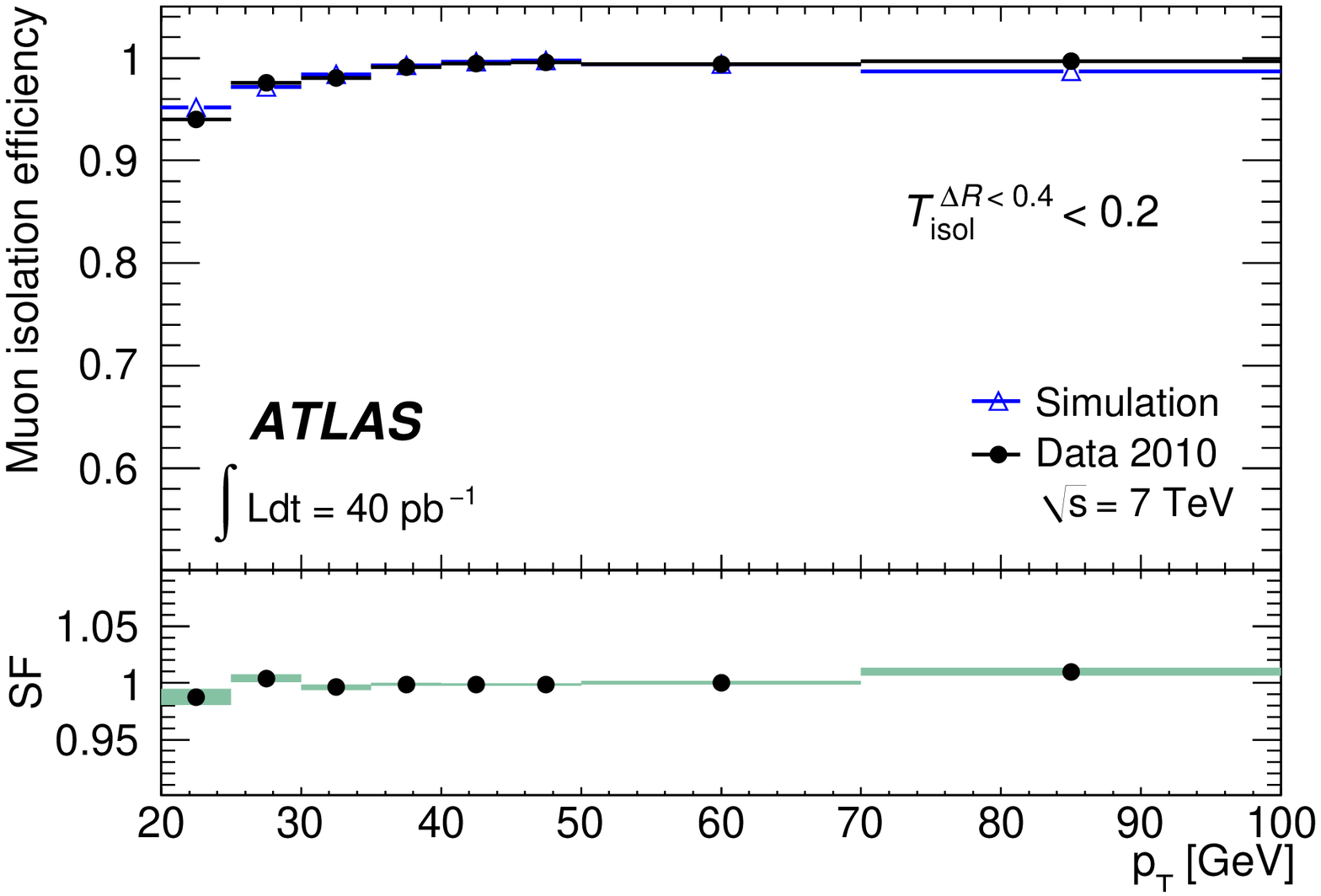} &
        \includegraphics[width=0.48\textwidth]{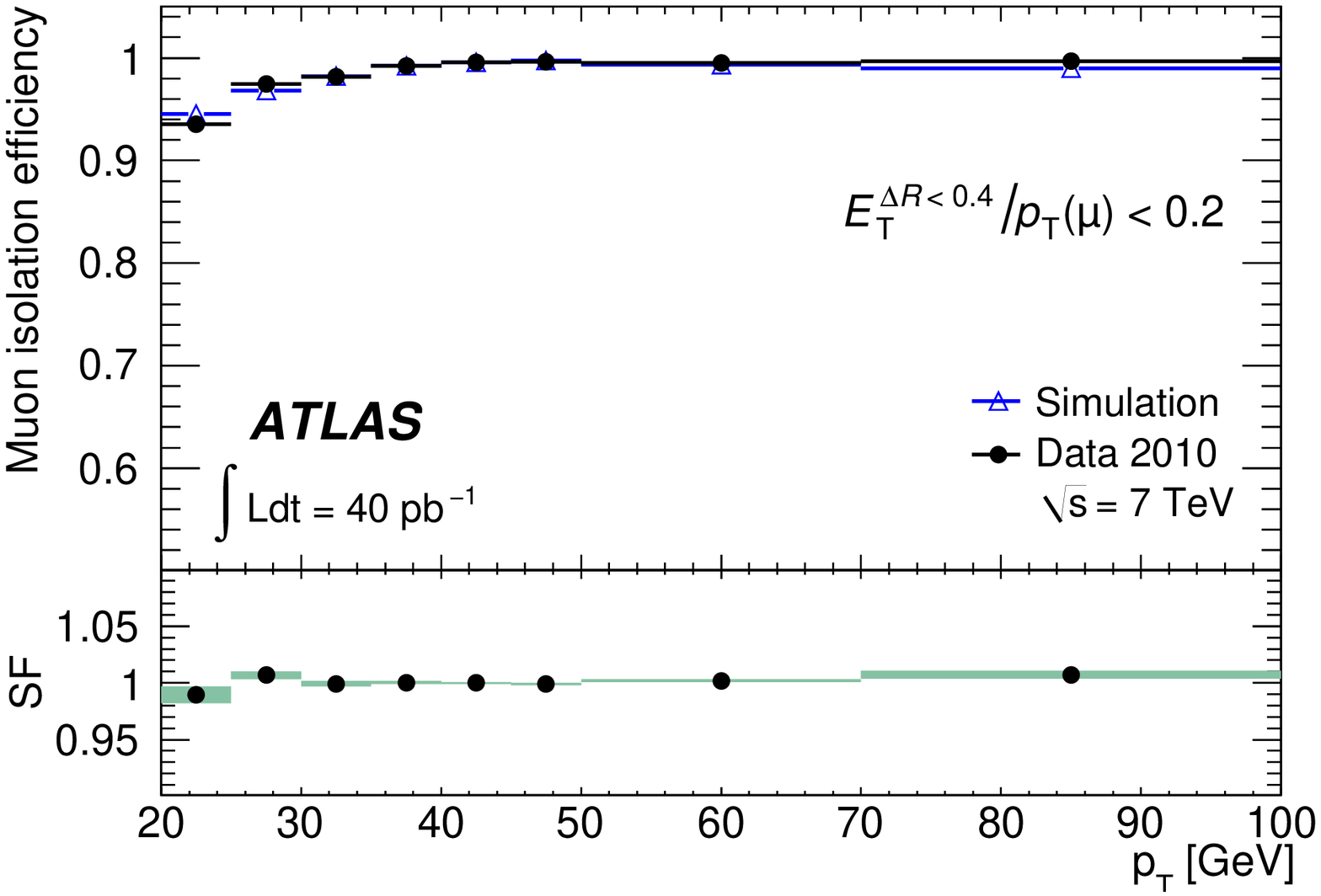} \\
        \includegraphics[width=0.48\textwidth]{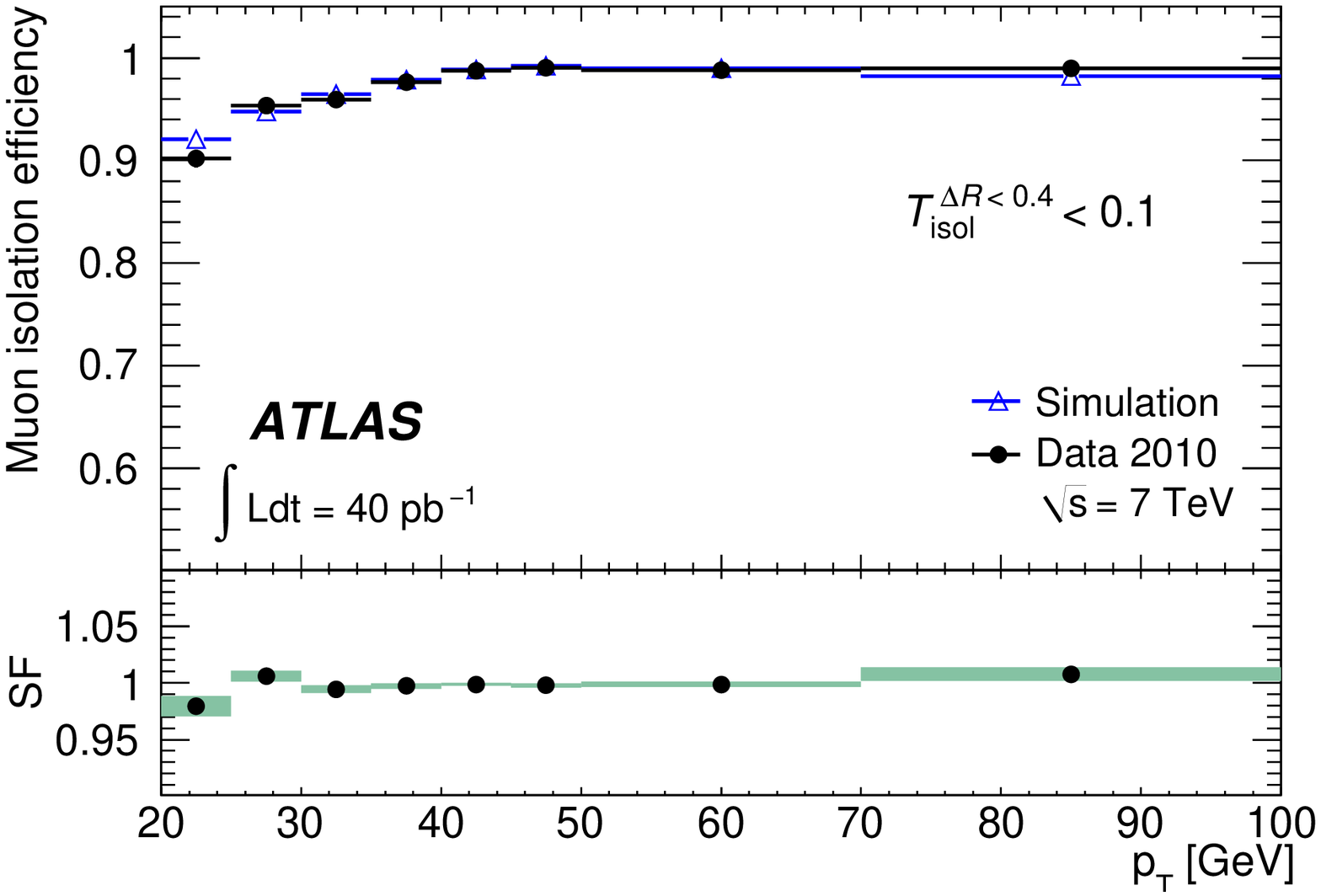} &
        \includegraphics[width=0.48\textwidth]{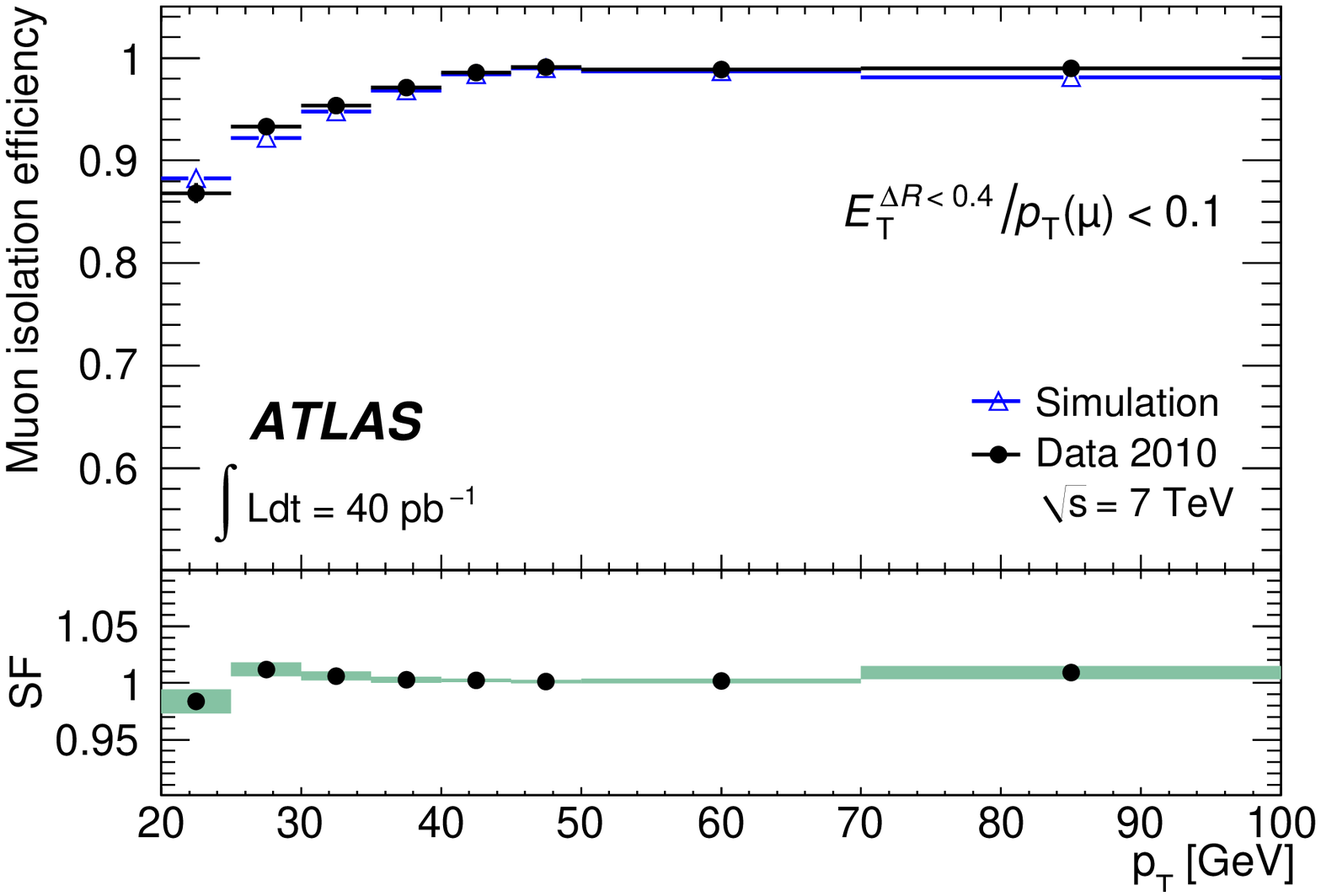} \\
        \includegraphics[width=0.48\textwidth]{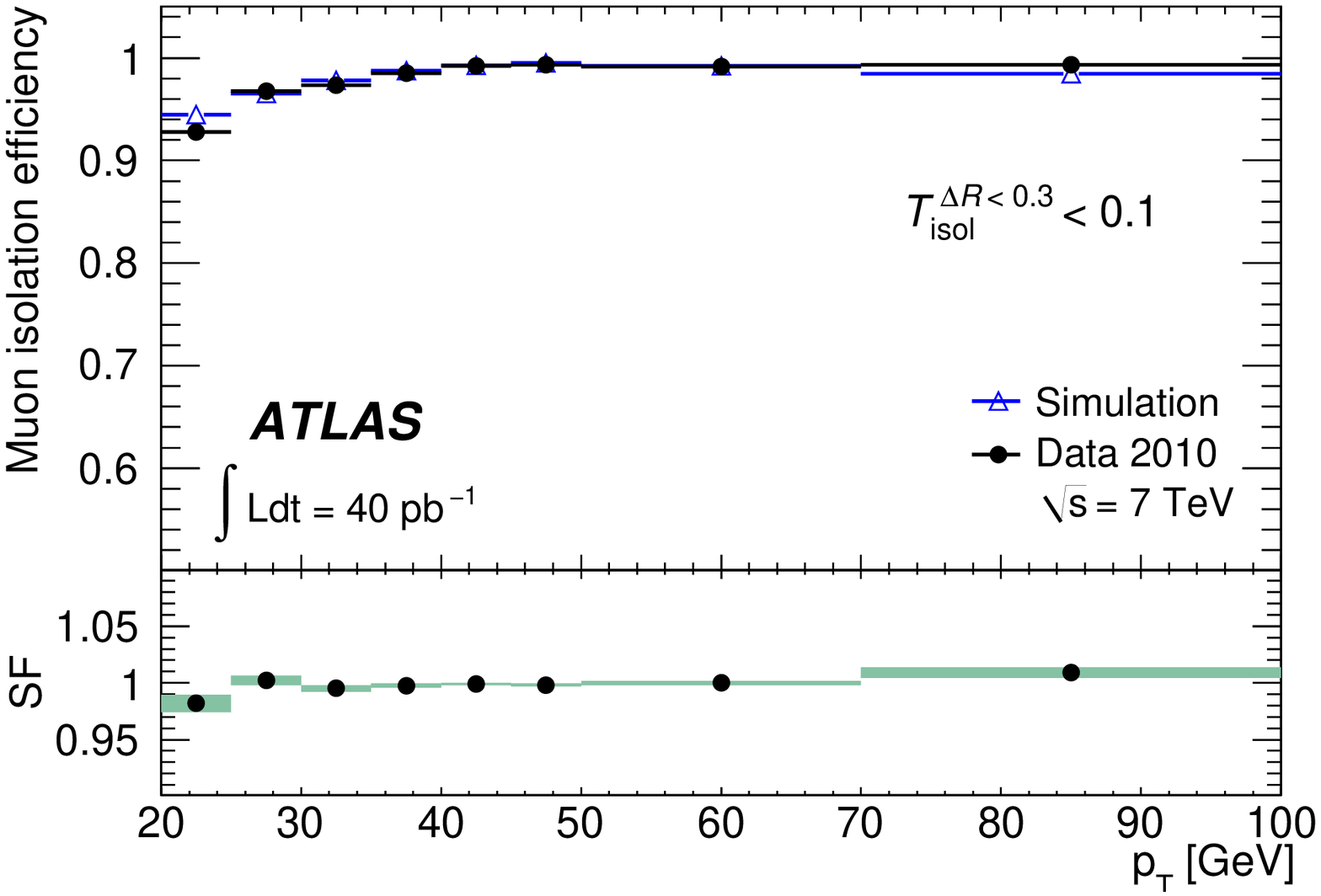}         & 
        \includegraphics[width=0.48\textwidth]{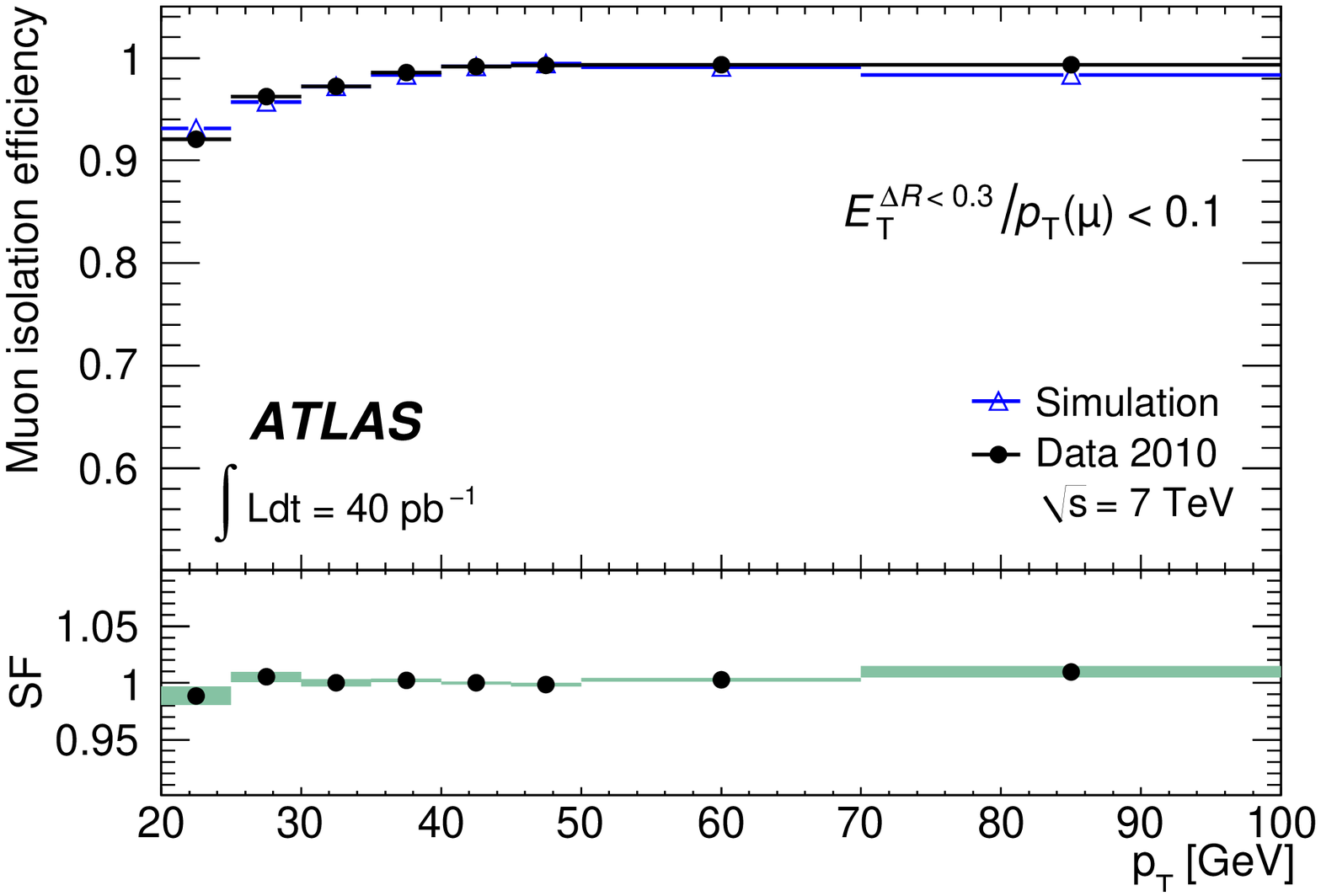}         \\
    \end{tabular}
  \end{center}
  \caption{Isolation efficiencies for muons from $Z$ decays as a function of \pT,
           for track isolation (left) and calorimeter isolation (right) requirements
           with different isolation cone radii, $\Delta R$, as described in the legend.
The Monte Carlo predictions include background processes as well as the $Z$ signal.
The uncertainties are statistical only.}
  \label{fig:isoeff}
\end{figure*}%
The measured isolation efficiencies and the corresponding Monte Carlo predictions are compared for \Staco in \myFig~\ref{fig:isoeff}; the results for \Muid are consistent. Experimental and simulated data agree within uncertainties. 
The lower efficiencies at low $\pt$ are mainly caused by the fact that the $\pt$  and $E_\mathrm{T}$ sums, which depend only weakly on the muon $\pt$, are divided by this quantity, leading to isolation variables that rise with decreasing muon $\pt$. They are also partially due to the background, which is non-negligible in the low-$\pt$ region.

\section{Measurement of the muon momentum resolution}
\label{sec::momres}
The muon momentum resolution of the ATLAS detector depends on the $\eta$, $\phi$, and $\pt$ of the muon~\cite{AtlasDetectorPaper}. In the ID, the $\pt$ dependence of the relative momentum resolution can be parameterised to a good approximation~\cite{CSCBook} by the quadratic sum of two terms,
\begin{align}
    \label{sec::momres::eqn1}
    \frac{\sigma_\mathrm{ID}(\pt)}{\pt} = a_\mathrm{ID}(\eta)\oplus
                                          b_\mathrm{ID}(\eta)\cdot\pt
&& \text{for $0<|\eta|<2.0$}~;\\
    \frac{\sigma_\mathrm{ID}(\pt)}{\pt} = a_\mathrm{ID}(\eta)\oplus
                                          \frac{b_\mathrm{ID}(\eta)\cdot\pt}
                                          {\mathrm{tan}^2(\theta)}
&& \text{for $2.0<|\eta|<2.5$}~. \notag
\end{align}
The first term describes the multiple scattering contribution, whilst the second term describes the intrinsic resolution caused by the imperfect knowledge of the magnetic field in the ID, by the spatial resolution of the detector components, and by any residual misalignment of the detector components. For $|\eta|>2.0$, the best parameterisation of the second term is given by ${b_\mathrm{ID}(\eta)\cdot\pt}/\mathrm{tan}^2(\theta)$. Measurements (from data) of the material distribution in the ID~\cite{ATLAS-CONF-2010-019,ATLAS-CONF-2010-058} constrain $a_\mathrm{ID}(\eta)$ to values which agree with the Monte Carlo prediction to within $5$\% in the barrel and $10$\% in the end-caps. The parameter $b_\mathrm{ID}(\eta)$ is derived from the dimuon invariant mass resolution in $Z\!\to\!\mu^+\mu^-$ decays.

The stand-alone muon resolution can be parameterised as follows:
\begin{equation}
  \label{sec::momres::eqn2}
    \frac{\sigma_\mathrm{SA}(\pt)}{\pt} =
      a_\mathrm{MS}(\eta,\phi)\oplus
                                   b_\mathrm{MS}(\eta,\phi)\cdot\pt\oplus
                                   \frac{c(\eta,\phi)}{\pt}~,
\end{equation}
where the first two terms parameterise the effect of the multiple scattering and the contribution of the intrinsic momentum resolution of the MS,  respectively. The third term parameterises the effect of the fluctuations of the muon energy loss in the calorimeters, but this is small for the momentum range under consideration and is fixed to the value predicted by MC simulation.

A special data set, recorded in 2011, with no toroidal magnetic field in the MS, was used to simulate high-momentum (i.e.~straight) tracks and estimate $b_\mathrm{MS}(\eta,\phi)$, yielding $b_\mathrm{MS}(\eta,\phi)\sim0.2$~TeV$^{-1}$ in the barrel and the MDT end-cap region (excluding the transition region) and $\sim0.4$~TeV$^{-1}$ in the CSC end-cap region, with a relative accuracy of about 10\% in both regions. This special data set made it possible to improve the alignment of the muon chambers, leading to $b_\mathrm{MS}(\eta,\phi)\lesssim0.2$~TeV$^{-1}$ everywhere in the MS in 2011.

\begin{figure*}[]
\begin{center}
    \begin{tabular}{@{\hspace*{-0.3ex}}cc}
        {\includegraphics[width=0.49\textwidth]{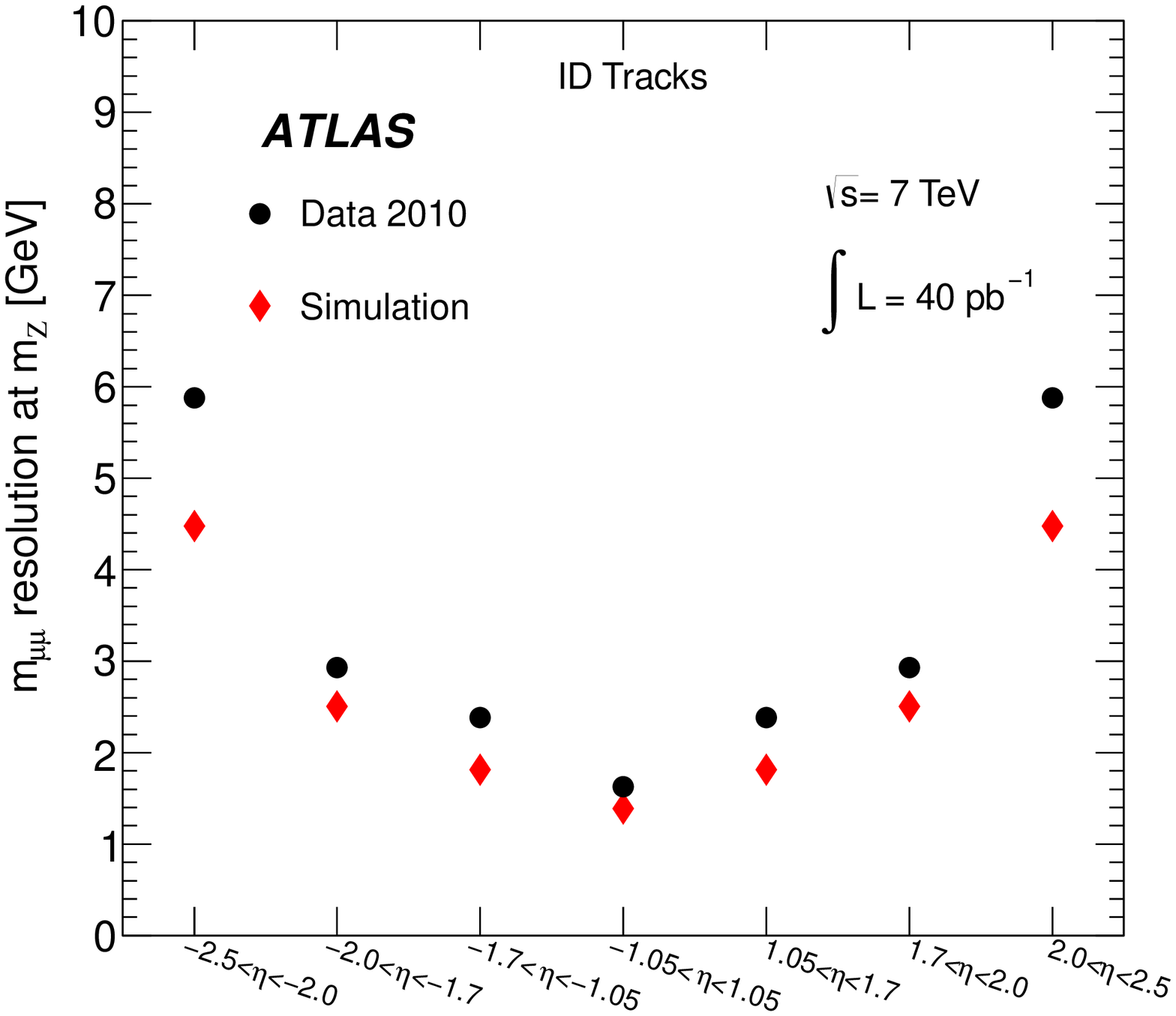}} &
        {\includegraphics[width=0.49\textwidth]{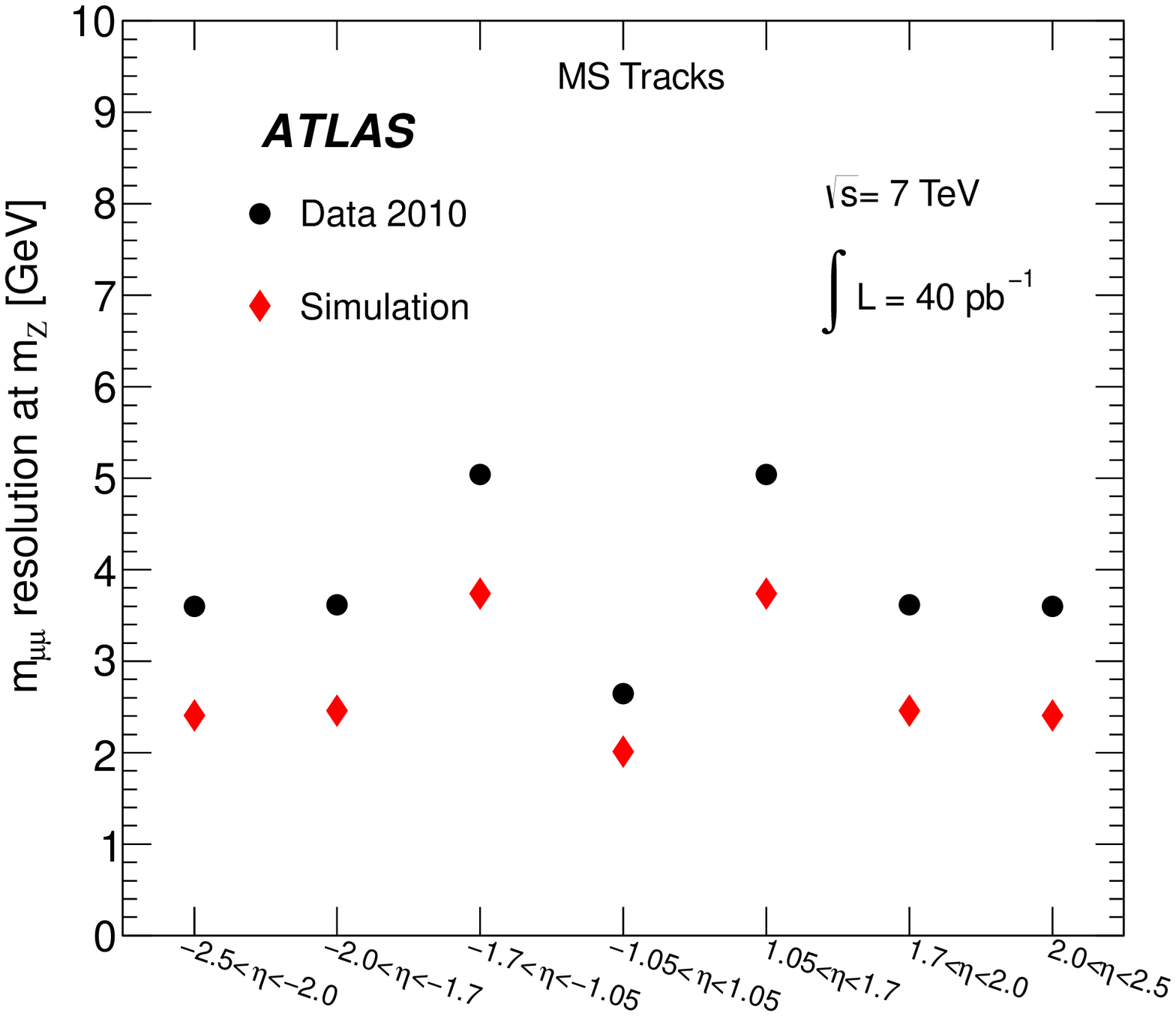}} \\
        \end{tabular}
\end{center}
\caption{The dimuon invariant mass ($m_{\mu\mu})$ resolutions in $Z\!\to\!\mu^+\mu^-$
         decays in the data and in the MC as a function of $\eta$ region with both
         decay muons in the same $\eta$ region, for the ID (left) and MS (right).
         The simulation assumes a perfectly aligned ATLAS detector.}
\label{sec_momres_fig1}
\end{figure*}%
Figure~\ref{sec_momres_fig1} shows the dimuon invariant mass resolution of the ID in $Z\!\to\!\mu^+\mu^-$ decays as a function of the pseudorapidity interval of the decay muons, where both are required to lie in the same interval. The mass resolution is the width of a Gaussian which, when convolved with the generator-level dimuon invariant mass, reproduces the dimuon invariant mass distribution observed in data. The ID dimuon invariant mass resolution is best in the barrel, where it is about 2~GeV, is better than 3~GeV for $|\eta|<2.0$ and degrades to about 6~GeV for $2.0<|\eta|<2.5$. The degradation of the mass resolution with increasing $|\eta|$  is primarily caused by the fact that as $|\eta|$ increases there is a lower field integral per track. That the dimuon invariant mass resolution measured in experimental data is worse than predicted (typically by about 30\%), is attributed to residual internal misalignments of the ID. The internal alignment of the ID was performed by minimising track residuals. This procedure has certain ambiguities which can be resolved by adding constraints such as the requirement that the energy/momentum ratio ($E/p$) distributions of electrons and positrons be the same. These constraints were only introduced into the alignment procedure for the 2011 data~\cite{ATLAS-CONF-2012-141}, in which a significantly improved dimuon invariant mass resolution is observed.

Due to the toroidal magnetic field, the relative momentum resolution of SA muons (and hence the corresponding dimuon invariant mass resolution -- as shown in \myFig~\ref{sec_momres_fig1}) is expected to be independent of the $\eta$ of the decay muons, except in the magnet transition region ($1.05<|\eta|<1.7$) where the magnetic field in the MS is highly non-uniform, with a field integral approaching 0 in certain ($\eta,\phi$) regions~\cite{AtlasDetectorPaper}. Furthermore, some chambers in the region $1.05<|\eta|<1.3$ were not yet installed,\footnote{This detector configuration was
also used for the 2011 data taking.}  which means that the momentum measurement relies on only two layers of chambers, causing a significant degradation in the momentum resolution. 

Figure~\ref{sec_momres_fig1} also shows that the MS dimuon invariant mass resolution is consistently worse in data than in simulation (typically between 30\% and 50\% worse, depending on $\eta$ region). Two sources for this effect were identified.
\begin{enumerate}
    \item   {\bf Asymmetry of the magnetic field}: 
            in the MC simulation, a perfectly aligned detector is assumed. In reality, the two end-cap toroid systems are not symmetric with respect to the plane orthogonal to the major axis of the ID, and situated at the centre of the detector. This small asymmetry translates
            into an asymmetry of the magnetic field integrals, in particular in the transition regions.
            The reconstruction of the 2010 data with a corrected field map improves the dimuon invariant mass resolution in the transition
            region by 0.4~GeV.
    \item   {\bf Residual misalignment of the muon chambers}:
            even after the MS alignment procedures are applied, residual misalignments remain, which limit the attainable momentum resolution. The analysis of a special set of 2011 data with no magnetic field
            in the MS was used to produce a Monte Carlo simulation of $Z\!\to\!\mu^+\mu^-$ events with the addition of a realistic residual misalignment of the MS. The
            results of this simulation are in agreement with the experimentally
            determined invariant mass resolutions.
\end{enumerate}

\begin{figure}[]
\begin{center}
    {\includegraphics[width=0.49\textwidth]{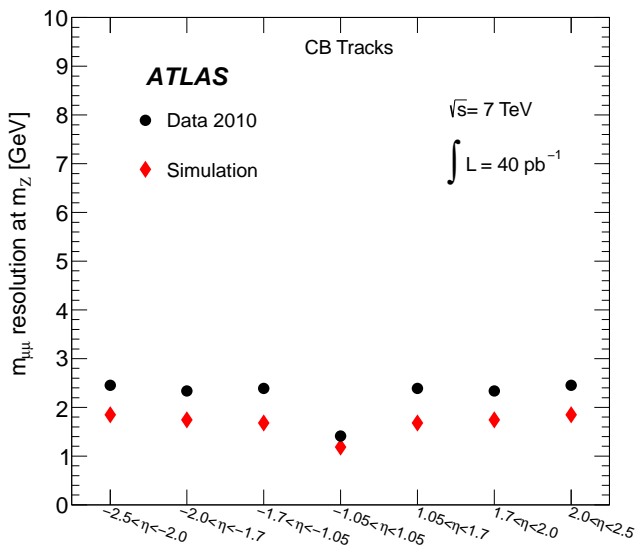}}
\end{center}
\caption{Dimuon invariant mass ($m_{\mu\mu})$ resolution for combined muons in $Z\!\to\!\mu^+\mu^-$ decays in the data and in the MC as a function of $\eta$ region with both decay muons in the same $\eta$ region. The simulation assumes a perfectly aligned ATLAS detector.}
\label{sec_momres_fig2}
\end{figure}

The dimuon invariant mass resolution obtained with CB muons profits from the complementary momentum measurements of the ID and MS. 
As shown in \myFig~\ref{sec_momres_fig2}, a dimuon invariant mass resolution between 1.4~GeV and 2.5~GeV is achieved, with little dependence on $\eta$.

\begin{figure*}[]
\begin{center}
    \begin{tabular}{@{\hspace*{-0.3ex}}cc}
        {\includegraphics[width=0.49\textwidth]{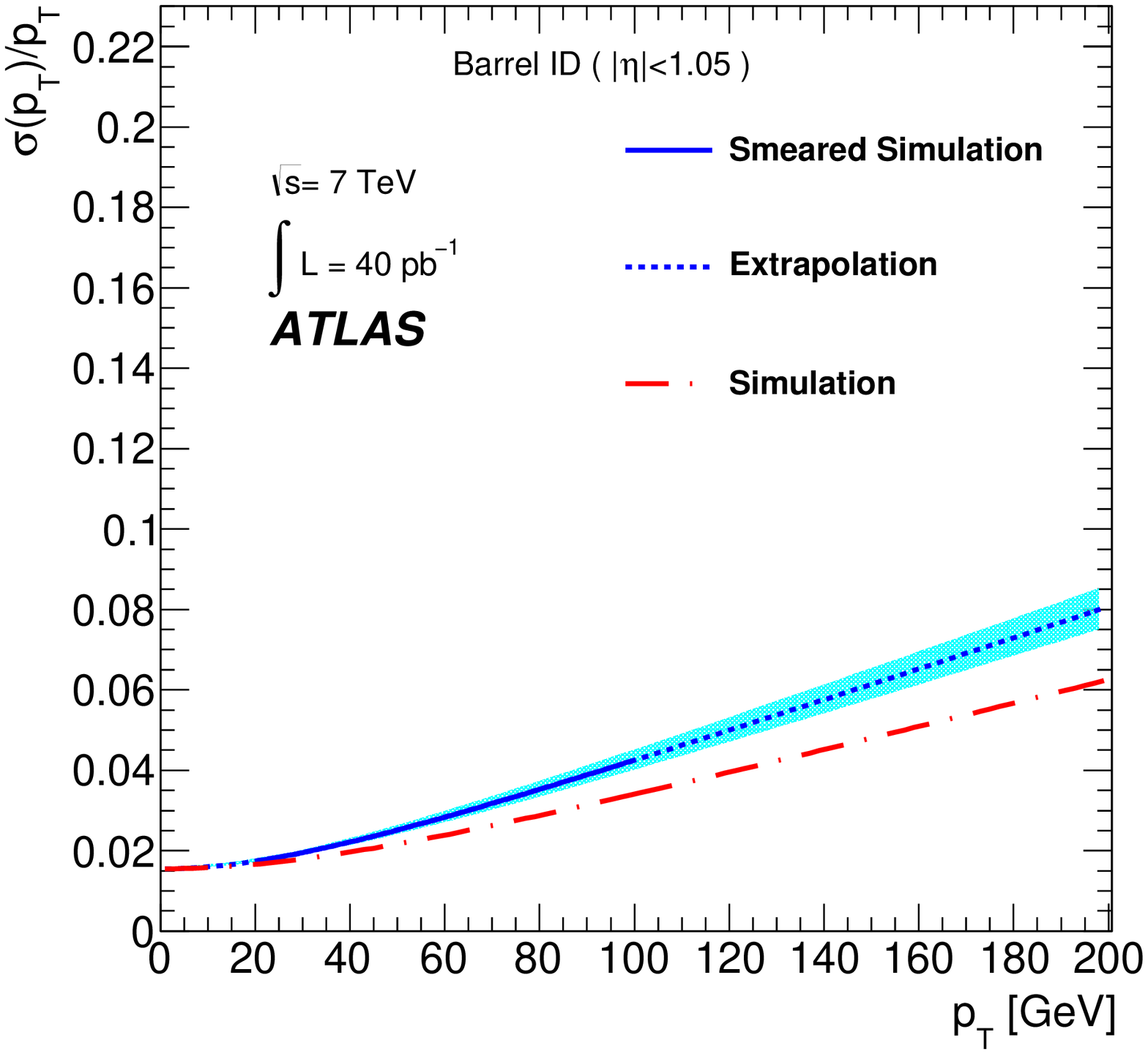}} &
        {\includegraphics[width=0.49\textwidth]{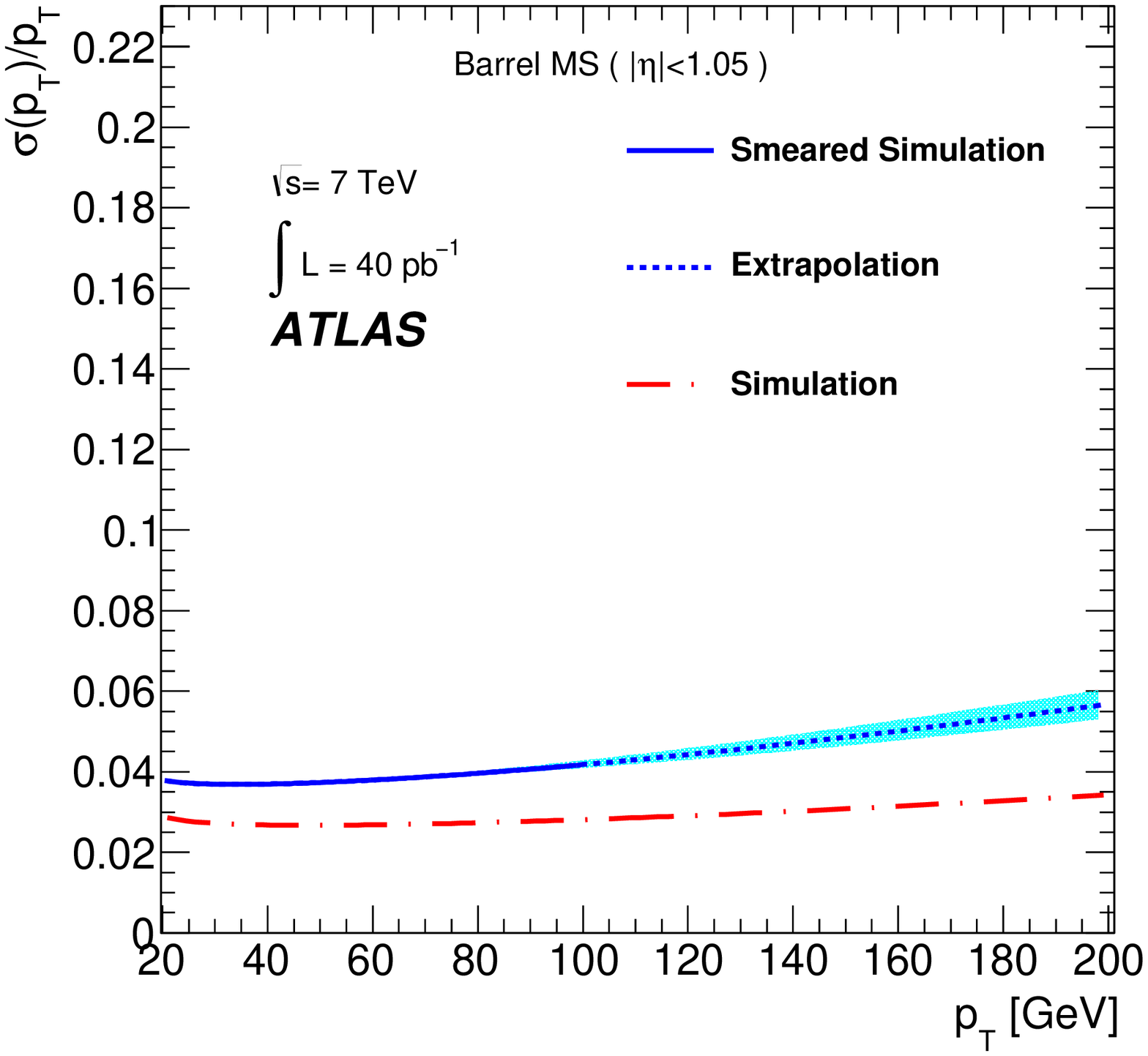}} \\
        {\includegraphics[width=0.49\textwidth]{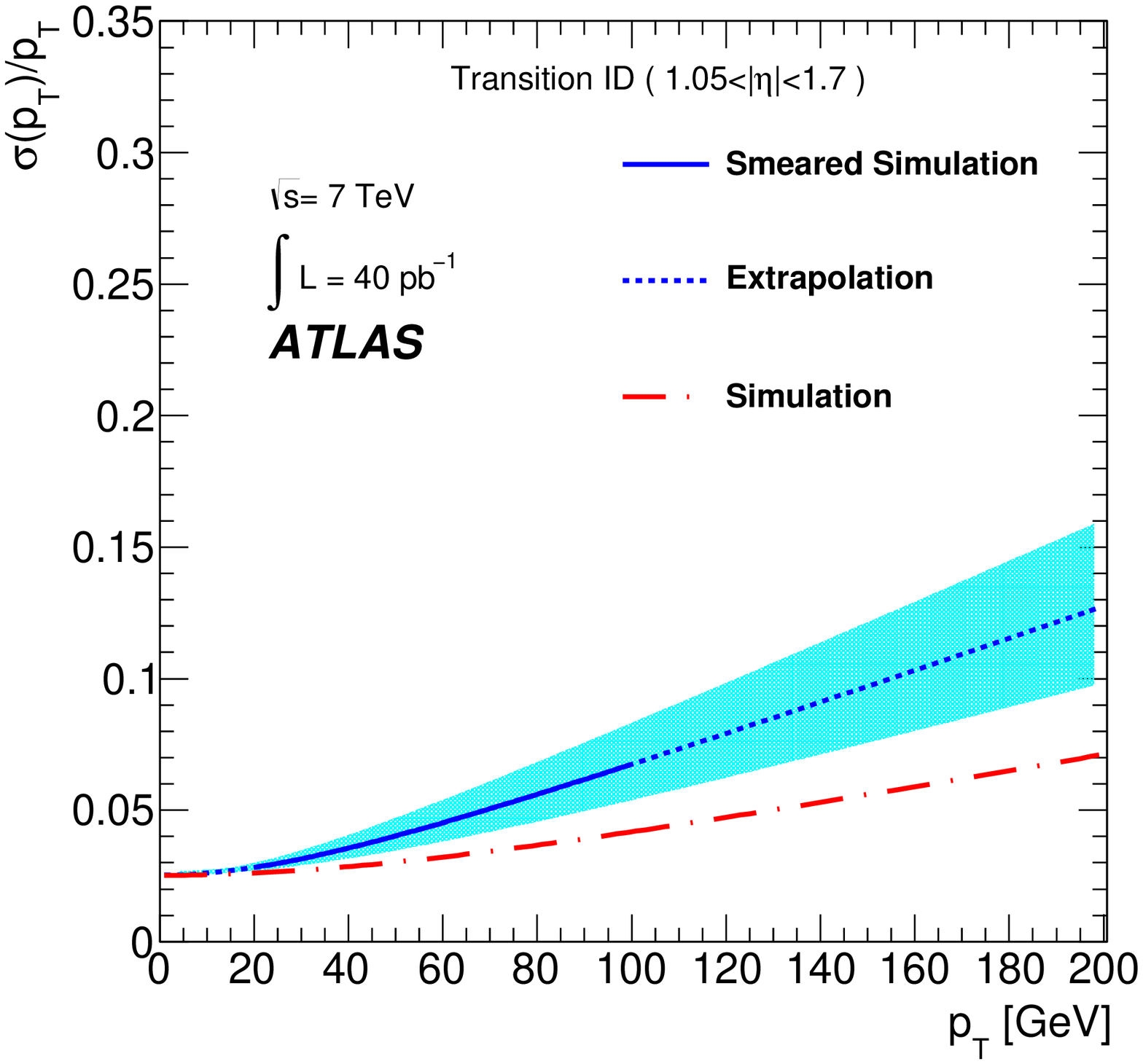}} &
        {\includegraphics[width=0.49\textwidth]{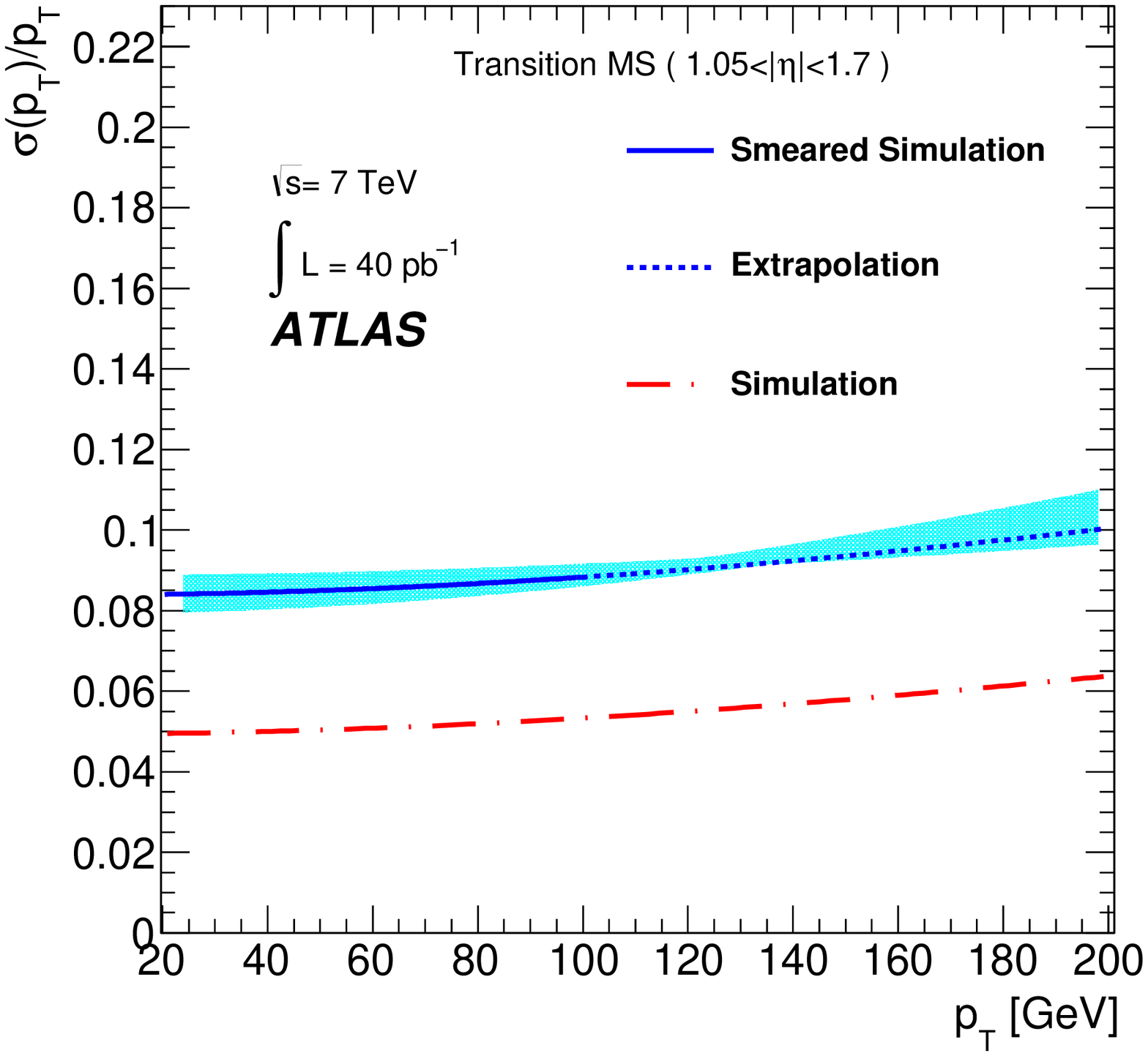}} \\
        \end{tabular}
        
\end{center}
\caption{Muon momentum resolution as a function of $\pt$ for different barrel and transition $|\eta|$ regions as denoted in the legend. The dot-dash line is from a  simulation which assumes perfect alignment of the ATLAS detector, whilst the  solid/dotted line shows simulation smeared to reproduce the invariant mass resolution measured in data. The solid section of the line shows the {\pT} range measured by $Z$ and $W$ decays, and the dotted section the `extrapolation' regions. The shaded bands show the uncertainty of the curves, computed from the uncertainties of the parameters derived in the resolution functions shown in \myEqs~(\ref{sec::momres::eqn1}) and (\ref{sec::momres::eqn2}).}
\label{sec::momres:fig3}
\end{figure*}%
\begin{figure*}[t]
\begin{center}
    \begin{tabular}{@{\hspace*{-0.3ex}}cc}
    {\includegraphics[width=0.49\textwidth]{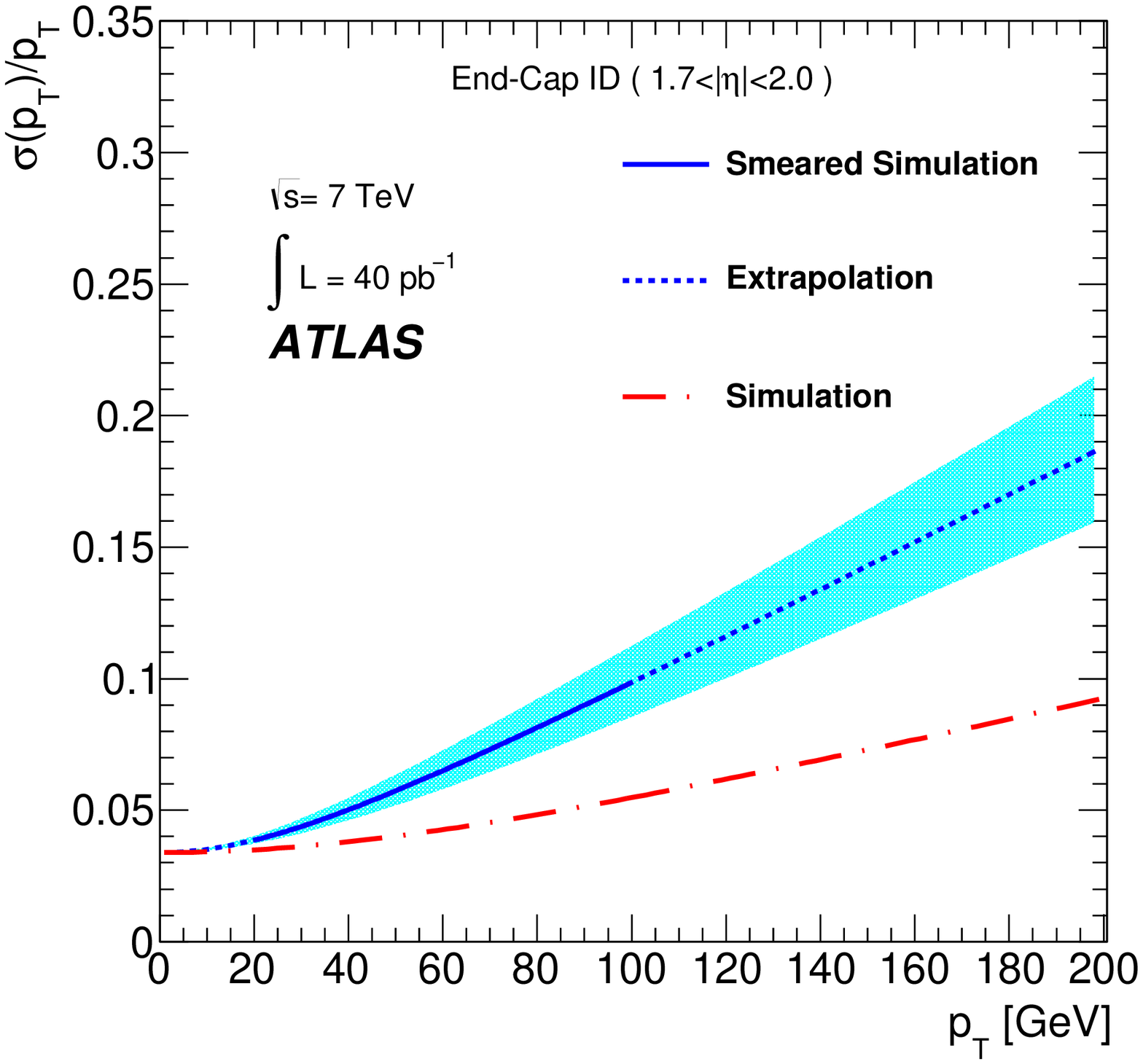}} &
    {\includegraphics[width=0.49\textwidth]{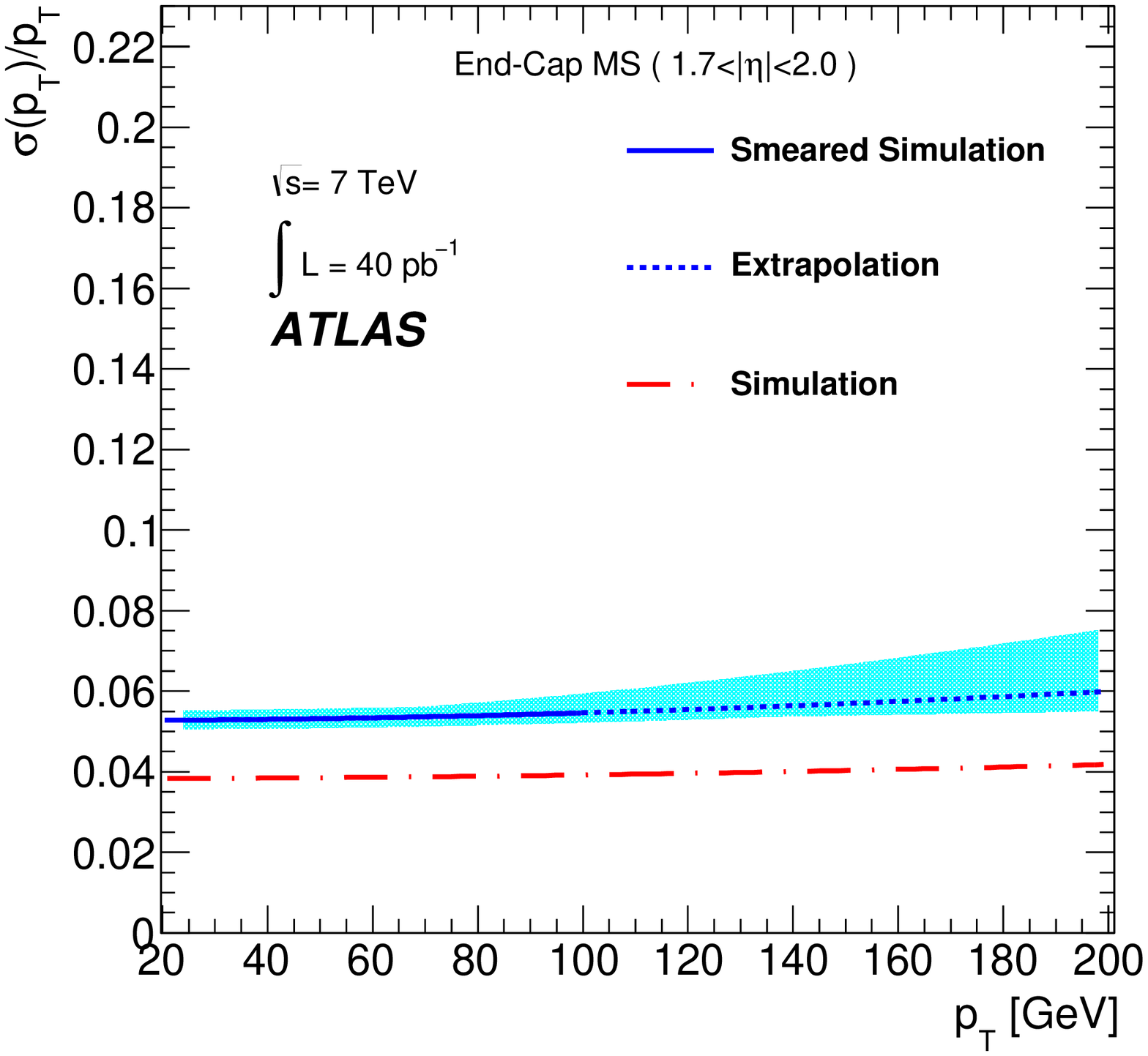}} \\ 
    {\includegraphics[width=0.49\textwidth]{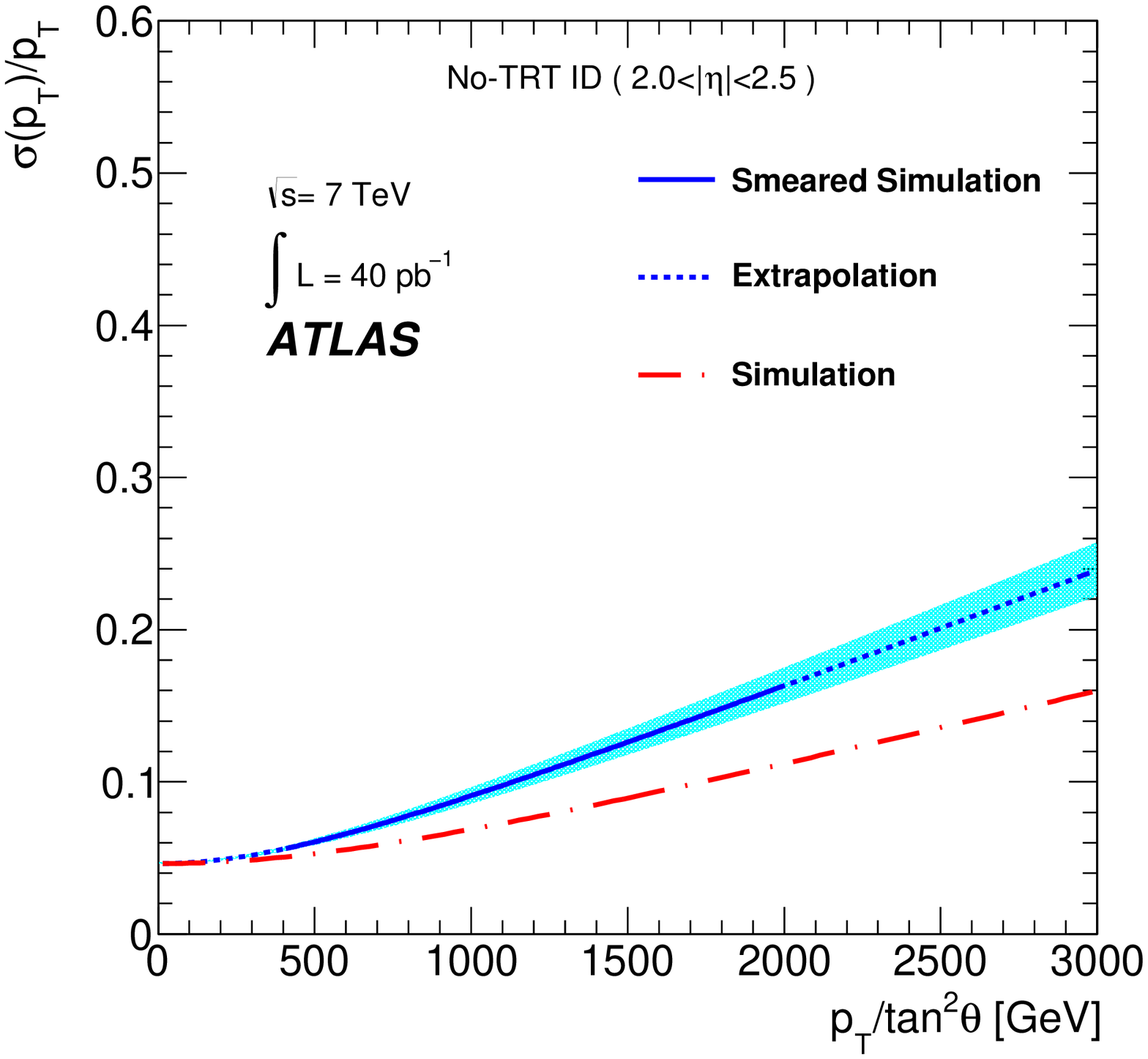}}  &
    {\includegraphics[width=0.49\textwidth]{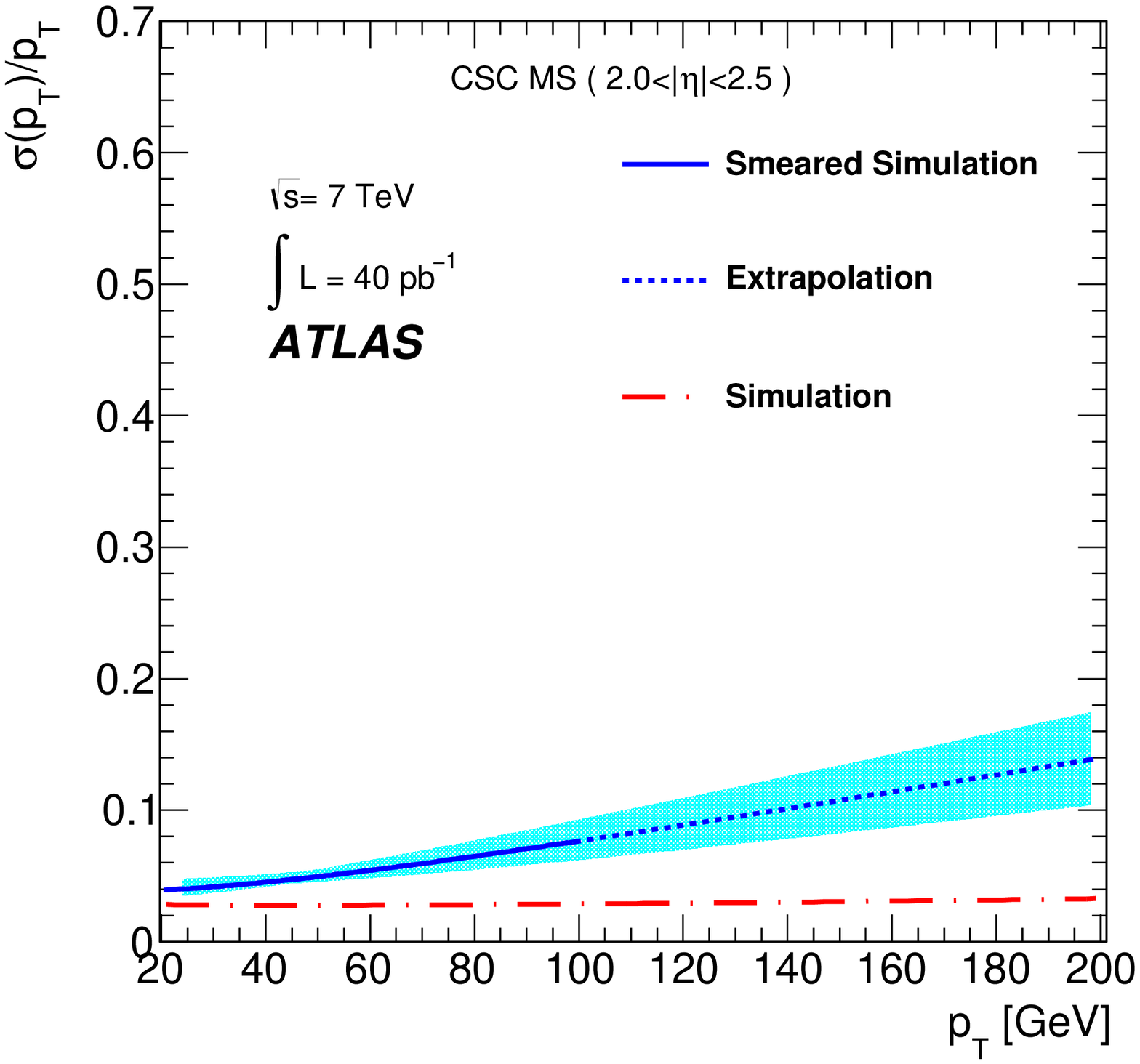}} \\
\end{tabular}
\end{center}
\caption{Muon momentum resolutions as a function of $\pt$ for different
         end-cap $|\eta|$ regions as denoted in the legend. For the ID region $|\eta|>2.0$ (bottom-left), the best parameterisation of the resolution depends on $\pt/\mathrm{tan}^2(\theta)$ instead of \pT. The dot-dash line is simulation which assumes perfect alignment of the ATLAS detector, whilst the solid/dotted line shows simulation smeared to reproduce the invariant mass resolution measured in data. The solid section of the line shows the {\pT} range measured by $Z$ decays, and the dotted section the `extrapolation' regions. The shaded bands show the uncertainty of the curves, which are computed from the uncertainties of the parameters in the resolution functions shown in \myEqs~(\ref{sec::momres::eqn1}) and (\ref{sec::momres::eqn2}).}
\label{sec::momres:fig4}
\end{figure*}%
The measured dimuon invariant mass resolutions can be translated into muon momentum resolutions. This was done by smearing the generated muon momenta, according to \myEqs~(\ref{sec::momres::eqn1}) and (\ref{sec::momres::eqn2}), by the amounts necessary to reproduce the measured dimuon invariant mass resolutions. Only the parameters $b_\mathrm{ID}(\eta)$ and $a_\mathrm{MS}(\eta,\phi)$ were varied during this procedure. The parameter $a_\mathrm{ID}(\eta)$ was set to the Monte Carlo prediction and varied within its uncertainty (see above) to evaluate the impact on the result for $b_\mathrm{ID}(\eta)$. The parameter $b_\mathrm{MS}(\eta,\phi)$ was set to the value derived from the special straight-track data set while $c(\eta,\phi)$ was set to its predicted value. In order to gain additional sensitivity to the momentum resolutions of the ID and MS, in addition to the dimuon mass spectrum of $Z$ boson decays, the distributions of the differences between the ID and
SA momenta of muons from $W\!\to\!\mu\nu_\mu$ decays were compared between the experimental and smeared MC data. The $W$ boson selection and the
MC samples for the analysis are the same as in Ref.~\cite{WZCrossSectionPaper}. As the use of $W$ boson decays correlates the SA and ID resolutions, those are extracted simultaneously in the fit.
The results are displayed for the different detector regions in \myFigs~\ref{sec::momres:fig3} and \ref{sec::momres:fig4}, with the uncertainty of the curves computed from the uncertainties of the parameters in the resolution functions (\myEqs~(\ref{sec::momres::eqn1}) and (\ref{sec::momres::eqn2})). Also shown is the expected resolution beyond the region in {\pt} probed by the $Z$-boson decays. As discussed earlier, the momentum resolution in experimental data is worse than in the Monte Carlo simulation, which is attributed, in part, to the residual misalignments of the ID and MS.

\section{Summary \label{sec::summary}}
The ATLAS muon reconstruction efficiencies were studied with $J/\psi\!\to\!\mu^+\mu^-$ and $Z\!\to\!\mu^+\mu^-$ decays using 40~pb$^{-1}$ of $\sqrt{s}=7$\,TeV $pp$ LHC collision data recorded in 2010.  
Samples of $\jpsi$ and $Z$ decays were used to access the transverse momentum regions of 
$\pt<10$~GeV and $20$~GeV${}<\pt<100$~GeV respectively. The muon reconstruction efficiency is found to be $>96\%$ and agrees with the MC prediction to better than 
1\%. The reconstructed quantities used to ensure muon isolation are 
shown to be well modelled in Monte Carlo simulations, and the corresponding muon isolation efficiencies are in excellent agreement with the MC predictions.

The muon momentum resolutions for $\pt>20$~GeV are derived from the dimuon mass resolutions in $Z\!\to\!\mu^+\mu^-$ decays and from the differences between the ID and SA momenta of muons from $W\!\to\!\mu\nu_\mu$ decays.
The resolutions are worse in data than in simulation for the entire momentum range considered. For instance, at $\pt\approx30$~GeV and $1.7<|\eta|<2.0$  the resolutions in experimental data are found to be about 30\% worse than predicted by the simulation. These differences are attributed to mis-modelling of the magnetic field and residual misalignments of the inner detector and muon spectrometer. An improved magnetic field map was used from 2011 onwards, and there have since been several iterations of the alignment. 

%
\section*{Acknowledgements}

We thank CERN for the very successful operation of the LHC, as well as the
support staff from our institutions without whom ATLAS could not be
operated efficiently.

We acknowledge the support of ANPCy\kern-0.1em{}T, Argentina; YerPhI, Armenia; ARC,
Australia; BMWF and FWF, Austria; ANAS, Azerbaijan; SSTC, Belarus; CNPq and FAPESP,
Brazil; NSERC, NRC and CFI, Canada;\break
CERN; CONICYT, Chile; CAS, MOST and NSFC,
Chi\-na; COLCIENCIAS, Colombia; MSMT CR, MPO CR and VSC CR, Czech Republic;
DNRF, DNSRC and Lundbeck Foundation, Denmark; EPLANET, ERC and NSRF, European Union;
IN2P3-CNRS, CEA-DSM/IR\-FU, Fra\-nce; GNSF, Georgia; BMBF, DFG, HGF, MPG and AvH
Foundation, Germany; GSRT and NSRF,\break
Greece; ISF, MINERVA, GIF, I-CORE and Benoziyo Center,
Israel; INFN, Italy; MEXT and JSPS, Japan; CNRST, Morocco; FOM and NWO,
Netherlands; BRF and RCN, Norway; MNiSW and NCN, Poland; GRICES and FCT, Portugal; MNE/IFA, Romania; MES of Russia and ROS\-ATOM, Russian Federation; JINR; MSTD,
Serbia; MSSR, Slovakia; ARRS and MIZ\v{S}, Slovenia; DST/NRF, South Africa;
MINECO, Spain; SRC and Wallenberg Foundation, Sweden; SER, SNSF and Cantons of
Bern and Geneva, Switzerland; NSC, Taiwan; TAEK, Turkey; STFC, the Royal
Society and Leverhulme Trust, United Kingdom; DOE and NSF, United States of
America.

The crucial computing support from all WLCG partners is acknowledged
gratefully, in particular from\break
CERN and the ATLAS Tier-1 facilities at
TRIUMF (Canada), NDGF (Denmark, Norway, Sweden), CC-IN2P3 (France),
KIT/GridKA (Germany), INFN-CNAF (Italy), NL-T1 (Netherlands), PIC (Spain),
ASGC (Taiwan), RAL (UK) and BNL (USA) and in the Tier-2 facilities
worldwide.

\FloatBarrier\newpage
\bibliographystyle{atlasnote}
\bibliography{muon_paper_2010}

\providecommand{\href}[2]{#2}\begingroup\raggedright\begin{thebibliography}{10}

\bibitem{CSCBook}
{ATLAS} Collaboration, {\em {Expected Performance of the ATLAS Experiment -
  Detector, Trigger and Physics}\/},
\href{http://arxiv.org/abs/0901.0512}{{\tt arXiv:0901.0512 [hep-ex]}}.

\bibitem{AtlasLumi2013}
{ATLAS Collaboration}, {\em {Improved luminosity determination in pp collisions
  at sqrt(s) = 7 TeV using the ATLAS detector at the LHC}\/},
  \href{http://dx.doi.org/10.1140/epjc/s10052-013-2518-3}{Eur. Phys. J. {\bf
  C73} (2013)  2518},
\href{http://arxiv.org/abs/1302.4393}{{\tt arXiv:1302.4393 [hep-ex]}}.

\bibitem{AtlasDetectorPaper}
{ATLAS} Collaboration, {\em {The ATLAS Experiment at the CERN Large Hadron
  Collider}\/},
\href{http://dx.doi.org/10.1088/1748-0221/3/08/S08003}{JINST {\bf 3} (2008)
  S08003}.

\bibitem{MuonEndcapAlignment}
{ATLAS} Collaboration, {\em {The Optical Alignment System of the ATLAS Muon
  Spectrometer Endcaps}\/},
\href{http://dx.doi.org/10.1088/1748-0221/3/11/P11005}{JINST {\bf 3} (2008)
  P11005}.

\bibitem{JPsiObservation}
{ATLAS} Collaboration, {\em {Measurement of the differential cross-sections of
  inclusive, prompt and non-prompt $J/\Psi$ production in p--p collisions at
  $\sqrt{s}=7$~TeV}\/},
  \href{http://dx.doi.org/10.1016/j.nuclphysb.2011.05.015}{Nucl. Phys. B {\bf
  850} (2011)  437}, \href{http://arxiv.org/abs/1104.3038}{{\tt arXiv:1104.3038
  [hep-ex]}}.

\bibitem{Pythia}
T.~Sjostrand, S.~Mrenna, and P.~Z. Skands, {\em {PYTHIA 6.4 Physics and
  Manual}\/},  \href{http://dx.doi.org/10.1088/1126-6708/2006/05/026}{JHEP {\bf
  05} (2006)  026}.

\bibitem{:2010wqa}
{ATLAS} Collaboration, {\em {The ATLAS Simulation Infrastructure}\/},
  \href{http://dx.doi.org/10.1140/epjc/s10052-010-1429-9}{Eur. Phys. J. {\bf
  C70} (2010)  823--874},
\href{http://arxiv.org/abs/1005.4568}{{\tt arXiv:1005.4568 [hep-ex]}}.

\bibitem{Geant42003}
{Geant4} Collaboration, S.~Agostinelli et al., {\em {Geant4: A simulation
  toolkit}\/},
\href{http://dx.doi.org/10.1016/S0168-9002(03)01368-8}{Nucl. Instrum. Meth.
  {\bf A506} (2003)  250}.

\bibitem{GEANT42006}
J.~Allison et al., {\em Geant4 developments and applications\/},
  \href{http://dx.doi.org/10.1109/TNS.2006.869826}{IEEE Trans. Nucl. Sci. {\bf
  53} (2006) no.~1, 270}.

\bibitem{WZCrossSectionPaper}
{ATLAS} Collaboration, {\em Measurement of the inclusive
  ${W}^{\ifmmode\pm\else\textpm\fi{}}$ and $Z/\gamma^{*}$ cross sections in the
  $e$ and $\mu$ decay channels in $pp$ collisions at
  $\sqrt{s}=7\text{\,}\text{\,}\mathrm{TeV}$ with the ATLAS detector\/},
  \href{http://dx.doi.org/10.1103/PhysRevD.85.072004}{Phys. Rev. {\bf D85}
  (2012)  072004}, \href{http://arxiv.org/abs/1109.5141}{{\tt arXiv:1109.5141
  [hep-ex]}}.

\bibitem{ATLAS-CONF-2010-019}
{ATLAS} Collaboration, {\em Study of the Material Budget in the ATLAS Inner
  Detector with $K_S^0$ decays in collision data at $\sqrt{s}=900$~GeV\/},
\newblock ATLAS-CONF-2010-019, 2010.
\newblock \url{http://cdsweb.cern.ch/record/1277651}.

\bibitem{ATLAS-CONF-2010-058}
{ATLAS} Collaboration, {\em Mapping the material in the ATLAS Inner Detector
  using secondary hadronic interactions in 7 TeV collisions\/},
\newblock ATLAS-CONF-2010-058, 2010.
\newblock \url{http://cdsweb.cern.ch/record/1281331}.

\bibitem{ATLAS-CONF-2012-141}
{ATLAS} Collaboration, {\em Study of alignment-related systematic effects on
  the ATLAS Inner Detector tracking\/},
\newblock ATLAS-CONF-2012-141, 2012.
\newblock \url{https://cds.cern.ch/record/1483518}.

\end{thebibliography}\endgroup

\onecolumn
\clearpage
\begin{flushleft}
{\Large The ATLAS Collaboration}

\bigskip

G.~Aad$^{\rm 47}$,
T.~Abajyan$^{\rm 20}$,
B.~Abbott$^{\rm 110}$,
J.~Abdallah$^{\rm 11}$,
S.~Abdel~Khalek$^{\rm 114}$,
A.A.~Abdelalim$^{\rm 48}$,
O.~Abdinov$^{\rm 10}$,
R.~Aben$^{\rm 104}$,
B.~Abi$^{\rm 111}$,
M.~Abolins$^{\rm 87}$
% ...
\bigskip
\\
% ...
$^{10}$ Institute of Physics, Azerbaijan Academy of Sciences, Baku, Azerbaijan\\
$^{11}$ Institut de F{\'\i}sica d'Altes Energies and Departament de F{\'\i}sica de la Universitat Aut{\`o}noma de Barcelona and ICREA, Barcelona, Spain\\
% ...
$^{20}$ Physikalisches Institut, University of Bonn, Bonn, Germany\\
% ...
$^{47}$ Fakult{\"a}t f{\"u}r Mathematik und Physik, Albert-Ludwigs-Universit{\"a}t, Freiburg, Germany\\
$^{48}$ Section de Physique, Universit{\'e} de Gen{\`e}ve, Geneva, Switzerland\\
% ...
$^{87}$ Department of Physics and Astronomy, Michigan State University, East Lansing MI, United States of America\\
% ...
$^{104}$ Nikhef National Institute for Subatomic Physics and University of Amsterdam, Amsterdam, Netherlands\\
% ...
$^{110}$ Homer L. Dodge Department of Physics and Astronomy, University of Oklahoma, Norman OK, United States of America\\
$^{111}$ Department of Physics, Oklahoma State University, Stillwater OK, United States of America\\
% ...
$^{114}$ LAL, Universit{\'e} Paris-Sud and CNRS/IN2P3, Orsay, France\\
% ...
\end{flushleft}

\end{document}